\documentclass{aa}
\usepackage{graphicx}
\usepackage{txfonts}
\usepackage{natbib}
\usepackage{rotating}
\usepackage{color}
\usepackage{longtable,lscape}
\usepackage{caption}
\DeclareCaptionLabelFormat{cont}{#1 ̃#2\alph{ContinuedFloat}}
\newcommand{\lxtx}{L_{\mathrm X}-T_{\mathrm X}}
\newcommand{\chandra}{{\sl Chandra}}

\newcommand{\wav}{{\tt wavdetect}}
\newcommand{\xspec}{{\it XSPEC}}

\begin{document}
\title{The Swift X-ray Telescope Cluster Survey. \\
II.  X-ray spectral analysis\thanks{Tables 1 and 2 and Appendix A are available in electronic form at http://www.anda.org}$^,$\thanks{Catalog and data products of SWXCS, constantly updated, are made available to the public through the websites http://www.arcetri.astro.it/SWXCS/ and http://swxcs.ustc.edu.cn/
}}

\subtitle{}


\author{P. Tozzi\inst{1},  A. Moretti\inst{2}, E. Tundo\inst{1},
  T. Liu\inst{3}, P. Rosati\inst{4}, S. Borgani\inst{5,6,7}, G. Tagliaferri\inst{8}, S. Campana\inst{8}, D. Fugazza\inst{8} , P. D'Avanzo\inst{8}}


\institute{INAF, Osservatorio Astrofisico di
  Firenze, Largo Enrico Fermi 5, I-50125, Firenze, Italy \and INAF, Osservatorio Astronomico di Brera, via Brera 28, I-20121 Milano, Italy \and 
  USTC, No.96, JinZhai Road Baohe District, Hefei, Anhui, 230026, P.R.China
  \and Universit\`a degli Studi di Ferrara, Dipartimento di Fisica e Scienze della Terra, Via Saragat 1 I-44121 Ferrara. Italy
  \and  INAF, Osservatorio Astronomico di
  Trieste, via G. B.  Tiepolo 11, I--34143, Trieste, Italy  \and
  Dipartimento di Fisica dell'Universit\`a di Trieste, via
  G.B. Tiepolo 11, I--34131, Trieste, Italy\and  INFN-National
  Institute for Nuclear Physics, via Valerio 2, I--34127, Trieste \and  INAF, Osservatorio Astronomico di Brera, via Bianchi 46, I-23807, Merate (LC), Italy }

\date{Received 9 October 2013 / Accepted 16 May 2014}


\abstract
{} 
{We present a spectral analysis of a new, flux-limited sample of 72 X-ray selected clusters of galaxies identified with the X-ray
Telescope (XRT) on board the Swift satellite down to a flux limit of $\sim 10^{-14}$ erg s$^{-1}$ cm$^{-2}$ (SWXCS).  
We carry out a detailed X-ray spectral analysis with the twofold aim of measuring redshifts and characterizing the properties of the 
Intra-Cluster Medium (ICM) for the majority of the SWXCS sources.}
{Optical counterparts and spectroscopic or photometric redshifts for some of the sources are obtained with a cross-correlation with the NASA/IPAC
Extragalactic Database.  Additional photometric redshifts are computed with a dedicated follow-up program with the Telescopio 
Nazionale Galileo and a cross-correlation with the Sloan Digital Sky Survey.  In addition, we also blindly search for the 
Hydrogen-like and He-like iron $K_\alpha$ emission line complex in the X-ray spectrum.  We detect the iron emission lines 
in 35\% of the sample, and hence obtain a robust measure of the X-ray redshift $z_X$ with typical rms error $1 - 5$\%.   We use $z_X$ whenever
the optical redshift is not available.  Finally, for all the sources with measured redshift, background-subtracted spectra are fitted with a single-temperature {{\tt mekal}} model 
to measure global temperature, X-ray luminosity and iron abundance of the ICM.  We perform extensive spectral simulations to accounts for fitting bias, and to assess the 
robustness of our results.  We derive a criterion to select reliable best-fit models and an empirical formula to account for fitting bias.  The bias-corrected values are then 
used to investigate the scaling properties of the X-ray observables. }  
{Overall, we are able to characterize the ICM of 46 sources with redshifts (64\% of the sample).  The sample is mostly constituted by clusters with 
temperatures between 3 and 10 keV, plus 14 low-mass clusters and groups with  temperatures below 3 keV.  The 
redshift distribution peaks around $z\sim 0.25$ and extends up to $z\sim 1$, with 60\% of the sample at $0.1<z<0.4$.  
We derive the Luminosity-Temperature relation for these 46 sources, finding good agreement with previous studies.}  
{{Thanks to the good X-ray spectral quality and the low background of Swift/XRT, we are able to measure 
ICM temperatures and X-ray luminosities for the 46 sources with redshifts.  Once redshifts are available for the remaining 26 sources, this sample will constitute a well-characterized, 
flux-limited catalog of clusters distributed over a broad redshift range ($0.1\leq z \leq 1.0$) providing a statistically complete view of the cluster 
population with a selection function that allows a proper treatment of any measurement bias.   The quality of the SWXCS sample is comparable to other samples 
available in the literature and obtained with much larger X-ray telescopes.  Our results have interesting implications
for the design of future X-ray survey telescopes, characterized by good-quality PSF over the entire field of view and low background.}}

\keywords{galaxies: clusters: general -- galaxies: high-redshift --  cosmology: observations -- X-ray: galaxies: clusters -- intergalactic medium}

\titlerunning{The {\sl Swift} X-ray Telescope Cluster Survey. II.}
\authorrunning{Tozzi et al.}  

\maketitle

\section{Introduction}



Clusters of galaxies are at the crossroads of astrophysics and cosmology \citep[see][]{Kravtsov2012}.  
Their baryon content is mostly in the form of a hot, diffuse plasma, 
the intracluster medium (ICM), which emits in the classic X-ray band $0.5-10$ keV.  In the last decade, observations with {\it Chandra} and XMM-Newton satellites revealed a wealth of complex phenomena going on in the ICM \citep[see][]{Markevitch2007,McNamara2007}, and allowed to trace its chemical and thermodynamical evolution up to redshift $z\sim 1.3$ \citep[see][]{2004Ettori,2007Balestra,2008Maughan,2009Anderson}.  Despite the rich physics involved, the hydrostatic equilibrium is typically satisfied and robust mass proxies can be obtained by X-ray observations \citep[see][]{kravtsov2006}.  This information, coupled with a well-defined selection function, allows one to compute the mass function of galaxy clusters over a broad range of redshifts.  Ultimately, one can use X-ray selected cluster samples to constrain the cosmological parameters and the spectrum of the primordial density fluctuations \citep[see][]{2002Rosati,2005Schuecker,2005Voit,2008Borgani,2009Vikhlinin,2010Mantz,2010Bohringer,2011Allen}.  
We note, however, that the main contribution of {\it Chandra} and XMM-Newton consists in
providing the detailed properties of clusters already identified in previous X-ray missions.  

The state--of--the--art of cosmological tests with clusters \citep{2009Vikhlinin,2011Allen} has been obtained thanks to deep follow-up
observations with {\it Chandra} of targets discovered in previous ROSAT wide-angle surveys or in ROSAT serendipitous fields.  
A key aspect is that the three major X-ray facilities existing today ({\sl Chandra}, XMM--Newton and Suzaku) 
have a limited field of view ($\sim 0.1-0.2 $ deg$^2$) and are mostly used in deep, targeted observations, 
which better fit their instrumental properties.  In the case of {\sl Chandra}, the process of assembling a wide
and deep survey is slow due to the small field of view (FOV) and the low collecting area.  On the other hand, its exquisite angular resolution 
($\sim 1$ arcsec at the aimpoint) makes it extremely easy to identify extended sources.  In the case of XMM-Newton the collecting area is
significantly higher, but the identification of extended sources, in particular at medium and high redshift, is more difficult due to the
larger size of the Point Spread Function (PSF, whose half energy width is $15"$ at the aimpoint) and its rapid degradation as a function of the
off--axis angle.  In addition, the relatively high and unstable background may hamper the proper characterization of low surface-brightness 
sources.  Finally, the large PSF of the X-ray Telescope on board Suzaku ($\sim 2$ arcmin) makes it
unfit to the detection of groups and clusters, despite its low background.  An obvious strategy is to exploit the entire archive of these telescopes to identify
serendipitously new X-ray cluster candidates.  However, the discovery data of X-ray clusters, in general, 
are not sufficient to provide a reliable mass proxy, except for the brightest ones, and therefore a 
time--consuming follow-up is always needed in order to measure at least the global temperature from their ICM emission.  

Despite these difficulties, several cluster surveys based on {\sl Chandra} and XMM--Newton are currently ongoing.  
These surveys are based on the compilation of serendipitous medium and deep--exposure observations not associated to
previously known X-ray clusters, or assembled with dedicated contiguous observations \citep[][]{2011Adami,suhada2012}.  An updated list of surveys as of mid-2012 is
presented in Table 1 of \citet{2012Tundo}.  Some of these projects are aiming mostly at high redshift ($z>1$) clusters \citep[XDCP,][]{2011fassbender} 
thanks to the exploitation of the large number of archival extragalactic fields (i.e., avoiding the Galactic plane).  
Among these surveys, the largest solid angle covered with sparse fields is $\sim 400$ {\rm deg}~$^2$ \citep[XCS,][]{2011lloyd-davies}, and the maximum number of clusters candidates is $\sim 1000$.  However, the typical size of a complete and well defined cluster sample amounts at best to $\sim 100$ objects, and, as already mentioned, in most cases X-ray data are not sufficient to properly characterize them (i.e., to have a robust mass proxy).  In conclusion, the design of current large X-ray telescopes is optimized to obtain detailed images
and/or spectra of isolated sources to explore the deep X-ray sky, while they are not efficient as survey instruments.  Therefore, a substantially
different mission strategy is required for surveys.

Looking at the near future, the planned eROSITA satellite \citep{2010Predehl,2012Merloni}, 
will finally provide a long-awaited all-sky survey at a depth more than one order of magnitude
larger than the ROSAT All-Sky Survey.  However, due to its low angular resolution the eROSITA discovery space 
is limited for distant clusters.  In  particular, the X-ray telescope hits the confusion limit at fluxes 
$\sim 10^{-14}$ erg s$^{-1}$ cm$^{-2}$, and, as a consequence, the cluster selection function will rapidly drop at high redshifts.   In addition,  the
low effective area above 2 keV severely limits the characterization  of the ICM in medium and high temperature clusters \citep{2014Borm}. 

The urge of a wide and deep all-sky survey in the X-ray band for clusters is even stronger if we consider 
what is happening at other wavelengths.  In particular,  Sunyaev-Zeldovich (SZ) surveys of clusters are providing the first exciting results.  
Recent results from the South Pole Telescope Survey \citep{Reichardt2013} and the Atacama Cosmology
Project \citep{act2012} showed that SZ selection is efficient and can identify 
clusters up to high redshift.   Recently, the all-sky survey of the Planck satellite 
provided a cluster catalog of about 1200 SZ-selected clusters \citep[][]{PlanckSZ1}. Cosmological constraints from 
the SZ, all-sky Planck survey are found to depend mostly on possible systematic bias in the relation between the total mass and the SZ parameter $Y$  \citep[][]{PlanckSZ2}.  Indeed, the X-ray follow--up of SZ
clusters is crucial to fully exploit the cosmological relevance of SZ surveys, either for narrowing down the uncertainties
on the cluster mass, or to firmly evaluate purity and completeness of
the sample.   However, in the future, SZ cluster surveys are expected to dominate the field 
of  ICM physics and cosmological tests with clusters of galaxies.  
Only a survey-optimized mission like the proposed Wide Field X-ray Telescope \citep[WFXT,][]{2010Murray,2011Rosati} can provide a 
large number of new detections well below the $10^{-14}$ erg s$^{-1}$ cm$^{-2}$ flux limit, therefore 
entering the unexplored realm of distant X-ray clusters, and, at the same time, provide X-ray mass proxies and redshifts for a large number of them.  
We also stress that the combination of future SZ surveys and deep, wide-angle X-ray surveys is the unique way to take full advantage
of galaxy clusters as tracers of the growth of cosmic structures.  To summarize, a wide-field, sensitive, X-ray mission will be the 
only means to bring our view of the X-ray sky, in terms of solid angle and sensitivity, at the level of the surveys in other wavelengths as
planned in the next decade.

In this framework, we are undertaking a new cluster survey using an X-ray facility that was never exploited before for this goal: the X-ray Telescope (XRT) 
on board the Swift satellite \citep{Burrows2005}.  Despite its low collecting area (about
one fifth of that of {\it Chandra} at $1$ keV), XRT has characteristics which are optimal for X-ray cluster surveys: a low background, an almost constant PSF across the 
FOV \citep{2007Moretti} and spectral capability up to $7$ keV.  Moreover, almost all the Swift/XRT pointings can be used to build a serendipitous survey 
based on the follow-up fields centered on Gamma Ray Bursts (GRB) detected by Swift.  The {\sl Swift-XRT cluster survey} (SWXCS) has a sky coverage ranging from 
total 40 ${\rm deg}^2$ to 1 ${\rm deg}^2$ at a flux limit of about $10^{-14}$ erg s$^{-1}$ cm$^{-2}$. 
The first catalog, including 72 sources, has been published in \citet{2012Tundo}, hereafter Paper I.  The SWXCS has been realized using 336 fields with galactic latitude $|b|>20^\circ$ 
centered on GRBs present in the XRT archive as of April 2010.  
This catalog is already one of the deepest samples of X-ray selected clusters, with a well defined completeness criterion and a negligible contamination.   
The SWXCS sample is expected to grow by a factor of at least $\sim 3$ in the eventual analysis of the full Swift/XRT archive, which will be based on a newly developed detection strategy 
\citep[][]{2013Liu}.

In this Paper we perform the X-ray spectral analysis of the 72 sources in the SWXCS catalog of Paper I to measure their temperature, total
luminosity, metal abundance and redshift. The paper is organized as follows. In Section 2 we briefly recall the properties of the XRT instrument, the
X-ray data reduction procedure and the properties of the survey.  In Section 3 we collect the redshift for the sources identified in
the NASA Extragalactic Database (NED) and in the SLOAN Digital Sky Survey (SDSS), and present new photometric redshifts for some of
our sources from a dedicated optical follow-up with the Telescopio Nazionale Galileo (TNG).  In Section 4 we present and discuss our X-ray spectral
analysis for all the sources with redshift information.  In Section 5 we present the X-ray spectral analysis of a serendipitous {\it Chandra} observation of one
SWXCS source, providing a simple but revealing comparison of the two instruments.  Finally, in Section 6, we summarize our results and discuss them
in the context of present and future X-ray cluster studies.  We assume the 9-year WMAP cosmological parameter results: $H_0 = 69.7$ km/s/Mpc, $\Omega_m = 0.28$, $\Omega_\Lambda = 0.72$ 
\citep[see][]{2013Hinshaw}.  Our results are clearly unaffected by the most recent {\sl Planck} cosmological results \citep{2013Planck}.



\section{Instrument, data reduction and survey properties}

In this Section we briefly recall the properties of the XRT
instruments, the data reduction, and the main characteristics of the SWXCS sample.  For more
details we refer to Paper I.

\subsection{The X-ray Telescope}

The XRT is part of the scientific payload of the Swift satellite \citep{2004gehrels}, a mission dedicated to the study of gamma-ray
bursts (GRBs) and their afterglows operating since November 2004.  GRBs are detected and localized by the Burst Alert Telescope
\citep[BAT,][]{Barthelmy05}, in the $15-300$ keV energy band and followed-up at
X-ray energies ($0.3-10$ keV) by the X-Ray Telescope (XRT).  The XRT \citep{Burrows2005} is an X-ray CCD imaging spectrometer which utilizes
the third flight mirror module originally developed for the JET-X telescope \citep{Citterio94} to focus X-rays onto a
XMM-Newton/EPIC MOS CCD detector \citep{Burrows2005}.  The effective FOV of the system is $\sim 24$ arcmin. The PSF, similar to XMM-Newton, is
characterized by a half energy width (HEW) of $\sim 18$ arcsec at $1.5$ keV \citep{2007Moretti} and, most important, is almost flat across the
entire FOV, with a negligible dependence on the photon energy.  Finally, XRT has the lowest background not associated to
astronomical sources among the currently operating X-ray telescopes due to the low orbit and its short focal length.  
This aspect has been recently exploited to make an unprecedented measure of the residual hard ($2-10$ keV) X-ray background, 
thanks to a combination of deep Chandra and XRT data \citep{2012Moretti}.  Clearly, the low X-ray background has a strongly positive impact on 
the detection of extended sources at all fluxes.  In Figure 2 of \citet{2012Moretti} we show that the ratio signal/background for extended
sources is a factor $\sim 10$ better in Swift-XRT than in {\sl Chandra}.

\subsection{Data reduction and analysis}

In Paper I we considered the Swift/XRT archive from February
2005 to April 2010, including 336 fields in the corresponding GRB
positions, with Galactic latitude $|b|>$20$^\circ$, to avoid crowded
fields and strong Galactic absorption.  We followed a standard data
reduction procedure by means of the {\tt xrtpipeline} task of the
current release of the HEADAS software (version v6.8) with the most
updated calibration (CALDB version 20111031, Nov 2011).  Then, we
proceeded with a customized data reduction, aimed at optimizing our data
to the detection of extended sources.  The most important aspect of
this step is the reduction of the background thanks to the exclusion
of the intense GRB emission at the beginning of each observation,
after a careful investigation of the background light curves.

The identification of the X-ray sources is performed with the
standard algorithm \wav\ within the CIAO software used successfully on
\chandra\ and XMM-Newton images \citep{2011lloyd-davies}.  We run the
algorithm on the images obtained in the soft ($0.5-2$ keV) band.  This
step leads us to identify a total of $\sim 10^4$ sources in the 336
GRB fields.  Among them, we select extended sources with a simple
criterion based on the measured Half Power Radius (HPR) effectively
defined as the 50\% encircled energy radius within a box of $45\times
45$ arcsec, which includes 80\% of the flux of a point source.  The choice
of a restricted region is motivated by the need of sampling the
growth curve with a high signal-to-noise ratio (S/N).

Extensive simulations spanning all the parameter range found in the
survey (in particular source fluxes and background levels) have been
used to estimate the expected distribution of the HPR for point
sources.  The simulations allow us to identify a threshold values HPR$_{th}$ above which a source is
inconsistent with being unresolved at the 99\% level.  This criterion does not depend on the off--axis angle
$\theta$, but significantly depends on the image background.  
The expected number of spurious sources in the entire survey surviving this criterion is $\sim 5$.   
Finally, our sample is constituted by all the sources satisfying our criteria
with more than $100$ net counts in the soft band within the extraction radius 
$R_{ext}$ (defined by the circular region where the source surface brightness is larger
than the background level).  In addition, the low number of sources in the final list allows us to
perform a careful visual inspection of each source candidate, hence
significantly reducing the number of spurious detections.

The sky coverage ranges from 40 {\rm deg}$^2$ at the high-flux end, to $1$ {\rm deg}$^2$ at $10^{-14}$ erg s$^{-1}$ cm$^{-2}$ 
(the flux limit of the survey).  The sky coverage is comparable to that of RDCS by \citet{1998Rosati}, while it is lower but deeper than that of the 400sd survey 
 by \citet{2007Burenin}, as shown in Figure 13 of Paper I.  We also characterize
the completeness of the sample, defined as the fraction of extended
sources actually selected as extended by our procedure as a function
of their soft-band net counts. In order to do that, we produce another set
of simulations with a few thousand extended sources, modeled from
ten real cluster images originally obtained with the
{\sl Chandra} satellite (therefore with a resolution much higher that that
of XRT), cloned at a typical redshift and re sampled at the XRT
resolution. This cloning procedure, already used to investigate the
evolution of cool core clusters at high redshift
\citep[see][]{2008Santos, 2010Santos} allows us to measure the
completeness of the sample in the assumption that
the cluster images used for cloning are representative of the entire 
cluster population.  We find that the completeness is a function of the input counts, starting from 
$\sim$70\% for sources just above our threshold of $\sim$100 net counts
within the extraction regions. This level increases to $\sim$90\% for
sources with $\sim$200 net counts, and reaches a completeness level 
$>$95\% for sources with $\sim$300 net counts (see Figure 9 of  Paper I).  Given the distribution of exposure times 
and the observed logN-logS of groups and clusters, this implies that we may have missed, in total,
about 10 extended sources, mostly with less than 200 counts, above the nominal flux limits associated to each field.  
From simulations, it appears that the major source of incompleteness is due to the blending of bright, unresolved sources. 


\subsection{The Swift-X-ray Cluster Survey: the GRB fields}

Our final catalog consists of 72 X-ray extended sources detected with $\geq 100$ soft net counts. The measured net counts
from each source are computed by aperture photometry within the extraction radius $R_{ext}$, after removing the emission associated to
point sources included in this region.  The net count rate measured for each source is obtained dividing the net
counts by the exposure time, after correcting for vignetting effects in the soft band. From the corrected net count rate, the
energy flux is computed simply by multiplying it by the energy conversion factor (ECF) computed at the aimpoint, whose value for a
thermal spectrum is typically $ 2.35 \times 10^{-11}$ erg s$^{-1}$ cm$^{-2}$/ (cts s$^{-1}$), with a maximum systematic uncertainty of
$0.1 \times 10^{-11}$ erg s$^{-1}$ cm$^{-2}$/ (cts s$^{-1}$).  This value has a
weak dependence on the ICM temperature and the redshift, and a significant dependence on the Galactic absorption. 
The effect of Galactic absorption in each source position is taken into account assuming the
Galactic $N_H$ values measured in the radio survey in the Leiden/Argentine/Bonn survey \citep{2005LAB}.  Finally, the total net flux is
obtained by accounting for the missed contribution beyond $R_{ext}$. In order to measure the lost signal, we fit the surface brightness ({\it SB}) of every extended source with a King profile. Then, we extend
the surface brightness profile up to $2 \times R_{ext}$.  
The photometric properties of the SWXCS sources are published in
Paper I, along with a preliminary optical identification for some of the sources obtained
with a simple cross-correlation with the NED database.  In this paper,
the energy fluxes obtained fitting the X-ray spectra will supersede the
fluxes obtained with the ECFs. 

\section{Redshift from optical counterparts}

We collect the redshifts for our sources as follows.  First we retrieve the spectroscopic or photometric
redshift from the literature, using the NASA Extragalactic Database (NED); then we compute the photometric redshift for
12 sources imaged by the TNG with a dedicated program.  Eventually, we use
the optical redshift (both spectroscopic and photometric) to assess the robustness of the redshift $z_X$ measured
by the X-ray spectral analysis through the identification of the $K_\alpha$ iron line complex.  This allows us to assess the
reliability of  X-ray redshift measurements for sources without optical redshift.  In this section, we present and discuss the optical redshifts.

\subsection{Cross correlation with optical catalogs}

A preliminary identification with optical counterparts in NED has been performed in Paper I (see their Table 3).  Here we briefly describe the adopted procedure
and the results, updating in few cases the results of Paper I.

First we checked for counterparts in optical cluster surveys, assuming a search radius of 1 arcmin from our X-ray centroid.  The choice of 1 arcmin is motivated by the following facts.  
The X-ray centroid of SWXCS sources is measured with an accuracy of about $\leq 7$ arcsec, depending on the 
S/N of the source and the surface brightness distribution.  We expect that the X-ray emission is often peaked
close to the central galaxy, even in the absence of a cool core, while a significant displacement of the X-ray centroid 
with respect to the brightest galaxy (i.e., several tens of arcsec) is expected only in the rare occurrence of ongoing massive mergers.  We are also aware that  the position of optically selected clusters could be uncertain up to $1$ arcmin, or even more in the case of sparse optical clusters.  Given that $1$ arcmin corresponds to 200 kpc at the typical redshift of $0.2$, and considering the strong dependence of the galaxy density from radius in clusters, we argue that for the optical counterparts of our clusters, the uncertainty on the center is always below $1$ arcmin. However, in order to include particular cases such as loose groups or strongly sub-structured clusters, we also search for counterparts up to 3 arcmin from our X-ray centroid.  For all our source, we did not find any convincing candidates besides those closer than $1$ arcmin.  

We find a total of nineteen previously known, optically identified clusters with measured redshift, many of them listed in more than one catalog.  Among them, nine
are found in the Wen+Han+Liu cluster sample \citep[WHL,][]{2009Wen,2012Wen}, which is an optical catalog of galaxy clusters obtained from an
adaptive matched filter finder applied to Sloan Digital Sky Survey DR6.   We report five clusters from the Gaussian
Mixture Brightest Cluster Galaxy \citep[GMBCG,][]{2010Hao} based on SDSS DR7, which is an extension of the maxBCG cluster catalog
\citep{Koester2007} to redshift beyond $0.3$ and on a slightly larger sky area $\sim 8000$ {\rm deg}$^2$ (one of these clusters is also in the MaxBCG catalog).  
We also find four clusters in the Abell catalogs \citep{1989Abell}, among them Abell 2141 which has been already studied in detail 
by \citet{2011Moretti} using XRT data.  We also find one cluster in each of the following catalogs:
the Northern Sky optical Cluster Survey \citep[NSCS,][]{Gal2003,Lopes2004}, the Zwicky Cluster Catalog \citep[CGCG,][]{1963Zwicky},
the Sloan Digital Sky Survey C4 Cluster Catalog \citep[SDSS-C4-DR3, based on DR3,][]{2005SDSS-C4-DR3, Linden2007}, the
Edinburgh-Durham Southern Galaxy Catalogue \citep[EDCC,][]{1992EDCC}, and the  ESO/Uppsala Survey of the ESO(B) Atlas \citep{1982Lauberts}.
Finally, we find four clusters in the \citet{2011Szabo} catalog (AMF clusters) which is also based on an adaptive matched filter finder applied to Sloan 
Digital Sky Survey DR6, but it is not included in the NED.   One of the four AMF clusters is not included in any of the previous catalogs found in the NED.  
Note that we removed the optical identification of SWXCS J232248+0548.1
reported in Paper I with the NSCS cluster at $z_{phot} =0.45$, whose distance from the centroid is $\sim 1$ arcmin, since our
new photo-$z$ and $z_X$ both indicates a lower value (see Section 4).  

To increase the number of available redshifts, we also search for galaxies with published redshift not associated to previously known
clusters within a search radius of 7 arcsec from the X-ray centroid of our sources. We find 11 galaxies, whose  redshift is
always consistent with the photometric redshift of the optical cluster counterpart when present.  This finding shows that our matching criterion 
is efficient in identifying the central cluster galaxy.  Therefore, in the 4 cases where no optical cluster counterpart is found, we assign the galaxy redshift 
to our X-ray source.

To summarize, we have 23 optical redshifts (spectroscopic or photometric) published in the literature and associated to our cluster
candidates.  Whenever a photometric redshift and the spectroscopic redshift are both present, we use the
spectroscopic one. The redshifts obtained from the literature are reported in the second column of Table \ref{tab:zobs}.  This table 
is consistent with its preliminary version presented in Paper I except for the removed identification of SWXCS J232248+0548.1, and two 
identifications with Abell clusters previously missed.  

We also remark that we now use the new designation "SWXCS J" for our clusters, as opposed to 
"SWJ" used in Paper I. This new acronym has been officially  accepted by the IAU Registry of a new acronym in
February 2013.  The format is SWXCS JHHMMSS+DDMM.m.  For clarity, we add, in the last column of Table \ref{tab:zobs}, the name
used in Paper I.

\subsection{Redshift from optical follow-up with TNG}

For twelve clusters we obtained photometric redshifts with a dedicated program at the TNG (AOT 22, PI A. Moretti).
Our strategy consisted  in supplementing SDSS data with  NICS $J$  and DOLORES $r$ deep observations (20 minutes each filter)  
when  SDSS data do not provide a clear counterpart (4 cases, plus  1 redundant).  In the regions not covered by the SDSS we performed
observations in the  NICS $J$  and DOLORES $r$ g bands (7 cases).  In all cases we performed the photometry of the brightest galaxy
by means of SExtractor \citep{bertin96}.  We measured the photometric redshifts  by means of the  HYPERZ software \citep{Bolzonella2000}
using galaxy templates which are built with the spectrophotometric code of \citet{bruzual03}, assuming exponentially declining star formation histories with
time scales 0.5 Gyr, Chabrier IMF and metallicity $0.5$ solar.   We verified that varying the metallicity content and the age  
of the input template  has a negligible impact on the redshift. 
We typically obtain a 1 $\sigma$ error of about 10\% on the photometric redshift.  The photometric redshift obtained by our TNG follow-up are listed in the third column of Table \ref{tab:zobs}.  

We also explored the possibility of obtaining photometric redshifts from VLT imaging of the GRB fields.  We note that, although most of
the GRB have been followed up by VLT and other optical/IR telescopes, in most cases the X-ray centroids of our serendipitous sources
fall outside the optical field of view.  In only one case SWXCS J232345-3130.8, we are able to use the VLT optical/IR follow-up
observation of GRB051001.  We apply our photometric redshift measurement to this cluster which turns out to be the most distant in
our subsample ($z=0.85_{-0.03}^{+0.07}$).  

To summarize, we have thirteen photometric redshifts which complement the information obtained by a cross-correlation with NED.  One of them (SWXCS J084749+1331.7) already had a
consistent spectroscopic $z_{opt}$.  Therefore, a total of 35 sources, about half the total X-ray sample, have at least one optical measurement of the redshift.

\section{Redshift from X-ray spectral analysis}

The results from the X-ray spectral analysis will be provided only for the clusters with a robust redshift measurement, either from
the optical or from the X-ray data.  Therefore, first we blindly search for the iron $K_\alpha$ line complex in order to measure the X-ray redshift $z_X$ in 
all cases where the S/N ratio allows us to do so.  The measurement of X-ray redshift from {\it Chandra} data has been extensively explored by
\citet{2011Yu}.  However, in the case of XRT we have a significantly different situation, due to the harder effective area of XRT with respect to that of {\it Chandra}, and to the typically lower S/N of our sources.  Therefore, we cannot directly apply the criteria found in  \citet{2011Yu} to our sample. 

To exploit at best the information contained in our X-ray spectra, we proceed in two steps.  First, we fit blindly all our X-ray spectra with a {\tt mekal} model, leaving all relevant parameters, including the 
redshift, free to vary during the minimization procedure.  Then, we  compare the optical and X-ray redshifts for the sources which have both, to establish a new robust criterion to measure the  redshift. 
Finally, we apply this criterion to the other sources with no optical redshift to obtain reliable $z_X$ measurements. 

\begin{figure}
\includegraphics[width=9cm] {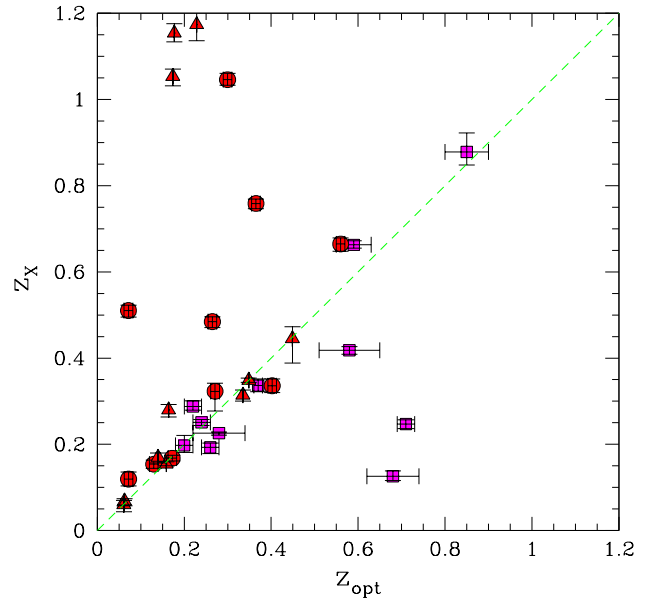}      
\caption{\label{zx_all}X-ray vs. optical redshift for the 35 sources with spectroscopic or photometric redshift (three sources do not appear
  since they have $z_X>1.2$).  Red symbols are for clusters whose redshift is found in the literature after inspecting the 
  NED database (circles and triangles are for photometric and spectroscopic redshifts, respectively).  Magenta squares are for clusters whose photometric redshift has been computed by us with a dedicated TNG
  program or with VLT archive images (one case).  Error bars on $z_X$ and $z_{opt}$ correspond to 1 $\sigma$.}
\end{figure}

\begin{figure}
\includegraphics[width=9cm]{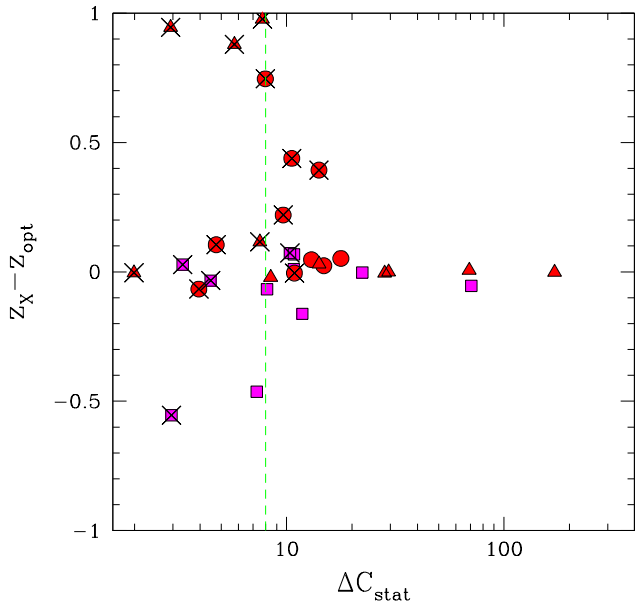}
\caption{Difference $z_X-z_{opt}$ for the 35 sources with spectroscopic or photometric redshift plotted as a function of the $\Delta C_{stat}$ value.  Red circles (triangles) are for clusters whose 
photometric (spectroscopic) redshift is found in the literature through a match with the NED, while the magenta squares are for clusters whose photometric redshift has been computed by us with a 
dedicated TNG program or with VLT archival image (in one case).  Crosses corresponds to $z_X$ values rejected after a visual inspection because a secondary minimum in $C_{stat}$ is found close to the 
minimum (see text for details).  Error bars on $z_X$ and $z_{opt}$ correspond to 1 $\sigma$.}
\label{dzx}
\end{figure}

\subsection{Method}

The spectra are extracted from circular regions with radius $R_{ext}$, defined as the radius where the surface brightness is equal to the background intensity.  Intervening point
sources are removed.  We sampled the background from the same observation, in a region typically three times larger by size than the extraction region of the source spectra.  Calibration files (rmf and
arf) are built for each extraction region.

The background subtracted spectra are analyzed with \xspec\ v.12.3.0 \citep{arn96}.  We use the {\sl C-statistic} to find the best-fit parameters for the adopted model \citep[see][]{bevington03,arnaud11}.  
The Cash statistic parameter $C_{stat}$ \citep{1979Cash} provides a better criterion with respect to the canonical $\chi^2$ analysis of binned data, particularly for low S/N spectra \citep{1989Nousek}. 
However, we first apply a {\tt group min 1} command in
{\tt GRPPHA}, to have at least 1 count per bin.  This step is necessary for \xspec\  to correctly calculate the {\sl C-statistic} (Arnaud, private communication; see also \citet{2009Evans}). 
All our sources are fitted with a single temperature {\tt mekal} model \citep{kaa92,lie95}.  The ratio between the elements are fixed to the solar values as in \citet{asp05}.  We model the Galactic absorption with {\tt tbabs} \citep{wilms00} fixing the Galactic neutral Hydrogen column density to the values measured in the Leiden/Argentine/Bonn radio survey \citep{2005LAB}.  

The fits were performed over the energy range 0.5-7.0 keV. We do not include photons with energies below 0.5 keV in order to avoid
uncertainties due to a rapidly increasing background below 0.5 keV. The cut at high energies, instead, is imposed by the rapidly
decreasing S/N, due to the combination of the lower effective area of XRT and of the exponential cut-off of the thermal spectra.
With this choice we also avoid the strong calibration lines present in the XRT background.

\subsection{Measurement of $z_X$}

In the first step, we focus only on the presence of the $K_\alpha$ iron line complex and therefore on the best--fit redshift $z_X$.  We fit the background-subtracted X-ray spectra assuming a {\tt mekal} model, with $N_H$ frozen to 
the Galactic value, while the redshift, metallicity, temperature and normalization parameters are left free.  Once the best fit is found, we freeze the metallicity and temperature parameters and vary the redshift
parameter only.  Then, we plot the {\tt $C_{stat}$} value as a function of the redshift and look for its minimum.  

In Figure \ref{zx_all} we plot the comparison of $z_X$ with $z_{opt}$ for the 35 sources with photometric or spectroscopic optical redshift.
While a large fraction of the $z_X$ values are consistent with the optical redshift, there is a significant number of catastrophic failures.
Therefore we need to apply a filter to select reliable measurements of $z_X$.  To do that, first we find the absolute
minimum in the parameter $C_{stat}$, then  we explore the redshift space with the {\tt steppar} command, covering the entire range of possible values from
$z=0$ to $z=2$ with a very small step $\delta z = 0.01$.  We then plot the difference $\Delta C_{stat}$ of the $C_{stat}$ value with respect to the minimum as a function of redshift.  In \cite{2011Yu} we investigated {\it Chandra} data to set a
first criterion on the $C-statistic$ value $\Delta C_{stat} >9$, corresponding to a confidence level of 3 $\sigma$ as tested {\sl a posteriori} with simulations.  However, we
also find that this criterion is not sufficient for sources with a low number of net detected counts.  In particular, below 1000 total net
counts (in the useful 0.5-7 keV band) the number of catastrophic failures rapidly increases, irrespective of $\Delta C_{stat}$.  Since most of the SWXCS sources have less than 1000 total net
counts, we should reject the large majority of $z_X$ measurements.  On the other hand, from Figure \ref{zx_all} we find that, given the large
fraction of the $z_X$ measurements found to be in good agreement with the optical one, we can relax the criteria in order to include most of the low S/N XRT spectra.
This is possible thanks to the larger ratio of the hard to soft effective area of XRT with respect to that of {\it Chandra}, as shown in Figure 1 in
Paper I.  This property implies that, for the same total number of net counts, a larger number of iron-line photons are found in XRT spectra than in {\it Chandra} spectra.  

\begin{figure}
\includegraphics[width=9cm] {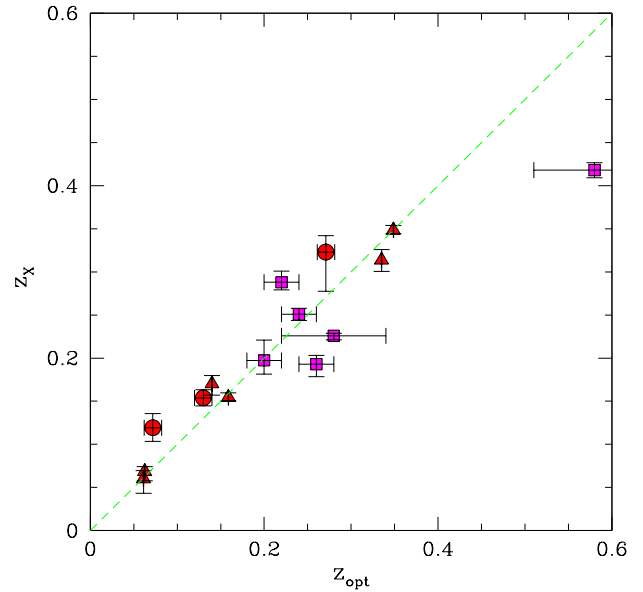}      
\caption{X-ray vs. optical redshift for the 15 sources with spectroscopic or photometric redshift and X-ray redshift satisfying the selection criteria.  Red circles (triangles) are for clusters whose 
photometric (spectroscopic) optical redshift is found in the literature through a match with the NED, while the magenta squares are for clusters whose photometric optical redshift has been computed by us with a 
dedicated TNG program or with VLT archival image (in one case).  Error bars on $z_X$ and $z_{opt}$ correspond to 1 $\sigma$.}
\label{zx_good}
\end{figure}

An obvious solution is simply to adopt a much larger $\Delta C_{stat}$ irrespective of the net counts.  However, we still find catastrophic failures with $\Delta C_{stat}$ as high as 
14.  This would strongly limit the number of new $z_X$ that we can measure for the remaining 37 sources with 
no optical redshift.  Therefore we consider a different strategy based on a visual inspection of the $\Delta C_{stat}$ plots.  We realize that additional
information can be provided by the secondary minima in the $\Delta C_{stat}$ vs. $z_X$ plots: whenever a significant number of secondary
minima with a depth close to the principal minimum is found, we reject the $z_X$ measurement.  We find that a good choice is to require
that the difference in $\Delta C_{stat}$ from the second deepest minimum must be larger than 2, irrespective of the $\Delta C_{stat}$
value of the absolute minimum.  In Figure \ref{dzx} we show the $\delta z = z_X-z_{opt}$ vs. $\Delta C_{stat}$, where we marked
with crosses the $z_X$ measurements which do not fulfill the requirement on the secondary minima.  We find that, after applying this
filter, we find a reasonable agreement between $z_X$ and $z_{opt}$ for $\Delta C_{stat}>8$, irrespective of the total net counts of the
source.  The $z_{opt}$-$z_X$ relation for the 15 spectra satisfying the selection criteria is finally shown in Figure \ref{zx_good}.  We note that
the discrepancy is often larger than expected on the basis of the formal 1 $\sigma$ error bars on $z_X$.  The largest discrepancy is for SWXCS J0926.0+3013.8 at $z_{opt} = 0.58\pm 0.07$, which has, 
however, the largest uncertainty on the optical redshift.  We postpone a more accurate investigation of the optimal criteria to measure $z_X$ to the final SWXCS catalog, 
when a larger source statistics will allow us to refine our strategy (Liu et al. in preparation).

\begin{figure}
\includegraphics[width=9cm] {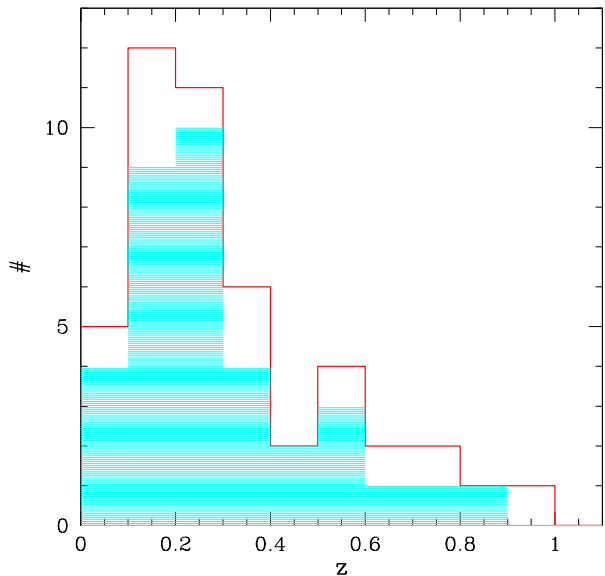}      
\caption{\label{fig:zhist}Redshift histogram of the 46 clusters with redshift values with the following priority:
i) optical spectroscopic redshift; ii) optical photometric redshift; iii) X-ray spectroscopic redshift. The shaded area corresponds to optical redshifts only
(both spectroscopic and photometric).}
\end{figure}

When we apply this criteria ($\Delta C_{stat} >8$ and no secondary minima within $\Delta C_{stat} =2$) to the rest of the sample, we are able to assign the redshift $z_X$ to 11 sources without optical redshift.
The 26 $z_X$ values satisfying our criteria are listed in the fourth column of Table \ref{tab:zobs}.  The total number of sources with redshift is
therefore 46.  We use the redshift values with the following priority: i) optical spectroscopic redshift; ii) optical photometric redshift; iii) X-ray spectroscopic redshift. The histogram distribution of the redshifts in our 
sample is shown in Figure \ref{fig:zhist}.  The distribution has a strong peak around $z\sim 0.2$, while about 1/4 of the sources are at $z>0.4$.
We note that X-ray redshift measurements contribute mostly at high redshift.  This is expected, since at higher redshift the Fe $K_\alpha$ line complex shift towards lower energies in the
observing frame, where the effective area of XRT is larger.  

%


\subsection{Redshift completeness}

\begin{figure}
\includegraphics[width=9cm] {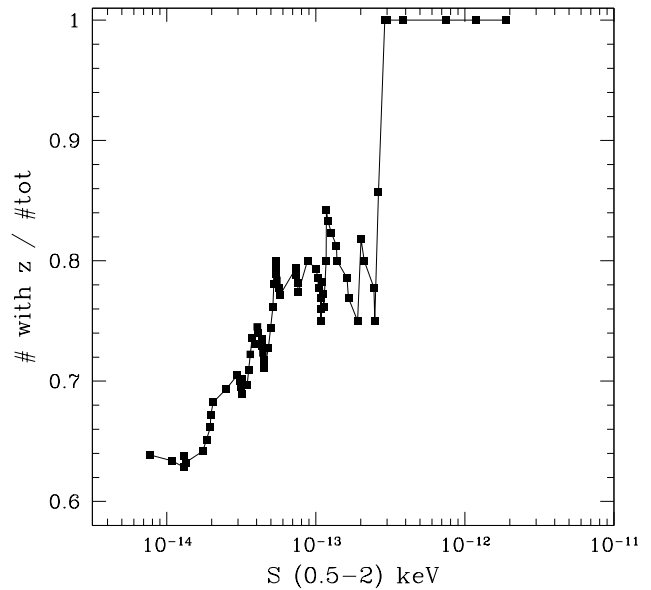}      
\caption{Redshift completeness (defined as the ratio of sources with redshift over the total number of sources above a given flux) as a function of the energy flux in the soft band. }
\label{completeness}
\end{figure}

We obtain the redshift information for 64\% of the sample.  Given the mixed criteria used to assign the redshift, we do not expect that the sources with redshift correspond 
to the bright end of the sample.  In addition, since our detection criteria is based on the soft net counts, and since the exposure times are distributed on a broad range, 
we find that several X-ray bright sources still do not have X-ray redshift measurement.  The redshift completeness, defined as the ratio of the number of sources with 
redshift to the total number of sources at fluxes larger than a given values, is shown in Figure \ref{completeness}.

We notice that the redshift completeness is not larger than 80\% for fluxes below $3 \times 10^{-13}$ erg s$^{-1}$ cm$^{-2}$.  This implies that if we want 
to extract a flux limited sample with a significant 
number of sources out of the total sample, either we are limited to few very bright sources, or we suffer a 20\% incompleteness \footnote{We also remind that this sample 
originally suffers from another source of incompleteness due to the X-ray selection process itself. The number of sources possibly missed by our X-ray selection has been 
estimated  with extensive simulations as described in Paper I, and it can reach about 15\% of the total sample (about ten sources). } Therefore, we conclude that 
a further effort to collect more redshifts must be done before this sample can be used to investigate the evolution of the X-ray properties, and, in the end, 
the mass function and the associated cosmological constraints.  In the rest of the paper we present and discuss the properties of the clusters with redshift.

\section{Spectral simulations  and robustness of spectral fits}

Before proceeding with the X-ray spectral analysis on the entire sample, and investigate scaling relations between observables, we thoroughly
explore possible systematic uncertainties in our best-fit temperature and abundance values associated to the fitting procedure. 
 This step is particularly relevant given the low S/N regime of our spectra.

\begin{figure}
\includegraphics[width=9cm] {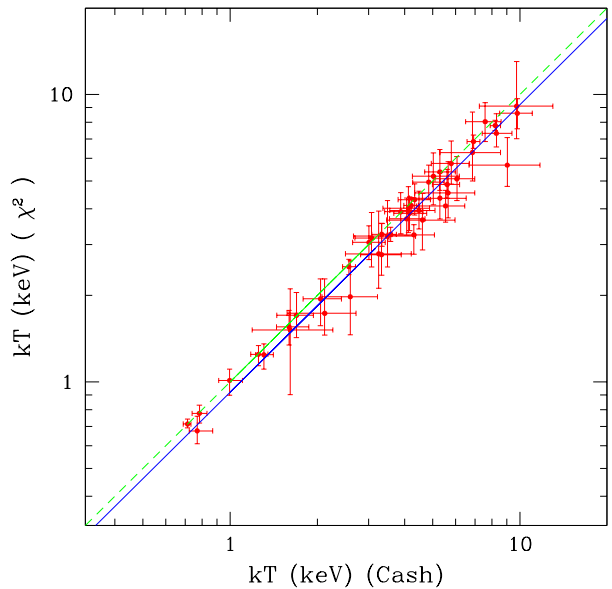}      
\caption{\label{chisq_test}Best-fit temperatures values obtained with $\chi^2$ vs those obtained with our reference analysis using Cash-statistics.  
The dashed line shows the $kT_{\chi^2}=kT_{Cstat}$ relation, while the solid line shows the $kT_{\chi^2}=0.925 \times kT_{Cstat}$ relation.}
\end{figure}

As a first check, we repeat all our fits adopting the $\chi^2$ statistics 
and a binning of 20 counts per bin.  This choice has a significant impact on the best fit values, since, given the low amount of counts, the binning has the effect of
smoothing the spectra.  We find a good agreement between the best-fit values of the temperatures obtained with the two methods, and also confirm that all the fits 
have an acceptable reduced $\chi^2$ (on average $\langle \tilde \chi^2 \rangle \sim 1.07$).  However, we also find that, on average, the best-fit values obtained
with the $\chi^2$ are 8\% lower than those obtained with the Cash statistics.  This can be visually appreciated in Figure \ref{chisq_test}.  We note that
\citet{2007bBranchesi} also found larger temperatures with Cash statistics with respect to $\chi^2$, but by a much smaller amounts $\sim 1$\%.  However, their results
are not directly comparable to ours given the different instrument ({\sl Chandra}) and the different S/N regime.
On the other hand, the best-fit values of $Z_{Fe}$ obtained with the $\chi^2$ is significantly larger than those obtained with the Cash statistics.   
This is not surprising, since the binning washes out the emission line features, hampering also the measurement of $z_X$.  As a further check, we also repeat our fit assuming an {\tt apec} model 
instead of the {\tt mekal} model, and we find no differences in the temperature best-fit values, while $Z_{Fe}$ values are lower by 20\% on average.  
Since the detection of the Fe emission line is a key step in our strategy, we keep using the Cash statistics with a {\tt mekal} model in our analysis.   

Then, we assess the robustness of our best-fit temperature values.  First, we notice that in our sample we have only 5 sources whose spectrum has more than 1500 counts 
in the total 0.5-7 keV band\footnote{We remind that the soft counts of sources SWXCS J000315-5255.2 and SWXCS J000324-5253.8 do not correspond to the actual counts in the
spectra, since the source are overlapping and the soft-band photometry actually refer to the reconstructed circular regions.  
As a consequence, they have less than 1500 total net counts in the 0.5-7 keV band despite the large soft band photometry reported in Table 2.}, which is traditionally considered a safe lower limit for spectral analysis.  
Moreover, we have 30 sources whose spectra have $100 < Cts < 500$, a regime which is usually not considered reliable for spectral analysis.  
Therefore, we carefully investigate the performance of our spectral fitting strategy in this regime by extensive spectral simulations.

We set up our simulations in order to reproduce at best the characteristics of the SWXCS, with a particular focus on the measurement of the average temperature.  
The parameters which can in principle affect the accuracy of the temperature measurements for a given number of net detected counts are: 
redshift, the temperature itself, the iron abundance, the Galactic absorption, the position on the detector, the size of the source and the exposure time 
(and hence the background level in the source region).  Clearly, the entire parameter space is too large to be fully investigated.   
Therefore we choose a typical situation for our survey, consisting in an exposure time of 150 ks (see Figure 5 in Paper I) and a circular 
extraction region with a radius of 1.7 arcmin, with average response matrix files.  The source size and the chosen exposure time set the 
intensity of the background, which is sampled from real XRT data with no source included.  The background level in each source spectrum typically 
amounts to about 300 net counts in the 0.5-7 keV band, and it is re-simulated for each spectrum.  
We also set a typical redshift of $\langle z \rangle = 0.25$ (see Figure \ref{fig:zhist}) and a typical 
Galactic absorption $N_{Hgal} = 2 \times 10^{20}$ cm$^{-2}$ (see Figure 12 in Paper I).
We vary the normalization of the input spectrum in order to have, on average, 1700, 1000, 750, 500, 300, 200, 150 and 100 net counts in the 0.5-7 keV band.  
We adopt the same fitting strategy (with a fixed redshift and a free abundance parameter) and compare the recovered best-fit temperature 
and iron abundance with the input values.  Each parameter set is simulated 1000 times.  

\begin{figure}
\includegraphics[width=9cm] {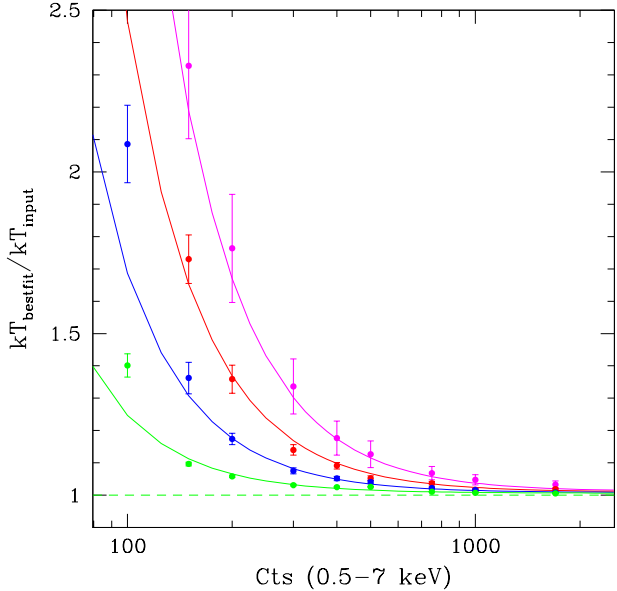}      
\caption{\label{tfitbias}Ratio of the best-fit over input temperature for our set of simulations, as a function of the total net counts in the 0.5-7 keV band.  
From top to bottom: magenta solid dots correspond to $kT=7.5$ keV; red dots to 5 keV; blue dots to 3 keV;  green dots to 1.5 keV.  
The corresponding lines show the best fit to the temperature bias function $R(T,cts)$ (see  equation \ref{tbias}).}
\end{figure}

\begin{figure}
\includegraphics[width=9cm] {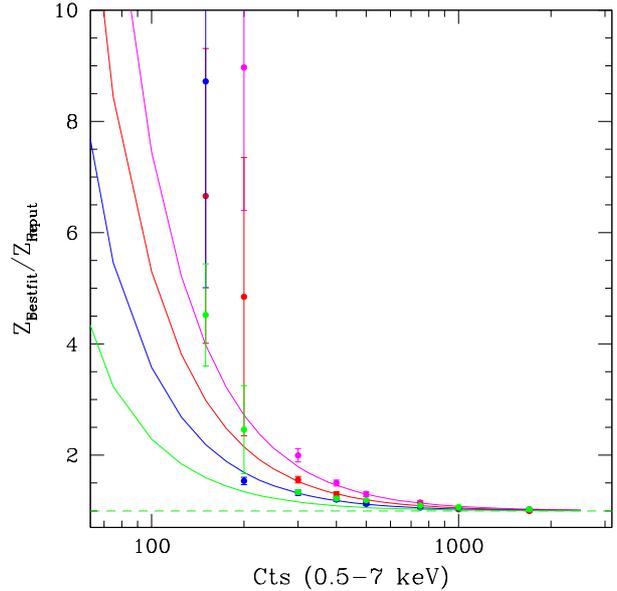}      
\caption{\label{zfitbias}Ratio of the best-fit over input abundance for our set of simulations, as a function of the total net counts in the 0.5-7 keV band.  
Colors as in Figure \ref{tfitbias}. The solid lines show the best fit to the abundance bias function $R_{abund}(Cts)$.}
\end{figure}

In Figure \ref{tfitbias} we show the ratio of the best-fit temperature to the input value as a function of the net detected counts, for four representative input 
temperatures (7.5, 5, 3 and 1.5 keV).  These simulation set is the best compromise to cover at best the parameter space of our data set.  Each point in Figure \ref{tfitbias} 
represents the ratio $kT_{best-fit}/kT_{input}$ averaged over 1000 realizations, while the error bars represent the 1 $\sigma$ rms on the mean.  
We can express the average best-fit temperature as a function of the true (input) temperature and the net detected counts as $kT_{best-fit} = R(kT_{input, Cts}) \times T_{input}$.  
With a simple minimization procedure, we find that the best approximation for the function $R(kT_{input, Cts}) $ is 

\begin{equation}
\label{tbias}
R(kT_{input},Cts) = 1+0.01 \times \big({kT_{inut}\over 5}\big)^{0.39}+1.45\times \big({kT_{input}\over 5}\big)^{1.5} \times \big({Cts\over {100}}\big)^{-2.0}\, .
\end{equation}

We find $\langle T_{best}\rangle/T_{input} \leq 1.10$ when the net counts are above 500 irrespective of the true temperature.  
However, we find that the best-fit values are significantly biased high with respect  to the input  values for low counts, particularly for high temperatures  $kT\geq 3$ keV.  
This study shows that we need to apply a significant correction to at least half of our sample.  In our spectral analysis we will invert equation \ref{tbias} to correct the best-fit temperature 
values obtained as a direct output of the spectral fits.  

In Figure \ref{zfitbias} we show the results for the iron abundance.  We notice that here there is no clear dependence on the temperature, while the abundance measurements is 
seriously compromised below 300 net counts.  Therefore, we simply decide to ignore the iron abundance measurements from spectra with less than 300 net counts, 
and provide this approximation for the function $R(Z_{Fe-input}, Cts)$ only above 300 counts:

\begin{equation}
\label{metbias}
R(Z_{Fe-input},Cts) = 1+4.3\times \big({kT_{input}\over 5}\big)^{1.0} \times \big({Cts\over {100}}\big)^{-1.9}\, .
\end{equation}

To summarize, our simulations show that it is possible to extend the X-ray spectral analysis down to very low counts, provided that one accounts for the fitting bias.
The set of simulations investigated in this work provides a good approximation to the bias affecting the best-fit spectral parameters for our sample. 
Ideally, massive spectral simulations should be run for each source, in order to accurately reproduce the actual S/N regime of the spectrum.  
However, this extremely time-consuming approach is mandatory if one wants to use large X-ray cluster samples in the faint regime to investigate the 
statistical properties of the ICM and constrain the cosmological models.   
As a final comment, one may argue that adopting the $\chi^2$ statistics may help in reducing the bias on the temperature found for the Cash statistics.  
However, we find that the $\chi^2$ statistics performs much worse as far as the iron abundance is concerned, confirming once again that the Cash statistics should be preferred if
the identification of the iron emission lines is a key step of the method.  
A detailed comparison of $\chi^2$ vs Cash statistics under different circumstances need to be addressed 
by a dedicated extensive study which goes far beyond the scope of this work.


\section{X-ray spectral analysis}

\subsection{Temperature and Fe abundance}

\begin{figure}
\centering
\includegraphics[angle=0, width=9.0 cm]{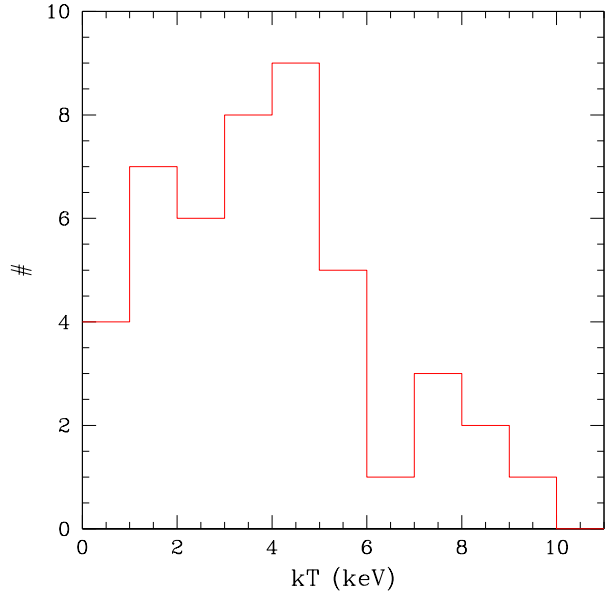}
\caption{ICM temperature distribution from X-ray spectral analysis, after accounting for fitting bias,  for the 46 sources with redshift.}
\label{fig:kthisto}
\end{figure}

We run the X-ray spectral analysis for all the 46 sources with available redshift, keeping fixed the redshift value.  We use Cash statistics after re-binning the spectra to a minimal value
of one count per bin \citep[see][]{2005Willis}.   We obtain a measure of the temperature for all our sources, irrespective of the S/N.  In fact, once the redshift is fixed, the steepness of 
the spectrum in the hard band (2-7 keV) provide a simple means of constraining the global temperature even with a low S/N.  Clearly, the presence of a multi-temperature ICM, 
in particular the presence of cold gas in a cool-core, would be practically unnoticed, and its presence may, in some cases, bias the value of the global temperature.  
This effect cannot be treated on each single source with XRT data, due to the limited angular resolution.
Another possible source of uncertainty is the presence of non-thermal emission due to a central AGN.  Due to the limited angular resolution, it is not possible to identify and remove a 
possible contribution from a central AGN.  Statistically, the contribution to the X-ray emission from AGN in clusters amounts to few percent 
(as found, for example, from the X-ray analysis of optically selected clusters, see \citet{2008Bignamini}, 
therefore we neglect this contribution in our analysis.  However, in at least two sources, SWXCS J094816-1316.7 and SWXCS J133055+4200.3, the best fit with a 
single {\tt mekal} model, fails to reproduce the emission above 2 keV.  For these two sources only we add an absorbed power law component with a fixed spectral slope 
$\Gamma = 1.8$, in order to properly account also for the hard-band emission.
 
Then, we apply the correction factor we derived in section 5.  We find that, on average, the corrected temperatures are about $10$\% lower than the best-fit values.
Global temperatures are measured with a typical $1\sigma$ error of 15\% for the lower errorbar 
and 20\% for the upper errorbar.  In Figure \ref{fig:kthisto} we show the distribution of the biased-corrected temperatures 
for the 46 sources with redshift.  Half of the sources have temperatures between 3 and 6 keV, and the highest temperature is about 10 keV.  
Only 17 sources are classified as groups or low-mass clusters, simply on the basis of their ICM temperature below 3 keV.  
This shows that our sample ranges from groups to hot, massive clusters. Direct best-fit temperatures and their values corrected for fitting bias, with 1 $\sigma$ error bars, are listed in the 
third and fourth columns of Table \ref{fit_results}.

\begin{figure}
\centering
\includegraphics[angle=0, width=9.0 cm]{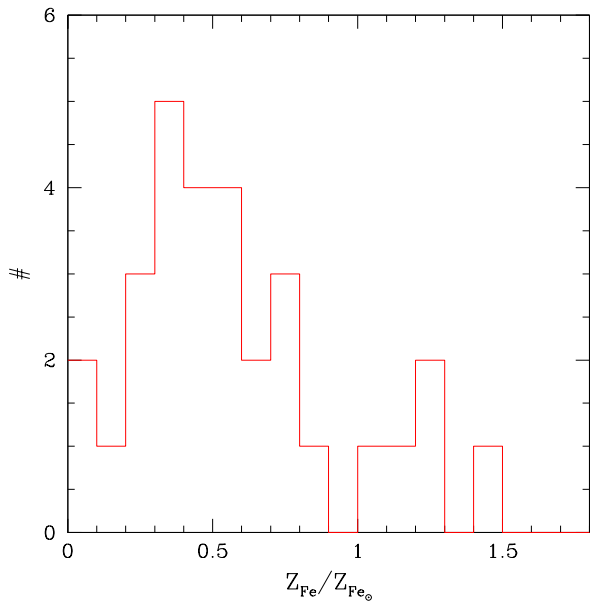}
\caption{Distribution of the bias-corrected values for the ICM iron abundance from X-ray spectral analysis for the 30 sources with redshift and more than 300 total net counts in the spectrum.  }
\label{fig:abundhisto}
\end{figure}

The presence of the {\sl K--shell} iron emission line complex at 6.7-6.9 keV rest-frame, allows us to measure an iron abundance larger than zero at 
2 $\sigma$ for about 17 sources out of 30 with more than 300 net counts.  The typical 1 $\sigma$ uncertainty on the iron abundance for the sources 
with iron line detection, is about 35\% for the lower error bar and 50\% for the upper error bar.  For the sources with global temperatures below 3 keV, the iron abundance is measured
also thanks to the increasing contribution of the {\sl L--shell} emission lines in the energy range $ 1-2$ keV.  
We remark that the iron abundance is slightly biased towards high values for clusters
whose redshift is provided by the X-ray spectral analysis \citep[see][]{2011Yu}.  The contribution of elements other than iron is negligible,
therefore we consider the best fit values as a measure of the iron abundance and not of the global metallicity, 
despite the abundance parameter in the {\tt mekal} model refers to the global metallicity.
As a robustness check, we repeat the fits with {\tt vmekal} setting all the other
elements to zero, and we find no changes with respect to the {\tt mekal} values.  The distribution of iron abundance values is shown in Figure \ref{fig:abundhisto}, 
and it is consistent with distribution broadly peaked around  $Z_{Fe} = 0.5 Z_{Fe\odot}$  in units of \citet{asp05}.   Best-fit iron abundances and the corrected 
values for sources with more than 300 net counts are listed in the fifth and sixth columns of Table \ref{fit_results}.

\begin{figure}
\centering
\includegraphics[angle=0, width=9.0 cm]{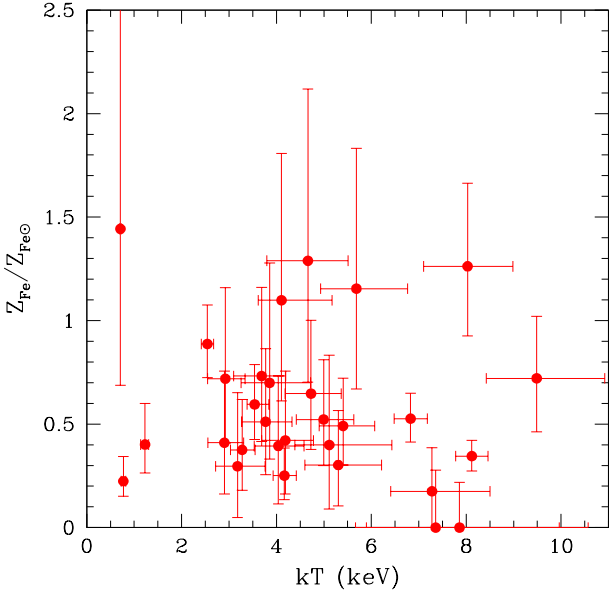}
\caption{Plot of iron abundance and global temperature for the 30 sources with X-ray spectral analysis and more than 300 net counts in the spectrum.}
\label{abund_kt}
\end{figure}

\begin{figure}
\centering
\includegraphics[angle=0, width=9.0 cm]{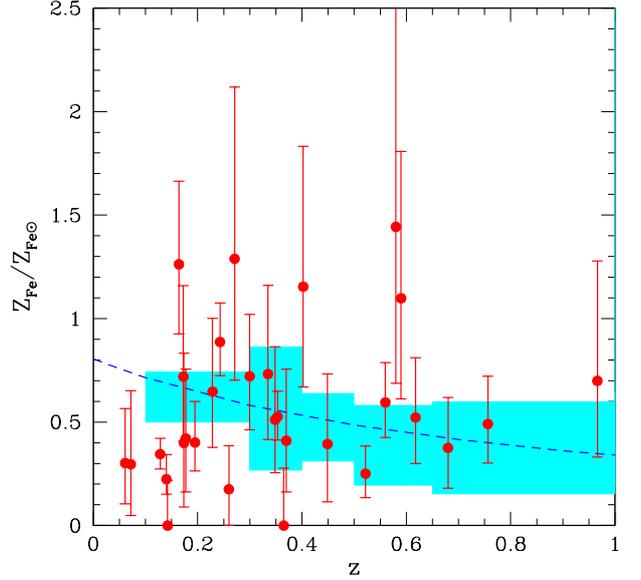}
\caption{Plot of iron abundance and redshift for the 30 sources with X-ray spectral analysis and more than 300 net counts in the spectrum. 
The shaded area correspond to the iron abundance average values measured 
with {\sl Chandra} in \citet{2007Balestra}, and the dashed line corresponds to their best fit.  Values from \citet{2007Balestra} are re-scaled to the solar abundance of \citet{asp05}.}
\label{abund_z}
\end{figure}

Despite the large statistical errors, we explore the correlation between measured iron abundance with temperature and redshift.  In Figure \ref{abund_kt} we
show the iron abundance versus the measured temperature for sources with more than 300 net counts. The data are in broad agreement 
with a constant $Z_{Fe}\sim0.5 Z_{Fe\odot}$ at temperature higher than 5 keV, a higher iron abundance between 3 and 5 keV, and lower values of $Z_{Fe}$
towards the group scales, as previously noticed by \citet{2005Baumgartner}  for local clusters and
by \citet{2007Balestra} at higher redshift.  However, the large uncertainties hampers us from drawing any conclusion.
When comparing the iron abundance versus redshift (Figure \ref{abund_z})
with \citet{2007Balestra} we find that the average $Z_{Fe}$ tends to be lower at redshift $z\leq 0.2$ (see also \citet{2008Maughan}).  Once again, 
the statistical uncertainties associated to our measurements of $Z_{Fe}$ prevent us from performing a 
meaningful statistical analysis.  We conclude that to investigate the behavior of the iron abundance with 
temperature and cosmic epoch, we would need a significantly higher S/N.

%

\subsection{Cluster luminosities}

\begin{figure}
\centering
\includegraphics[angle=0, width=9.0 cm]{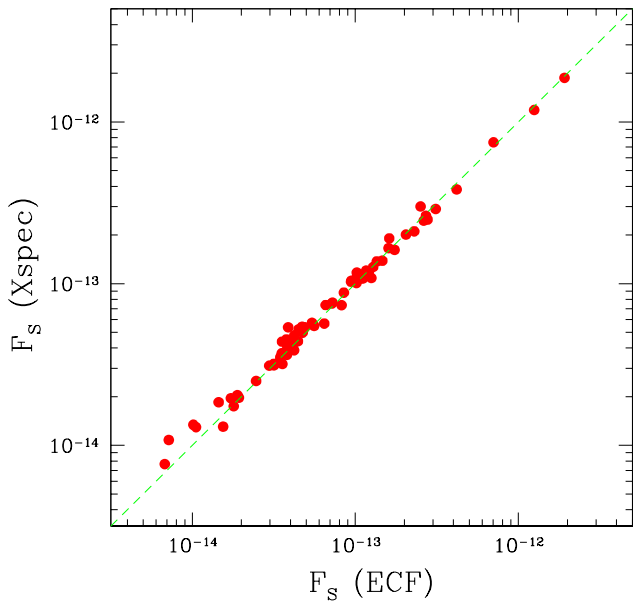}
\caption{Soft-band fluxes obtained directly from the X-ray spectral analysis, compared to the soft band fluxes measured in Paper I assuming a simple energy conversion factor.
Dashed line shows the $F_S(Xspec) = F_S(ECF)$ relation.}
\label{fluxcheck}
\end{figure}

Our spectral fits with a single {\tt mekal} model provide also a measurement of the observed flux and the rest-frame luminosity, whose values are only marginally
affected by the $\sim 10$\% correction on the best-fit temperatures.  These soft-band flux values are more accurate than those obtained in Paper I, 
which were estimated on the basis of a simple energy conversion factor after accounting for the Galactic absorption in the corresponding field.  
In Figure \ref{fluxcheck} we plot our soft-band fluxes from X-ray spectral fits versus those obtained with the conversion factor, finding a very good agreement. 
We note that we do not use the values shown in Table 2 of Paper I, which refer to the flux within $2 \times R_{ext}$, but we consistently use those
corresponding to $R_{ext}$, typically $5\%$ lower.  In addition, we do not include the two clusters
SWXCS J000315-5255.2 and SWXCS J000324-5253.8, since for both objects, the spectral fit has been performed in the non-overlapping regions, while
in Paper I the flux within $R_{ext}$ has been reconstructed by extrapolating the surface brightness profile in the overlapping sectors.
We find that the difference between the fluxes estimated with ECFs and the more accurate values obtained from the spectral fits, amounts on average to 1-2\%, 
confirming that the use of ECFs is a very reliable method to estimate the fluxes in the soft band for both groups and clusters.

Finally, we computed the X-ray soft band and bolometric rest-frame luminosities within $R_{ext}$ integrating over the entire X-ray band, adopting a
$\Lambda$CDM cosmology with $\Omega_m = 0.28$, $\Omega_{\Lambda} = 0.72$ and $H_0 = 69.7$ km s$^{-1}$ Mpc$^{-1}$.  
We also compute the luminosity within $R_{500}$ (defined as the radius
within which the average density contrast is 500) extrapolating the surface brightness model at $R>R_{ext}$.  
The radius $R_{500}$ is estimated from the corrected temperature using the empirical relation found by \citet{2006Vikhlinin}:
\begin{equation}
R_{500} = 1.14 \times h^{-1}  \times (\Omega_m (1+z)^3 +\Omega_\Lambda)^{0.5}\times (kT/10 \, \, {\rm keV})^{0.53}\, {\rm Mpc} \, ,
\end{equation}
\noindent 
 where $h \equiv H_0/(100 \, {\tt  m s}^{-1}\,  {\tt Mpc}^{-1})$.
In general the estimated $R_{500}$ is larger than the extraction radius $R_{ext}$ by a factor ranging from 1 to 4, with an average 
of $\sim 2.9$\footnote{We remark that the values of $R_{ext}$ listed in Table 2 of Paper I are in pixels and not in arcsec as erroneously written in the caption of the Table.  
Here we report the correct values in arcmin, obtained assuming the conversion factor of 2.36 arcsec per pixel.}.   The luminosity outside the extraction region is computed
extrapolating the fit to the surface brightness distribution as in Paper I.  Errors on the luminosities are taken from the Poissonian error on the net detected counts, plus the uncertainty on 
the factor accounting for the missing flux outside $R_{ext}$.  
Soft-band and bolometric luminosities are listed in the last two columns of Table \ref{fit_results}.

\subsection{Luminosity-Temperature relation}

\begin{figure}
\includegraphics[width=9cm] {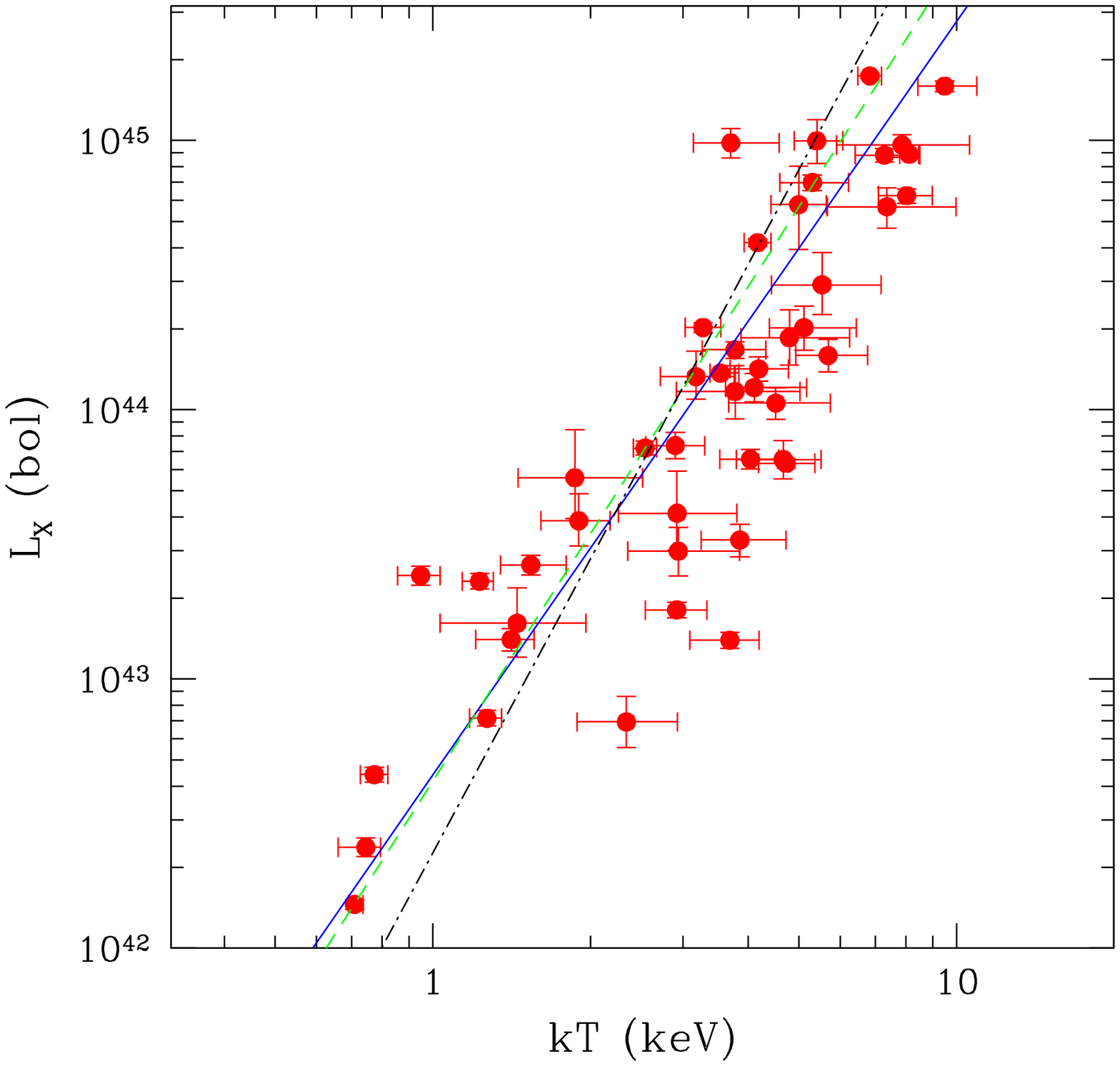}      
\caption{\label{lt}$\lxtx$ relation for the 46 sources with spectral analysis. The solid blue line is the best fit of the $\lxtx$ for our sample. 
The dashed green line is the best fit for the combined cluster sample analyzed 
by \citet{2007Branchesi} computed for $\langle z \rangle = 0.25$, while the black, dot-dashed line is for the "all cluster" sample presented in \citet{2012Maughan}
and computed for $\langle z \rangle = 0.25$. Luminosities are consistently computed at the estimated radius $R_{500}$.  
Error bars correspond to 1 $\sigma$ error both in temperature and luminosity.  }
\end{figure}

In Figure \ref{lt}, we plot the relation of bolometric luminosity versus the corrected temperature for the 46 sources with X-ray spectral analysis.  
We perform a simple linear regression in the log-log space, 
which is dominated by the error on the temperature. Our fitting function is 

\begin{equation}
log(L_{500}) = C_0 + \alpha \times log(kT/5 \, \, {\rm keV}) \, .
\end{equation}

\noindent
Despite the data present a significant intrinsic scatter, we consider the entire range of temperature and luminosities to search for a best fit.  
We find $C_0 = 44.6\pm 0.03$ and $\alpha = 2.8 \pm 0.12$.  
Our best-fit relation is compared to the $\lxtx$ relation obtained by \citet{2007Branchesi} and by \citet{2012Maughan}.  We adopt the $\lxtx$ relation obtained for 
all the clusters, including the core regions, in \citet{2012Maughan}, and correct for the $E(z)$ factor (which is not used in our analysis) 
by assuming as a representative redshift for our sample $\langle z \rangle = 0.25$.   We also use the $\lxtx$ relation obtained for the combined sample in \citet{2007Branchesi} 
and computed for the same average redshift $\langle z \rangle = 0.25$.  The slope of our best fit $\lxtx$ relation is somewhat shallower but still in
agreement within 1 $\sigma$ with the value  $\alpha = 3.05\pm 0.23$ found by \citet{2007Branchesi}, while is significantly shallower with respect to $\alpha = 3.63 \pm 0.27$ found by \citet{2012Maughan}.
The normalization of the best-fit relation is in good agreement with both works.
We note, however, that our sample includes many more low-mass clusters and groups, while the sample in the two quoted papers are dominated by clusters with $L_X > 10^{44}$ erg s$^{-1}$.
If we fit only the bright part of our sample at luminosities higher than   $10^{44}$ erg s$^{-1}$, we find a 15\% lower normalization and a steeper slope $\alpha = 3.4 \pm 0.4$.
We also remark that selection effects are difficult to model for a sample drawn directly from the {\sl Chandra} archive, without a well defined selection function.  
A meaningful comparison of the scaling properties of the X-ray observables between the SWXCS and previous studies would require a larger statistics, which we expect to reach with the 
spectral analysis of the final SWXCS sample, taking advantage of its accurately modeled selection function.



\section{A Chandra observation of an SWXCS cluster}

\begin{figure*}
\includegraphics[width=18cm]{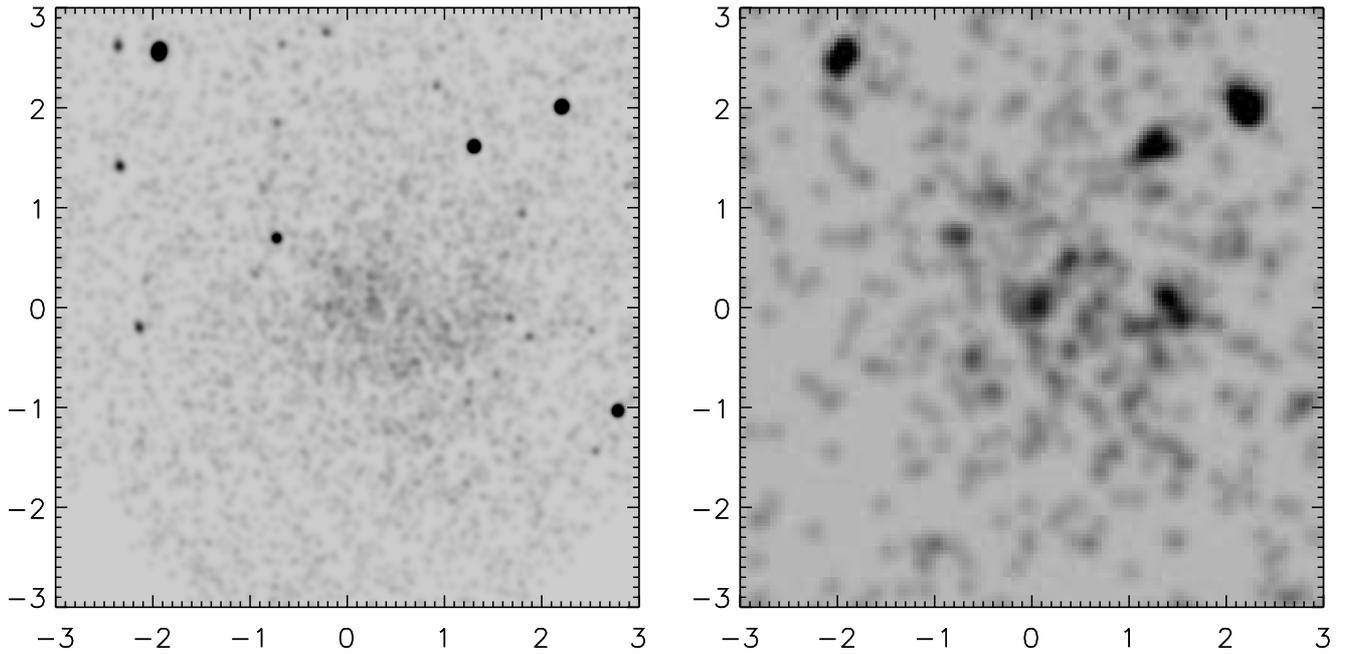}      
\caption{ACIS-S Chandra 50 ks image (left panel) and XRT 34 ks image (right panel) of SWXCS J123630+2859.1 at $z_{spec}=0.23$. 
Both images are in the soft [0.5-2] keV band and Gaussian smoothed with a FWHM of 10 and 5 arcsec respectively.  
The scale of the image is 7 arcmin. The difference of the image quality is striking due to the two order of magnitude difference in angular resolution, 
however the quality of the spectral analysis based on these data is comparable owing to the low XRT background and good hard energy response (see text).}
\label{Chandra_XRT}
\end{figure*}

We take advantage of a set of {\sl Chandra} follow-up observations of Swift GRB to search for {\it Chandra} images of our extended sources. Unfortunately, the use of 
ACIS-S, which has a significantly smaller FOV with respect to XRT, further reduces the probability of having {\sl Chandra} data for our sources.   
We find two Swift GRBs observed with {\it Chandra} whose fields contain a SWXCS cluster: 
GRB050509 (SWXCS J123620+2859.1) and GRB052120 (SWXCS J215507+1647.3).  Only the first one, however, covers our source with the ACIS S3 chip, 
while in the second case our source is at very large off-axis angle ($\sim 10 $ arcmin) and it lies in an ancillary CCD, therefore its imaging and spectral quality are not 
sufficient for a robust analysis.  In the following we will discuss only the {\sl Chandra} observation of SWXCS J123620+2859.1.

SWXCS J123620+2859.1 has a 50 ks observation with {\sl Chandra} ACIS-S (ObsId 5588).  We collect about 1700 net 
photons in the 0.5-7 keV band, corresponding to a total flux of $1.3 \times 10^{-13}$ erg s$^{-1}$ 
cm$^{-2}$  within the same extraction radius of 90 arcsec, consistent with the flux measured with the XRT data.
Our spectral fit with free redshift returns a values $z_X = 0.20\pm 0.02$, in very 
good agreement with the spectroscopic redshift $z_{spec}=0.23 \pm 0.01$ (see Table \ref{tab:zobs}).  We obtain $kT = (3.6 \pm 0.6) $ keV and 
$Z_{Fe} = 0.6_{-0.3}^{+0.4} Z_\odot$, in agreement within 1 $\sigma$ with the value found with the XRT analysis, based only on  $\sim200$ net counts 
in the 0.5-7 keV.  From XRT we obtain $kT = 4.6_{-1.1}^{+1.5}$ keV directly from the best-fit, and $kT = 3.8_{-0.9}^{+1.2}$ keV after the correction
for the fitting bias.  As for the iron abundance, we only obtain a loose upper limit of $Z_{Fe} < 0.40 Z_\odot$ at 1 $\sigma$ c. l.  Therefore, using about eight times fewer 
photons than those collected by {\it Chandra} in the same band, we obtain a consistent temperature value with error bars larger by a factor of two.  This 
simple comparison practically shows the good performance of a small instrument like XRT for the study of faint extended sources, despite the 
poor image quality when compared to the Chandra soft-band image (see Figure \ref{Chandra_XRT}).  
Both the low background and the relatively high response function in the hard band (see Figure 1 in Paper I) contribute 
to provide a reliable estimate of the temperature for this cluster despite the low S/N.  We argue that an instrument built as a larger version of the XRT, with a 
comparably low background and with a better angular resolution, would be extremely efficient in finding and characterizing groups and clusters of galaxies 
down to very low fluxes.  

\section{Conclusions}

We present detailed spectral analysis for the X-ray groups and clusters of galaxies in the SWXCS catalog \citep{2012Tundo}.  
We retrieve the optical spectroscopic or photometric redshift for 35 sources out of
72, thanks to the cross-correlation with the NASA/IPAC Extragalactic Database and a dedicated TNG follow-up program.  
A blind search of the iron $K_\alpha$ emission line complex in the X-ray spectra of 
our sources with optical redshift, allows us to set the criteria for a reliable measurement of the redshift from the X-ray spectra alone.  
Therefore, we are able to extend our X-ray redshift measurement to the entire sample, finding another 11 
reliable redshifts, for a total of 46 out of 72 sources in total.  The redshift distribution is peaked around $z\sim 0.2$, with a tail extending up to $z\sim 0.9$.  
We find that 12 sources (25\% of those with redshift) have redshift larger than 0.4.

Despite the relatively low S/N of our sample, we are able to measure the temperature for all our sources with redshift, with typical 1 $\sigma$ 
errorbars of 15-20\%.  We explore the robustness of our X-ray spectral analysis with extensive spectral simulations, deriving correction factors for temperature and 
abundance depending on the actual temperature of the source and on the total (0.5-7 keV) net counts in the source spectrum.  After accounting for this fitting bias, we find that the corrected
temperatures are on average $\sim 10$\% lower than the values obtained directly from {\sl Xspec}.  We also find that abundance measurements are not reliable
when the total net counts in the spectrum is lower than 300.   Despite this, we are also able to estimate the iron abundance in several cases, 
although the large statistical uncertainties do not allow us 
to draw conclusions on the correlation of iron abundance with temperature or redshift.  We find that about 60\% of our sources have global temperatures 
in the range of massive clusters ($kT>3$ keV), with only 17 sources with temperatures typical of groups ($kT<3$ keV).  
We also measure the rest-frame, soft and bolometric luminosities up to $R_{500}$, where we estimated $R_{500}$ from the temperature itself, 
and extrapolated the surface brightness profile beyond the extraction radius up to $R_{500}$
to evaluate the missing fraction of the flux.  We are then able to compute the $\lxtx$ relation, finding a slightly shallower slope with respect to previous studies, but overall
consistent normalization.

The major result of our study, is that for the first time we are able to characterize the majority of a flux-limited, X-ray cluster sample on the basis of 
the X-ray discovery data, without time-consuming follow-up observations, if not for a partial photometric campaign with the TNG.  The characterization of the 
sample is not complete yet, since 26 sources still lack a redshift  measurement.  In the near future, we plan to complete the identification of the sources 
in the SWXCS catalog of Paper I, and eventually to extend our sample to lower fluxes and to a larger portion of the Swift-XRT 
archive, using a detection algorithm based on Voronoi tessellation and developed {\sl ad hoc} for XRT data \citep[see][for a description of the algorithm]{2013Liu}.  
The final goal is to assemble a sizable sample with a good characterization, including mass proxies for the majority 
of the sample, and to use it  for a robust determination of the cluster mass function at $M\lesssim M_*$ and $z\sim 0.5$.

To summarize, the SWXCS is a practical example of how to build an extremely simple and effective X-ray cluster sample, and it can be useful to design efficient 
surveys with future X-ray facilities.  This result, despite the small size of the XRT telescope, has been achieved thanks to the well defined selection function associated 
to our detection strategy, and to some specific properties of the XRT, namely: the low 
background, the constant angular resolution across the field of view, and the good spectral response in the hard band (2-7 keV), which is crucial to measure temperature and iron emission lines.  
These properties are mandatory for assembling large sample of X-ray clusters for precision cosmology without time-prohibitive follow-up campaigns in other wavebands.


\acknowledgements 
We thank Heinz Handernach for pointing out the overlap of some of our sources with Abell clusters, and for suggesting to register the acronym SWXCS to the IAU registry.   
We acknowledge support from ASI-INAF I/088/06/0 and ASI-INAF I/009/10/0.  This work is also supported at OAB-INAF by ASI grant I/004/11/0.  
PT, AM, ET and TL  received support from the "Exchange of Researchers" program for scientific and technological cooperation between Italy and the People's Republic of China for the years 
2013-2015 (code CN13MO5).  SB acknowledges partial financial support from the PRIN-MIUR 2009 grant "tracing the growth of cosmic structures" and from the PD51 INFN grant.  
This research has made use of the NASA/IPAC Extragalactic Database (NED) which is 
operated by the Jet Propulsion Laboratory, California Institute of Technology, under contract with the National Aeronautics and Space Administration.  We also thank the anonymous Referee for
comments and suggestions which significantly improved the paper.

\bibliography{ref_SWXCSBright}


\clearpage

\longtabL{1}{
\begin{landscape}
\begin{longtable}{|c|c|c|c|c|c|c|c|}
\caption{\label{tab:zobs} Cluster sample. (i) Cluster name; (ii) optical redshift
 from cross-correlation with the NED; (iii) photometric redshift from
follow-up with TNG (one case, marked with $^*$, obtained with VLT archival data); (iv) X-ray
redshift; (v) reference for optical redshift; (vi) name of the optical counterpart (cluster); (vii) name of the optical counterpart (galaxy); (viii) name of the X-ray source as in Paper I.  The format of
 the name of the sources is  SWXCS JHHMMSS+DDMM.m.  This new designation supersedes that used in Paper I, and it has been officially accepted by the IAU register of acronym.
References for the redshifts from the Literature: $^1$ \citet{Shectman1996}; $^2$\citet{Colless2003}; $^3$\citet{Adelman2006}; $^4$ \citet{Wen2010}; $^5$ \citet{Adelman2008}; $^6$ \citet{Koester2007};
 $^7$ \citet{Abazajian2005}; $^8$ \citet{2010Hao}; $^9$ \citet{Linden2007}; $^{10}$ \citet{Hucra2012}; $^{11}$ \citet{Gal2003}; $^{12}$\citet{Adelman2007};  $^{13}$\citet{Estrada2007}; $^{14}$\citet{Frazer1995}; 
 $^{15}$\citet{Struble1999}; $^{16}$\citet{Schwope2000}; $^{17}$\citet{dinella1996}; $^{18}$ redshift obtained from Literature by the NED Team prior to November 1992; $^{19}$\citet{2011Szabo}.
 The source marked with $^*$ has two other cluster counterparts at distances larger than 1 arcmin from the X-ray centroid: ZwCl 1234.0+2916 and WHL J123620.6+285756.  These counterparts are not 
 considered here.}\\
\hline
      &      &      &     		& 		&    &  &   \\      
name              &   $z_{opt}$ (NED)      & $z_{opt}$ (TNG)  &      $z_X$		&  Ref. for $z_{opt}$	&  Optical counterpart (cluster) &  Optical counterpart (galaxy)  & Name as in Paper I \\
      &      &      &     		& 		&  &  &   \\      
\hline
\endfirsthead
\caption{continued.}\\
\hline
      &      &      &     		& 		&   &   &   \\      
 name              &   $z_{opt}$ (NED)      & $z_{opt}$ (TNG)  &      $z_X$		&  Ref for $z_{opt}$	&  Optical counterpart (cluster) &  Optical counterpart (galaxy)  & name as in Paper I \\	
       &      &      &     		& 		&     &    &   \\      
\hline
 \endhead
 \endfoot
 SWXCS J000345-5301.8 &                        &					& 				&    		&            &          &     SWJ000345-530149   \\
 SWXCS J000315-5255.2 &    	 		           &                	& $0.62\pm 0.1$  	&               &                  &       &      SWJ000315-525510     \\
 SWXCS J000324-5253.8 &  	 		           &           	   	& $0.76 \pm 0.01$  	&               &                    &       &     SWJ000324-525350    \\
 SWXCS J002437-5803.9 & 			              &           	   	& $0.195 \pm 0.012$  	&               &                  &       &     SWJ002437-580353      \\
 SWXCS J003316+1939.4 &                       &             	&          		&             &          &        &       SWJ003316+193922   \\
 SWXCS J005500-3852.4 & $0.164127\pm  0.000213$  (s)   &          	&   	&    LCRS$^{1}$       &                               &   LCRS B005239.6-390844  &   SWJ005500-385226      \\       
                                           &                            &          	&   	&                  &            EDCC493     &                                            &                                        \\       
  SWXCS J011432-4828.4 &                        &           	   	& $0.97\pm 0.02$    	&               &                     &      &   SWJ011432-482824      \\
  SWXCS J012210-1304.4 &                        &           	   	&			&               &                     &       &    SWJ012210-130422    \\
  SWXCS J012302+3756.3 &                        &           	   	&    			&               &                   &        &       SWJ012302+375615   \\
  SWXCS J015753+1659.6 &                        &           	   	&   			&               &                   &        &     SWJ015753+165933     \\
  SWXCS J020744+0020.9 &                        &           	   	&    			&               &                     &      &   SWJ020744+002055      \\
  SWXCS J021705-5014.2 &      		   &           	   	& $0.52 \pm 0.01$    	&               &              &             &    SWJ021705-501409     \\
  SWXCS J021747-5003.4 &                        &           	   	&$0.206\pm 0.01$  	&               &             &          &      SWJ021747-500322        \\
  SWXCS J022344+3823.2 &                        &           	   	&     			&               &                   &        &   SWJ022344+382311      \\
  SWXCS J022546-1855.9 &       		   &           	   	& 			&               &                    &       &    SWJ022546-185553     \\
  SWXCS J023224-7120.3 &       		   &           	   	&   			&               &                 &          &     SWJ023224-712020    \\
  SWXCS J023302-7116.6 &      		   &           	   	& 			&               &                    &       &   SWJ023302-711634      \\
  SWXCS J023924-2505.1 & $0.1737 \pm 0.000213$  (s)   &           	   	&   	       		&     2dFGRS$^2$    &            & 2dFGRS S159Z145      &     SWJ023924-250504      \\
  SWXCS J024010-2511.3 &                        &           	   	&     			&               &                    &       &   SWJ024010-251121  \\
  SWXCS J035259-0043.7 &  $0.334925 \pm 0.000293$  (s)  &           	   	&$0.31_{-0.01}^{+0.02}$ &     SDSS$^3$   	&                                          &     SDSSJ035259.38-004337.8  &     SWJ035259-004342    \\
                                            &     $0.325\pm 0.01$ (p)            &           	   	&$0.31_{-0.01}^{+0.02}$ &     WHL$^4$    & WHLJ035259.4-004337     &                                                 &     SWJ035259-004342    \\
  SWXCS J035310+2133.6 &  	                   &           	   	&    			&               &                  &          &   SWJ035310+213335     \\ 
  SWXCS J044144-1115.6 &       		   & $0.26 \pm 0.02$	&$0.193_{-0.015}^{+0.01}$&       	&              &            &   SWJ044144-111536      \\
  SWXCS J062155-6228.6 &       		   & 	     		& $0.35_{-0.01}^{+0.02}$&               &                    &       &      SWJ062155-622834    \\
  SWXCS J082113+3200.1 &      		   & $0.68 \pm 0.06$ 	&   	    		&     	     	&                  &        &    SWJ082113+320004       \\
  SWXCS J083340+3311.0 &                        &           		&     			&               &                      &      &      SWJ083340+331102    \\
  SWXCS J084749+1331.7 & $0.363\pm 0.01$ 	 (p)         & $0.35\pm 0.01$     &  $0.349 \pm 0.005$	&     WHL$^4$  & WHLJ084749.3+133140   &                                             &    SWJ084749+133141    \\
                                             & $0.348628 \pm 0.000147$ (s) & $0.35\pm 0.01$     &  $0.349 \pm 0.005$	&     SDSS$^5$ &                                         & SDSSJ084749.32+133140.4  &    SWJ084749+133141    \\
  SWXCS J085524+1102.0 &           		   & $0.37 \pm 0.01$    & 			&     	 	&                     &       &     SWJ085524+110201   \\
  SWXCS J090946+4157.2 & $0.14855\pm 0.01$ (p)             &           		& $0.171\pm 0.02$  	&     WHL$^6$       	&  WHL J090945.9+415721     &                           &      SWJ090946+415713    \\
                                             & $0.140048 \pm 0.000175$ (s)   &           		& $0.171\pm 0.02$  	&     SDSS $^7$	    &                                             &    B3 0906+421  &      SWJ090946+415713    \\
  SWXCS J092619-0905.8 &   	       		   &           		&   	    		&               &                     &       &      SWJ092619-090546    \\
 SWXCS J092729+3010.8 & $0.365 \pm 0.01$ (p)   &           		&   	    		&     WHL$^4$   	& WHLJ092730.0+301045  &                                                &      SWJ092729+301048  \\
                                             &  $0.293 \pm 0.01$ (p)  &           		&   	    		&     SDSS$^8$    &                                         &  SDSS J092730.01+301045.4   &      SWJ092729+301048  \\
  SWXCS J092650+3013.8 &            	     	   & $0.58 \pm 0.07$    &  $0.418\pm 0.008$ 	&     		&                      &       &       SWJ092650+301345   \\
  SWXCS J092719+3013.7 & $ 0.29975 \pm 0.01$ (p)  &           		&  			&     WHL$^6$   	& WHLJ092719.8+301355    &      &       SWJ092719+301342    \\
  SWXCS J093045+1659.5 & $0.177278\pm 0.000165$ (s)         &           		&   			&     2MASX$^5$     &            	          &  2MASX J09304539+1659294    &    SWJ093045+165931     \\
  SWXCS J093749+1535.7 & $0.271 \pm 0.01$ (p)        &    &$0.322_{-0.045}^{+0.020}$ &    GMBCG$^8$    &   GMBCG J144.45579+15.59434   &             &     SWJ093749+153540    \\
											 & $0.294\pm 0.01$ (p)          &    & $0.322_{-0.045}^{+0.020}$ &    AMF$^{19}$   &   AMFJ144.4580+15.5998             &             &     SWJ093749+153540    \\
                                             & $0.271 \pm 0.01$ (p)        &    &$0.322_{-0.045}^{+0.020}$&     SDSS $^8$      &                                           &   SDSS J093749.38+153539.6  &     SWJ093749+153540    \\
 SWXCS J094816-1316.7 &          		   &           		& $0.13 \pm 0.02$       &               &                       &      &    SWJ094816-131644     \\
 SWXCS J101341+4306.9 & $0.44886 \pm  0.000217 $ (s)         &           		&   	   		&     SDSS$^7$      &                      &  SDSS J101341.77+430656.6    &       SWJ101341+430655   \\
 SWXCS J105946+5348.2 & $0.072\pm 0.01$ (p)                 &           		& $0.12\pm 0.02$  	&     SDSS$^9$        &       SDSS-C4-DR3  3205    &                                   &     SWJ105946+534809    \\
                                            & $0.071199 \pm 0.000097 $ (s) &           		& $0.12\pm 0.02$  	&     2MASS$^{10}$  &                                          &    MCG +09-18-064   &     SWJ105946+534809    \\
  SWXCS J115811+4529.1 & $0.4024 \pm 0.01$ (p)           &           		&   			&    WHL$^4$   &     WHLJ115814.4+452930    &      &       SWJ115811+452906  \\
                                            & $0.4051 \pm 0.01$ (p)              &           		&   			&    AMF$^{19}$   &     AMFJ179.5554+45.4806   &      &       SWJ115811+452906  \\
SWXCS J123620+2859.1$^*$ & $0.243 \pm 0.01$ (p)              &           		&   			&     GMBCG$^8$     &  GMBCG J189.08767+28.99149   &                                &       SWJ123620+285905  \\ 
                                                  & $0.2305 \pm 0.01$ (p)              &           		&   			&     AMF$^{19}$     &  AMFJ189.0796 +28.9838   &                                &       SWJ123620+285905  \\ 
                                             &  $ 0.228573\pm  0.00017$ (s)  &           		&   			&     SDSS$^5$         &                                                     &    2MASXJ12362010+2859080 &                    SWJ123620+285905\\ 
 SWXCS J124312+1704.9 & $0.1424\pm 0.01$ (p)           &           		&   			&     NSCS$^{11}$  	&              NSCS J124308+170537       &   &      SWJ124312+170451   \\  
 SWXCS J131300+0803.0 & $0.5598 \pm 0.01$ (p)  &           		&   	  		&      WHL$^5$   	& WHLJ131300.1+080303    &     &    SWJ131300+080259     \\
 SWXCS J131522+1641.8 &                        &           		&   			&               &                       &          &     SWJ131522+164145    \\
  
SWXCS J133055+4200.3 &$0.061154 \pm  0.000197 $ (s)  &         		&$0.06_{-0.02}^{+0.01}$ &     SDSS$^{12}$  	&             &   2MASXJ13305564+4200176     &          SWJ133055+420017       \\
SWXCS J133051+4206.8 &                        &         		&			&              	&                    &      &      SWJ133051+420647    \\
SWXCS J140637+2743.8 &  			   &  $0.59 \pm 0.04$   &   	    		&          	&               &     	   &         SWJ140637+274349    \\ 
SWXCS J140639+2735.8 & $0.24305 \pm 0.01$ (s)   &         		& 			&      SDSS$^6$	&             GMBCG J211.66393+27.59979    &    &        SWJ140639+273546   \\
SWXCS J140728+2749.3               & $0.17255\pm 0.01$ (p)  &      		&   		&    Abell$^{13}$   	& Abell 1861    &    &       SWJ140728+274917   \\
                                                          & $0.17015\pm 0.01$ (p)  &     		&   			&      SDSS$^6$ 	&    WHLJ140726.6+274742     &    &       SWJ140728+274917   \\
SWXCS J143646+2752.0 & $0.2648\ 0.01$ (p)           &         		&    			&      AMF${19}$  	&        AMFJ219.2030 +27.8593      &     &      SWJ143646+275157      \\
SWXCS J151324+3057.6 & $0.0717\pm 0.01$  (p)          &         		&    			&      NED$^{18}$  	&        CGCG 165-032     	      &     &        SWJ151324+305738   \\
SWXCS J155742+3530.4 & $0.169\pm 0.01$ (p) &         		&$0.154\pm 0.004$ 	&     SDSS$^8$ 	&  GMBCG J239.42668+35.50827   &               &       SWJ155742+353023   \\
                                           & $0.1549\pm 0.01$ (p) &         		&$0.154\pm 0.004$ 	&       SDSS    	&  MaxBCG J239.42665+35.50827        &       &       SWJ155742+353023   \\
                                           & $0.1459\pm 0.01$ (p) &         		&$0.154\pm 0.004$ 	&        SDSS$^4$ 	&  WHL J155746.1+352954 239.44208  &        &       SWJ155742+353023   \\
                                           & $0.16\pm 0.01$ (s) &         		&$0.154\pm 0.004$ 	&      	&     Abell2141     &       &       SWJ155742+353023   \\
                                           & $0.1579\pm 0.0004$ (s) &         		&$0.154\pm 0.004$ 	&     B2$^{14}$     	&     &    B2 1555+35B 239.42558   &       SWJ155742+353023   \\
                                           & $0.158877\pm 0.000176 $ (s) &         		&$0.154\pm 0.004$ 	&     SDSS$^{12}$ 	&     &    2MASX J15574240+3530292  &       SWJ155742+353023   \\
 SWXCS J164956+3130.3 &                        &        		&    			&               &                          &       &     SWJ164956+313021    \\
 SWXCS J173721+4618.6 & 	 		   & $0.22\pm 0.02$ 	& $0.29\pm 0.01$   	&            	&                     &         &         SWJ173721+461834  \\
 SWXCS J173932+2720.9 &  			   &          		& 	   		&               &                  &             &        SWJ173932+272055   \\
 SWXCS J175640+3329.5 &                        &         		&$0.195_{-0.03}^{+0.02}$&        	&                  &           &     SWJ175640+332928       	 \\
 SWXCS J181053+5815.5 &                        &         		&    	    		&               &                        &       &  SWJ181053+581527         \\
 SWXCS J194004+7824.3 &                        &         		&    	    		&               &                       &        &   SWJ194004+782419        \\
 SWXCS J203723-4401.7 &                        &         		&   	    		&               &                       &        &        SWJ203723-440141   \\
 SWXCS J215507+1647.4 &                        &        		&    			&               &                          &      &      SWJ215507+164725     \\
 SWXCS J215354+1653.8 &   			   & $0.23\pm 0.02$   	&   			&           	&                      &       &      SWJ215354+165348       \\
 SWXCS J222600-5712.8 & $0.13\pm 0.01$ (p) 	   &          		& $0.15\pm 0.01$  	&     Abell$^{15}$    	&             Abell3875     &    &      SWJ222600-571248    \\
 SWXCS J222443-0220.5 &                        &         		&    			&               &                          &     &         SWJ222443-022031  \\
 SWXCS J222516-0208.5 &                        &         		&   			&               &                         &       &       SWJ222516-020827    \\
  SWXCS J222437-0222.5 &                        &         		&			&               &                           &       &         SWJ222437-022230\\
  SWXCS J222917-1101.1 &   			   &$0.20_{-0.02}^{+0.01}$& $0.197\pm 0.02$ 	&             	&                       &         &       SWJ222917-110106    \\
  SWXCS J222953+1943.9 & 			   & $0.24\pm 0.02$     &$0.25\pm 0.01$  	&          	&                         &       &      SWJ222953+194354     \\
 SWXCS J224207+2333.9 &   			   & $0.71 \pm 0.02$    &   			&        	&                           &      &        SWJ224207+233354 \\
 SWXCS J230754-6815.1 &                        &           		&$0.097\pm 0.02$     	&               &                       &          &    SWJ230754-681505      \\
 SWXCS J230650-6804.0 &    			   &        		&  	   		&               &                          &      &      SWJ230650-680401    \\
 SWXCS J232248+0548.2 &		   	   & $0.28_{-0.07}^{+0.04}$&$0.226\pm 0.006$    &               &             &           &       SWJ232248+054809  	  \\
 SWXCS J232345-3130.8 & 	   		   & $0.85_{-0.03}^{+0.07}$ $^*$&   			&         VLT &                       &        &       SWJ232345-313048    \\
 SWXCS J233518-6621.7 &  			   &        		& $0.383 \pm 0.009$  	&               &                     &     &    SWJ233518-662139      \\
 SWXCS J233617-3136.5 & $0.0623\pm 0.01$ (p)   &         		& $0.068\pm 0.006$  	&      RBS$^{16}$     	 &            AbellS1136     &   &       SWJ233617-313626    \\
                                          & $0.06255\pm 0.000163  $ (s)   &         		& $0.068\pm 0.006$  	&       ESO$^{17}$  	 &           ESO 470- G 020     &   &       SWJ233617-313626    \\
\hline
\end{longtable}
\end{landscape}
 }

\longtabL{2}{
\begin{landscape}
\begin{longtable}{|c|c|c|c|c|c|c|c|c|c|c|c|c|c}
\caption{\label{fit_results}Results of X-ray spectral analysis. (i) Source Name; (ii) radius of the circular region used for spectroscopic analysis; (iii)
  temperature directly obtained from the spectral analysis; (iv) temperature corrected for fitting bias; (v) iron abundance directly obtained from the spectral analysis; (vi) iron abundance corrected for the 
  fitting bias; (vii) soft band flux measured within $R_{ext}$; (viii) soft band (0.5-2 keV) luminosity measured within $R_{ext}$; 
  (ix) bolometric X-ray luminosity measured within $R_{ext}$; (x) soft net counts; (xi) estimated $R_{500}$ in kpc; (xii) estimated $R_{500} $ in arcmin; (xiii) estimated 
  soft band (0.5-2 keV) luminosity within $R_{500}$; (xiv) estimated bolometric X-ray luminosity within $R_{500}$.}\\
\hline
              &               &           &     		            & 		                                  &  	                         & 		                     &                         	& 	          & 	      	& 	          & 	 & 	       & 	                  \\      
name      &$R_{ext}$ &  $kT$ &    $kT_{corr}$ &  $Z/Z_\odot$	&  $Z_{corr}/Z_\odot$	&  $F_S$ 	                          &  $L_{0.5-2 keV}$ & $L_{bol}$ & Cts & $R_{500}$  & $R_{500}$ &  $L_{0.5-2 keV}$ & $L_{bol}$ \\
              &  arcmin   &   keV   &  keV    &     		         &     		                & erg s$^{-1}$ cm$^{-2}$  &  	$10^{44}$ erg s$^{-1}$     & 	$10^{44}$ erg s$^{-1}$  	 &     (0.5-2 keV)         	& 	 kpc    &  arcmin    &  	$10^{44}$ erg s$^{-1}$     & 	$10^{44}$ erg s$^{-1}$  \\      
\hline
\endfirsthead
\caption{continued.}\\
\hline
             &               &           &     		            & 		                                  &  	                         & 		                     &                         	& 	          & 	      	& 	          & 	 & 	       & 	                  \\      
name      &$R_{ext}$ &  $kT$ &    $kT_{corr}$ &  $Z/Z_\odot$	&  $Z_{corr}/Z_\odot$	&  $F_S$ 	                          &  $L_{0.5-2 keV}$ & $L_{bol}$ & Cts & $R_{500}$  & $R_{500}$ &  $L_{0.5-2 keV}$ & $L_{bol}$ \\
              &  arcmin   &   keV   &  keV    &     		         &     		                & erg s$^{-1}$ cm$^{-2}$  &  	$10^{44}$ erg s$^{-1}$     & 	$10^{44}$ erg s$^{-1}$  	 &     (0.5-2 keV)         	& 	 kpc    &  arcmin    &  	$10^{44}$ erg s$^{-1}$     & 	$10^{44}$ erg s$^{-1}$  \\      
      &      &      &     		& 		&  	& 		&  	& 	       & 	 & 	       & 	         \\      
\hline
\endhead
\endfoot
      &      &      &     		& 		&  	& 		&  	& 	       & 	& 	       & 	          \\      
SWXCS J000315-5255.2   & 1.73  &  $5.29_{-0.61}^{+0.67}$  &  $4.99_{-0.57}^{+0.64}$  & $0.61_{-0.26}^{+0.34}$ & $0.52_{-0.22}^{+0.29}$  &  2.93e-14   &  $0.381$  &  $1.235$  &  1487.5   &  $820\pm 103 $   & $1.99\pm 0.25 $  & $ 1.78_{-0.83}^{+0.70}$    & $ 5.78_{-2.68}^{+2.27} $    \\
SWXCS J000324-5253.8   & 2.05  &  $5.60_{-0.53}^{+0.69}$  &  $5.41_{-0.51}^{+0.66}$  & $0.54_{-0.21}^{+0.25}$ & $0.49_{-0.19}^{+0.23}$  &  4.49e-14   &  $0.921$  &  $3.053$  &  1847.4   &  $789\pm 90 $    & $1.76\pm 0.20 $  & $ 3.01_{-0.64}^{+0.60}$    & $ 9.97_{-2.12}^{+1.98} $    \\
SWXCS J002437-5803.9   & 1.69  &  $1.25_{-0.09}^{+0.08}$  &  $1.23_{-0.09}^{+0.08}$  & $0.44_{-0.15}^{+0.22}$ & $0.40_{-0.14}^{+0.20}$  &  1.01e-13   &  $0.111$  &  $0.187$  &  315.3    &  $492\pm 38 $    & $2.52\pm 0.19 $  & $ 0.137_{-0.009}^{+0.009}$ & $ 0.23_{-0.016}^{+0.016} $    \\
SWXCS J005500-3852.4   & 1.42  &  $2.12_{-0.47}^{+0.74}$  &  $1.87_{-0.41}^{+0.65}$  & $0.40_{-0.28}^{+0.47}$ & -  			  &  7.38e-14   &  $0.054$  &  $0.114$  &  104.8    &  $623\pm 179 $   & $3.67\pm 1.05 $  & $ 0.264_{-0.110}^{+0.134}$ & $ 0.56_{-0.233}^{+0.284} $    \\
SWXCS J011432-4828.6   & 1.85  &  $4.30_{-0.65}^{+1.02}$  &  $3.70_{-0.56}^{+0.88}$  & $1.09_{-0.49}^{+0.71}$ & -  			  &  1.12e-13   &  $4.247$  &  $12.262$ &  158.2    &  $572\pm 113 $   & $1.18\pm 0.23 $  & $ 3.39_{-0.46}^{+0.45}$    & $ 9.80_{-1.34}^{+1.29} $    \\
SWXCS J021705-5014.2   & 1.73  &  $8.29_{-0.96}^{+0.99}$  &  $8.03_{-0.93}^{+0.96}$  & $1.34_{-0.36}^{+0.43}$ & $1.26_{-0.34}^{+0.40}$  &  1.37e-13   &  $1.161$  &  $4.891$  &  749.9    &  $1115\pm 137 $  & $2.94\pm 0.36 $  & $ 1.48_{-0.10}^{+0.09}$    & $ 6.24_{-0.42}^{+0.36} $    \\
SWXCS J021747-5003.4   & 1.14  &  $3.08_{-0.39}^{+0.44}$  &  $2.92_{-0.38}^{+0.41}$  & $0.86_{-0.37}^{+0.53}$ & $0.72_{-0.31}^{+0.44}$  &  5.38e-14   &  $0.062$  &  $0.155$  &  207.6    &  $774\pm 108 $   & $3.79\pm 0.53 $  & $ 0.073_{-0.005}^{+0.005}$ & $ 0.180_{-0.013}^{+0.013} $    \\
SWXCS J023924-2505.1   & 1.49  &  $4.12_{-0.64}^{+0.92}$  &  $3.86_{-0.60}^{+0.86}$  & $0.86_{-0.45}^{+0.71}$ & $0.70_{-0.37}^{+0.58}$  &  5.68e-14   &  $0.045$  &  $0.130$  &  323.4    &  $911\pm 176 $   & $5.12\pm 0.99 $  & $ 0.114_{-0.017}^{+0.017}$ & $ 0.33_{-0.050}^{+0.048} $    \\
SWXCS J035259-0043.7   & 1.73  &  $4.22_{-0.24}^{+0.26}$  &  $4.17_{-0.24}^{+0.25}$  & $0.25_{-0.12}^{+0.14}$ & $0.25_{-0.12}^{+0.13}$  &  3.82e-13   &  $1.277$  &  $3.668$  &  1461.4   &  $874\pm 60 $    & $3.01\pm 0.21 $  & $ 1.45_{-0.04}^{+0.04}$    & $ 4.18_{-0.12}^{+0.13} $    \\
SWXCS J044144-1115.6   & 1.42  &  $5.54_{-0.78}^{+1.43}$  &  $5.11_{-0.72}^{+1.32}$  & $0.50_{-0.39}^{+0.54}$ & $0.40_{-0.31}^{+0.43}$  &  1.08e-13   &  $0.200$  &  $0.670$  &  311.5    &  $1013\pm 205 $  & $4.16\pm 0.84 $  & $ 0.602_{-0.128}^{+0.123}$ & $ 2.02_{-0.43}^{+0.41} $    \\
SWXCS J062155-6228.6   & 0.51  &  $3.88_{-0.62}^{+0.53}$  &  $3.69_{-0.59}^{+0.51}$  & $0.86_{-0.37}^{+0.50}$ & $0.73_{-0.32}^{+0.43}$  &  7.66e-15   &  $0.029$  &  $0.081$  &  286.9    &  $811\pm 125 $   & $2.69\pm 0.41 $  & $ 0.050_{-0.004}^{+0.004}$ & $ 0.139_{-0.010}^{+0.010} $    \\
SWXCS J082113+3200.1   & 1.38  &  $7.58_{-0.91}^{+1.27}$  &  $7.28_{-0.87}^{+1.22}$  & $0.19_{-0.19}^{+0.23}$ & $0.18_{-0.18}^{+0.21}$  &  1.27e-13   &  $1.964$  &  $7.494$  &  597.0    &  $966\pm 143 $   & $2.25\pm 0.33 $  & $ 2.31_{-0.14}^{+0.13}$    & $ 8.83_{-0.55}^{+0.50} $    \\
SWXCS J084749+1331.7   & 2.12  &  $6.91_{-0.35}^{+0.36}$  &  $6.83_{-0.34}^{+0.35}$  & $0.53_{-0.11}^{+0.12}$ & $0.53_{-0.11}^{+0.12}$  &  1.18e-12   &  $4.152$  &  $15.537$ &  3824.0   &  $1126\pm 70 $   & $3.78\pm 0.23 $  & $ 4.65_{-0.09}^{+0.09}$    & $ 17.39_{-0.34}^{+0.33} $    \\
SWXCS J085524+1102.0   & 1.30  &  $5.65_{-1.08}^{+1.72}$  &  $4.79_{-0.92}^{+1.46}$  & $0.72_{-0.50}^{+0.73}$ & -  			  &  5.48e-14   &  $0.222$  &  $0.756$  &  204.1    &  $923\pm 231 $   & $2.98\pm 0.74 $  & $ 0.544_{-0.144}^{+0.147}$ & $ 1.85_{-0.49}^{+0.50} $    \\
SWXCS J090946+4157.2   & 1.81  &  $1.69_{-0.21}^{+0.28}$  &  $1.54_{-0.19}^{+0.26}$  & $0.66_{-0.32}^{+0.53}$ & -  			  &  2.45e-13   &  $0.126$  &  $0.238$  &  133.5    &  $569\pm 86 $    & $3.82\pm 0.58 $  & $ 0.140_{-0.013}^{+0.012}$ & $ 0.265_{-0.024}^{+0.023} $    \\
SWXCS J092729+3010.8   & 1.34  &  $3.34_{-0.25}^{+0.28}$  &  $3.28_{-0.25}^{+0.27}$  & $0.39_{-0.20}^{+0.25}$ & $0.38_{-0.20}^{+0.24}$  &  1.62e-13   &  $0.669$  &  $1.719$  &  692.7    &  $757\pm 66 $    & $2.46\pm 0.21 $  & $ 0.787_{-0.035}^{+0.033}$ & $ 2.02_{-0.09}^{+0.08} $    \\
SWXCS J092650+3013.8   & 0.90  &  $4.07_{-0.55}^{+0.60}$  &  $3.77_{-0.51}^{+0.55}$  & $0.65_{-0.32}^{+0.45}$ & $0.51_{-0.26}^{+0.35}$  &  4.05e-14   &  $0.473$  &  $1.336$  &  254.8    &  $722\pm 105 $   & $1.81\pm 0.26 $  & $ 0.592_{-0.044}^{+0.043}$ & $ 1.67_{-0.13}^{+0.12} $    \\
SWXCS J092719+3013.7   & 1.10  &  $3.01_{-0.36}^{+0.42}$  &  $2.90_{-0.35}^{+0.40}$  & $0.46_{-0.28}^{+0.39}$ & $0.41_{-0.25}^{+0.35}$  &  5.41e-14   &  $0.144$  &  $0.355$  &  312.1    &  $735\pm 99 $    & $2.73\pm 0.37 $  & $ 0.299_{-0.035}^{+0.037}$ & $ 0.736_{-0.085}^{+0.091} $    \\
SWXCS J093045+1659.5   & 1.53  &  $3.25_{-0.74}^{+0.98}$  &  $2.92_{-0.67}^{+0.88}$  & $0.16_{-0.15}^{+0.47}$ & -  			  &  4.38e-14   &  $0.037$  &  $0.094$  &  160.5    &  $786\pm 210 $   & $4.34\pm 1.16 $  & $ 0.161_{-0.060}^{+0.070}$ & $ 0.412_{-0.153}^{+0.179} $    \\
SWXCS J093749+1535.7   & 0.83  &  $1.60_{-0.23}^{+0.17}$  &  $1.41_{-0.20}^{+0.15}$  & $1.36_{-0.77}^{+1.80}$ & -  			  &  3.13e-14   &  $0.068$  &  $0.115$  &  107.5    &  $509\pm 66 $    & $2.03\pm 0.27 $  & $ 0.083_{-0.008}^{+0.008}$ & $ 0.140_{-0.014}^{+0.014} $    \\
SWXCS J094816-1316.7   & 1.14  &  $0.78_{-0.05}^{+0.05}$  &  $0.77_{-0.05}^{+0.05}$  & $0.24_{-0.08}^{+0.13}$ & $0.22_{-0.07}^{+0.12}$  &  4.26e-14   &  $0.019$  &  $0.031$  &  279.7    &  $398\pm 28 $    & $2.87\pm 0.20 $  & $ 0.027_{-0.002}^{+0.002}$ & $ 0.044_{-0.003}^{+0.003} $    \\
SWXCS J101341+4306.9   & 0.90  &  $6.85_{-1.37}^{+2.03}$  &  $5.53_{-1.10}^{+1.64}$  & $0.56_{-0.43}^{+0.63}$ & -  			  &  3.11e-14   &  $0.191$  &  $0.712$  &  170.5    &  $953\pm 238 $   & $2.73\pm 0.68 $  & $ 0.782_{-0.227}^{+0.250}$ & $ 2.91_{-0.84}^{+0.93} $    \\
SWXCS J105946+5348.2   & 2.12  &  $2.05_{-0.31}^{+0.30}$  &  $1.90_{-0.29}^{+0.28}$  & $0.75_{-0.34}^{+0.53}$ & -  			  &  2.11e-13   &  $0.026$  &  $0.054$  &  177.7    &  $657\pm 101 $   & $7.94\pm 1.22 $  & $ 0.188_{-0.046}^{+0.048}$ & $ 0.387_{-0.094}^{+0.100} $    \\
SWXCS J115811+4529.1   & 1.30  &  $5.29_{-0.99}^{+1.44}$  &  $4.52_{-0.85}^{+1.23}$  & $0.81_{-0.49}^{+0.74}$ & -  			  &  4.10e-14   &  $0.201$  &  $0.664$  &  183.7    &  $879\pm 204 $   & $2.69\pm 0.62 $  & $ 0.321_{-0.049}^{+0.045}$ & $ 1.06_{-0.16}^{+0.15} $    \\
SWXCS J123620+2859.1   & 1.50  &  $4.61_{-1.05}^{+1.52}$  &  $3.78_{-0.86}^{+1.24}$  & $<0.36$ & -  			  &  1.02e-13   &  $0.146$  &  $0.437$  &  140.1    &  $877\pm 246 $   & $3.97\pm 1.11 $  & $ 0.390_{-0.103}^{+0.096}$ & $ 1.17_{-0.31}^{+0.29} $    \\
SWXCS J124312+1704.9   & 0.83  &  $3.49_{-0.69}^{+1.08}$  &  $2.94_{-0.58}^{+0.91}$  & $<0.61$ & -  			  &  3.71e-14   &  $0.019$  &  $0.051$  &  127.5    &  $802\pm 206 $   & $5.31\pm 1.36 $  & $ 0.113_{-0.027}^{+0.026}$ & $ 0.298_{-0.071}^{+0.068} $    \\
SWXCS J131300+0803.0   & 1.14  &  $9.04_{-2.08}^{+3.20}$  &  $7.36_{-1.69}^{+2.61}$  & $0.01_{-0.01}^{+0.44}$ & $<0.277$  		  &  7.54e-14   &  $0.741$  &  $3.068$  &  209.8    &  $1041\pm 306 $  & $2.65\pm 0.78 $  & $ 1.37_{-0.28}^{+0.24}$    & $ 5.68_{-1.14}^{+1.00} $    \\
SWXCS J133055+4200.3   & 1.85  &  $0.71_{-0.02}^{+0.03}$  &  $0.71_{-0.02}^{+0.03}$  & $1.46_{-0.76}^{+1.46}$ & $1.44_{-0.76}^{+1.44}$  &  8.20e-14   &  $0.007$  &  $0.010$  &  755.7    &  $392\pm 19 $    & $5.50\pm 0.26 $  & $ 0.011_{-0.001}^{+0.001}$ & $ 0.015_{-0.001}^{+0.001} $    \\
SWXCS J140637+2743.8   & 1.85  &  $9.79_{-1.10}^{+1.48}$  &  $9.49_{-1.06}^{+1.43}$  & $0.76_{-0.27}^{+0.32}$ & $0.72_{-0.26}^{+0.30}$  &  2.90e-13   &  $3.167$  &  $14.151$ &  935.4    &  $1171\pm 159 $  & $2.91\pm 0.40 $  & $ 3.57_{-0.17}^{+0.16}$    & $ 15.93_{-0.78}^{+0.69} $    \\
SWXCS J140639+2735.8   & 1.85  &  $4.33_{-0.51}^{+0.61}$  &  $4.19_{-0.50}^{+0.59}$  & $0.46_{-0.28}^{+0.37}$ & $0.42_{-0.26}^{+0.34}$  &  1.66e-13   &  $0.269$  &  $0.795$  &  458.0    &  $919\pm 124 $   & $3.97\pm 0.53 $  & $ 0.480_{-0.053}^{+0.052}$ & $ 1.42_{-0.16}^{+0.15} $    \\
SWXCS J140728+2749.3   & 2.12  &  $5.02_{-0.93}^{+0.91}$  &  $4.66_{-0.87}^{+0.85}$  & $1.60_{-0.73}^{+1.03}$ & $1.29_{-0.59}^{+0.83}$  &  1.08e-13   &  $0.083$  &  $0.273$  &  324.9    &  $1009\pm 189 $  & $5.69\pm 1.06 $  & $ 0.199_{-0.035}^{+0.035}$ & $ 0.653_{-0.116}^{+0.116} $    \\
SWXCS J143646+2752.0   & 0.75  &  $1.61_{-0.46}^{+0.57}$  &  $1.45_{-0.42}^{+0.51}$  & $0.12_{-0.12}^{+0.26}$ & -  			  &  1.31e-14   &  $0.028$  &  $0.057$  &  134.4    &  $518\pm 167 $   & $2.10\pm 0.68 $  & $ 0.081_{-0.027}^{+0.029}$ & $ 0.161_{-0.054}^{+0.057} $    \\
SWXCS J151324+3057.6   & 0.55  &  $2.60_{-0.51}^{+0.66}$  &  $2.34_{-0.46}^{+0.59}$  & $<0.04$ & -  			  &  1.08e-14   &  $0.001$  &  $0.003$  &  116.6    &  $734\pm 166 $   & $8.91\pm 2.02 $  & $ 0.029_{-0.007}^{+0.007}$ & $ 0.069_{-0.017}^{+0.017} $    \\
SWXCS J155742+3530.4   & 2.12  &  $8.22_{-0.34}^{+0.35}$  &  $8.12_{-0.34}^{+0.34}$  & $0.34_{-0.07}^{+0.08}$ & $0.35_{-0.07}^{+0.08}$  &  1.87e-12   &  $1.194$  &  $4.840$  &  11320.4  &  $1362\pm 74 $   & $8.23\pm 0.45 $  & $ 2.19_{-0.05}^{+0.05}$    & $ 8.89_{-0.21}^{+0.19} $    \\
SWXCS J173721+4618.6   & 1.46  &  $4.16_{-0.52}^{+0.56}$  &  $4.04_{-0.51}^{+0.54}$  & $0.43_{-0.30}^{+0.37}$ & $0.39_{-0.28}^{+0.34}$  &  8.80e-14   &  $0.116$  &  $0.334$  &  475.3    &  $913\pm 122 $   & $4.25\pm 0.57 $  & $ 0.227_{-0.020}^{+0.019}$ & $ 0.655_{-0.057}^{+0.056} $    \\
SWXCS J175640+3329.5   & 1.26  &  $1.00_{-0.10}^{+0.09}$  &  $0.95_{-0.09}^{+0.09}$  & $0.13_{-0.06}^{+0.10}$ & -  			  &  1.04e-13   &  $0.119$  &  $0.214$  &  138.2    &  $429\pm 43 $    & $2.19\pm 0.22 $  & $ 0.135_{-0.012}^{+0.012}$ & $ 0.242_{-0.021}^{+0.021} $    \\
SWXCS J215354+1653.8   & 1.89  &  $3.33_{-0.48}^{+0.61}$  &  $3.18_{-0.46}^{+0.58}$  & $0.35_{-0.29}^{+0.41}$ & $0.30_{-0.25}^{+0.36}$  &  1.39e-13   &  $0.204$  &  $0.528$  &  337.6    &  $800\pm 134 $   & $3.60\pm 0.60 $  & $ 0.513_{-0.111}^{+0.124}$ & $ 1.323_{-0.289}^{+0.32} $    \\
SWXCS J222600-5712.8   & 2.12  &  $6.06_{-0.80}^{+1.15}$  &  $5.68_{-0.75}^{+1.08}$  & $1.36_{-0.57}^{+0.80}$ & $1.15_{-0.49}^{+0.68}$  &  3.00e-13   &  $0.125$  &  $0.458$  &  398.2    &  $1144\pm 189 $  & $8.17\pm 1.35 $  & $ 0.434_{-0.065}^{+0.065}$ & $ 1.59_{-0.24}^{+0.24} $    \\
SWXCS J222917-1101.1   & 0.79  &  $1.31_{-0.10}^{+0.09}$  &  $1.27_{-0.09}^{+0.09}$  & $0.29_{-0.11}^{+0.16}$ & -  			  &  3.20e-14   &  $0.037$  &  $0.067$  &  216.3    &  $499\pm 39 $    & $2.50\pm 0.20 $  & $ 0.040_{-0.003}^{+0.003}$ & $ 0.072_{-0.005}^{+0.005} $    \\
SWXCS J222953+1943.9   & 2.05  &  $4.84_{-0.56}^{+0.65}$  &  $4.73_{-0.55}^{+0.64}$  & $0.68_{-0.28}^{+0.37}$ & $0.65_{-0.27}^{+0.35}$  &  1.20e-13   &  $0.189$  &  $0.597$  &  679.7    &  $982\pm 128 $   & $4.28\pm 0.56 $  & $ 0.200_{-0.008}^{+0.008}$ & $ 0.632_{-0.025}^{+0.025} $    \\
SWXCS J224207+2333.9   & 1.53  &  $9.74_{-2.43}^{+3.36}$  &  $7.86_{-1.96}^{+2.71}$  & $<0.35$ & $<0.22$  		  &  1.18e-13   &  $1.950$  &  $8.322$  &  237.8    &  $989\pm 296 $   & $2.26\pm 0.68 $  & $ 2.26_{-0.25}^{+0.20}$    & $ 9.63_{-1.07}^{+0.87} $    \\
SWXCS J230754-6815.1   & 0.90  &  $0.77_{-0.09}^{+0.05}$  &  $0.74_{-0.09}^{+0.05}$  & $0.28_{-0.11}^{+0.25}$ & -  			  &  4.40e-14   &  $0.011$  &  $0.017$  &  147.1    &  $396\pm 38 $    & $3.63\pm 0.35 $  & $ 0.015_{-0.001}^{+0.001}$ & $ 0.024_{-0.002}^{+0.002} $    \\
SWXCS J232248+0548.2   & 1.53  &  $3.58_{-0.16}^{+0.30}$  &  $3.54_{-0.16}^{+0.30}$  & $0.60_{-0.17}^{+0.19}$ & $0.60_{-0.17}^{+0.19}$  &  2.01e-13   &  $0.452$  &  $1.212$  &  1549.4   &  $825\pm 61 $    & $3.21\pm 0.24 $  & $ 0.510_{-0.014}^{+0.014}$ & $ 1.37_{-0.04}^{+0.03} $    \\
SWXCS J232345-3130.8   & 1.34  &  $5.79_{-0.77}^{+0.99}$  &  $5.30_{-0.70}^{+0.91}$  & $0.38_{-0.25}^{+0.33}$ & $0.30_{-0.20}^{+0.26}$  &  7.35e-14   &  $1.961$  &  $6.554$  &  338.0    &  $740\pm 116 $   & $1.59\pm 0.25 $  & $ 2.09_{-0.15}^{+0.14}$    & $ 6.99_{-0.50}^{+0.45} $    \\
SWXCS J233518-6621.7   & 1.18  &  $4.48_{-0.54}^{+1.17}$  &  $4.11_{-0.49}^{+1.07}$  & $1.43_{-0.63}^{+0.93}$ & $1.10_{-0.49}^{+0.71}$  &  5.37e-14   &  $0.239$  &  $0.731$  &  226.2    &  $844\pm 163 $   & $2.66\pm 0.51 $  & $ 0.395_{-0.051}^{+0.051}$ & $ 1.21_{-0.16}^{+0.15} $    \\
SWXCS J233617-3136.5   & 2.12  &  $2.57_{-0.13}^{+0.13}$  &  $2.54_{-0.13}^{+0.13}$  & $0.89_{-0.16}^{+0.19}$ & $0.89_{-0.16}^{+0.19}$  &  7.48e-13   &  $0.069$  &  $0.157$  &  1400.1   &  $770\pm 47 $    & $10.60\pm 0.65$  & $ 0.316_{-0.019}^{+0.020}$ & $ 0.721_{-0.044}^{+0.046} $    \\
\hline
\end{longtable}
\end{landscape}
}


\appendix

\section{Spectral analysis of single sources}

In this Section we present the folded spectra for the 46 sources with redshift analyzed in this paper. The best fit has been obtained assuming a single-temperature {\tt mekal} model, as described in the text.  
Note that the best-fit parameters of the ICM have been obtained by freezing the redshift parameter to the
optical value when available, and to the X-ray redshift $z_X$ in the other cases.  In the labels we indicate the redshift value used for the X-ray spectral analysis rounded to the third decimal digit or
the last significant digit.

\begin{figure*}
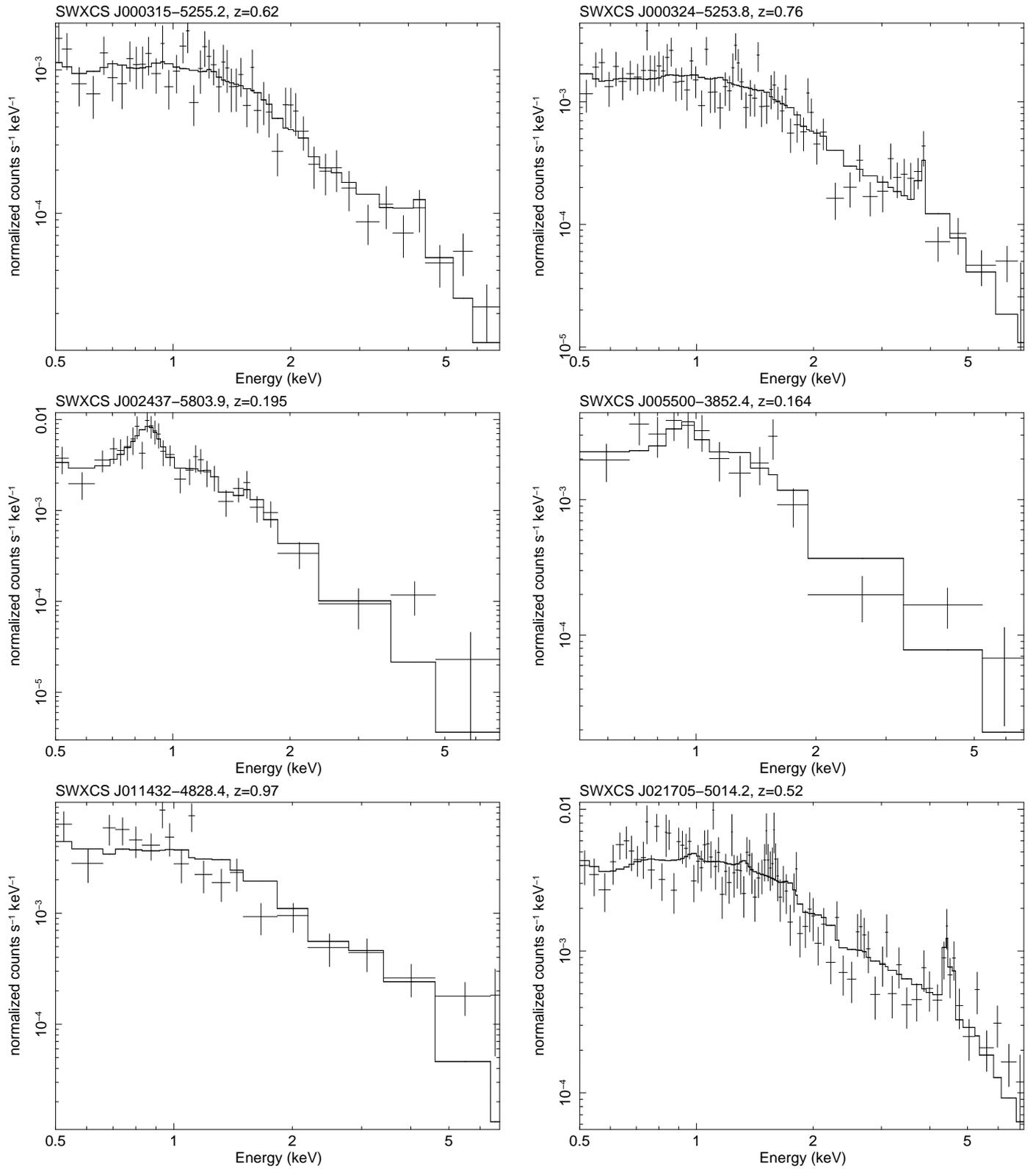

\includegraphics[width=7cm,angle=-90]{cl_data_SWJ000315-525510.ps}
\includegraphics[width=7cm,angle=-90]{cl_data_SWJ000324-525350.ps}
\includegraphics[width=7cm,angle=-90]{cl_data_SWJ002437-580353.ps}
\includegraphics[width=7cm,angle=-90]{cl_data_SWJ005500-385226.ps}
\includegraphics[width=7cm,angle=-90]{cl_data_SWJ011432-482824.ps}
\includegraphics[width=7cm,angle=-90]{cl_data_SWJ021705-501409.ps}
\caption{Folded spectrum with best-fit model for the 46 sources with redshift analyzed in this paper.  
The redshift in the label is rounded to the third decimal digit or to the last significant digit.}
\label{spectra1}
\end{figure*}

\renewcommand{\thefigure}{A.\arabic{figure}}
\addtocounter{figure}{-1}

\begin{figure*}
\includegraphics[width=7cm,angle=-90]{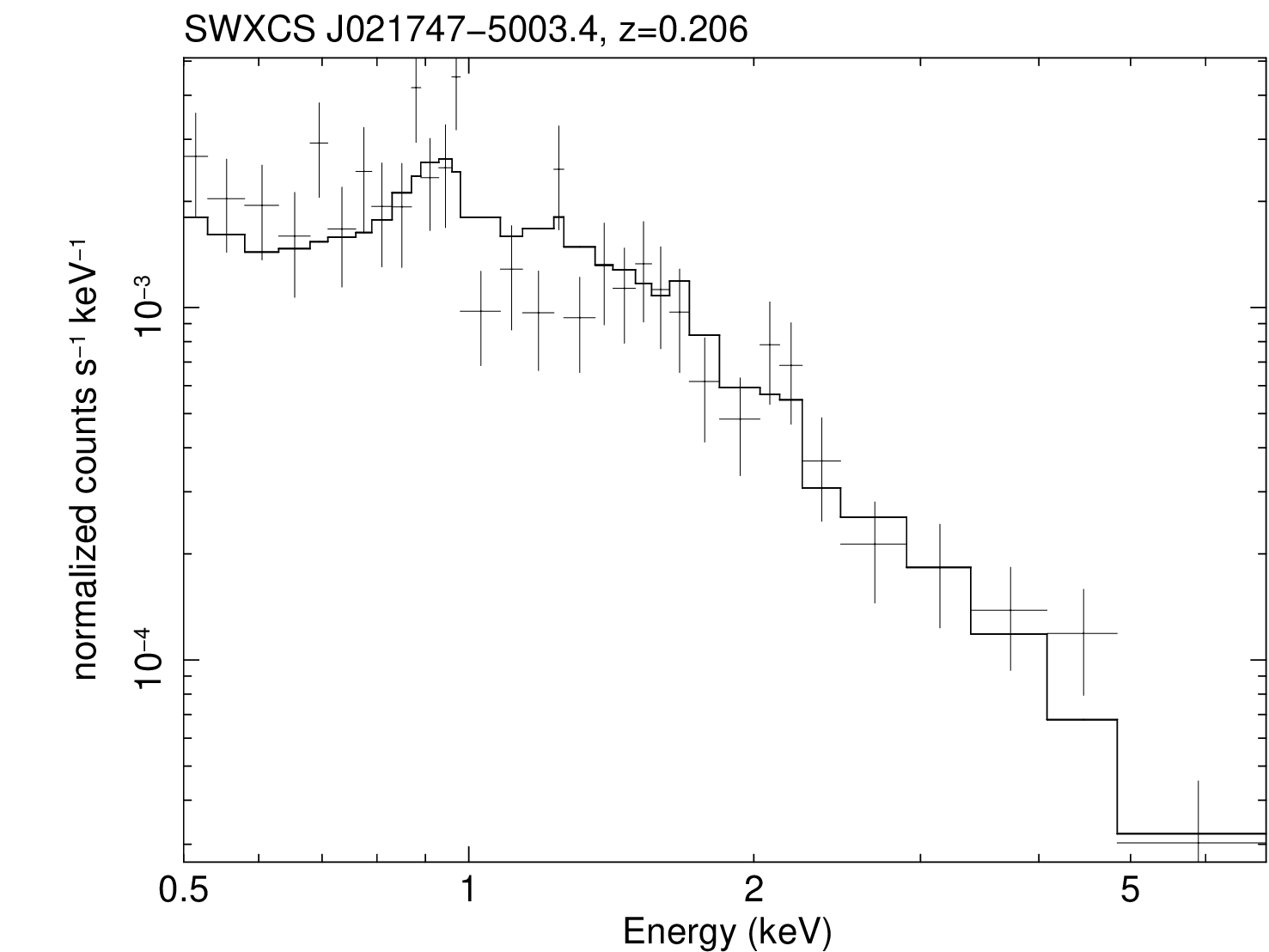}
\includegraphics[width=7cm,angle=-90]{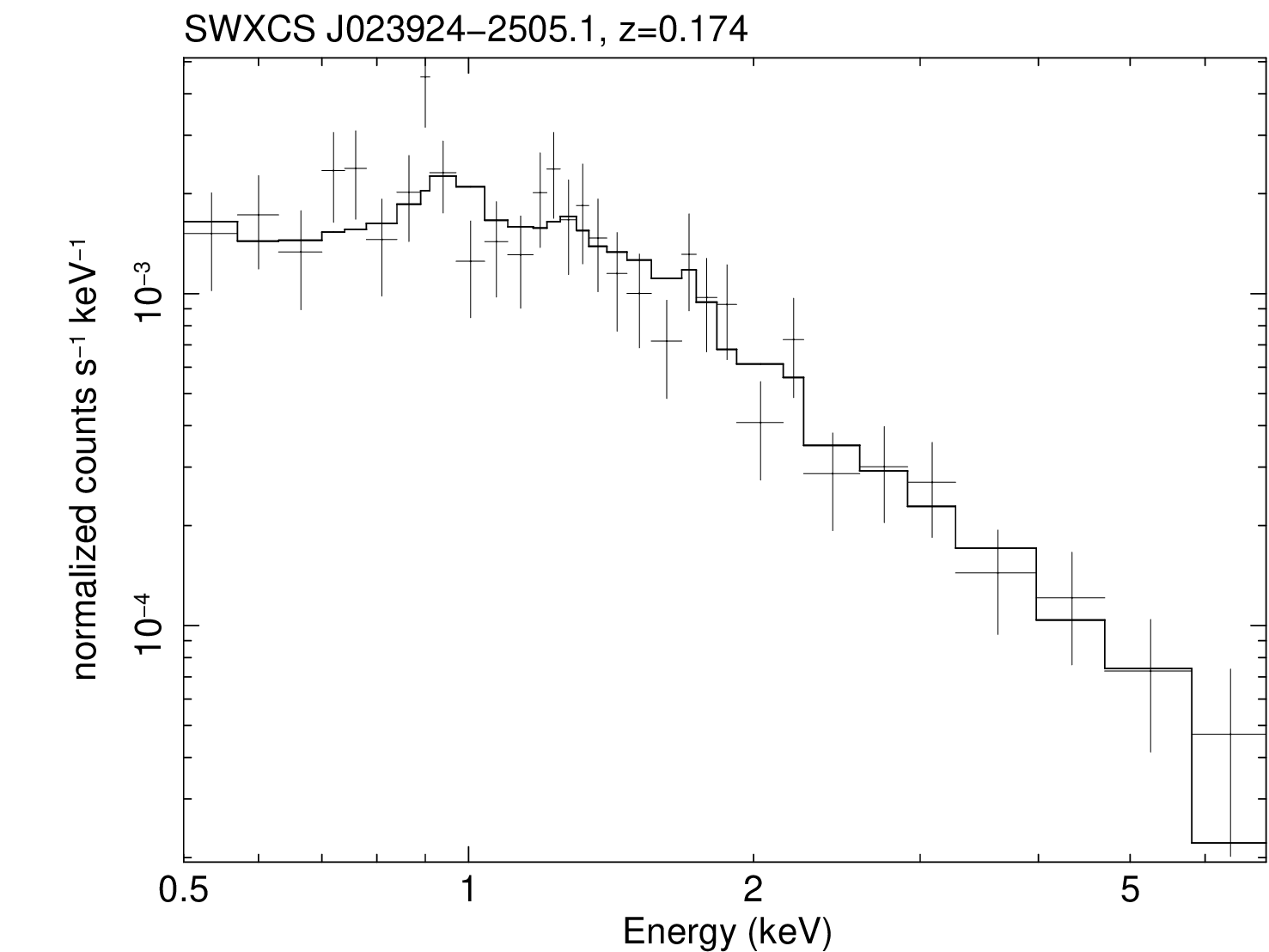}
\includegraphics[width=7cm,angle=-90]{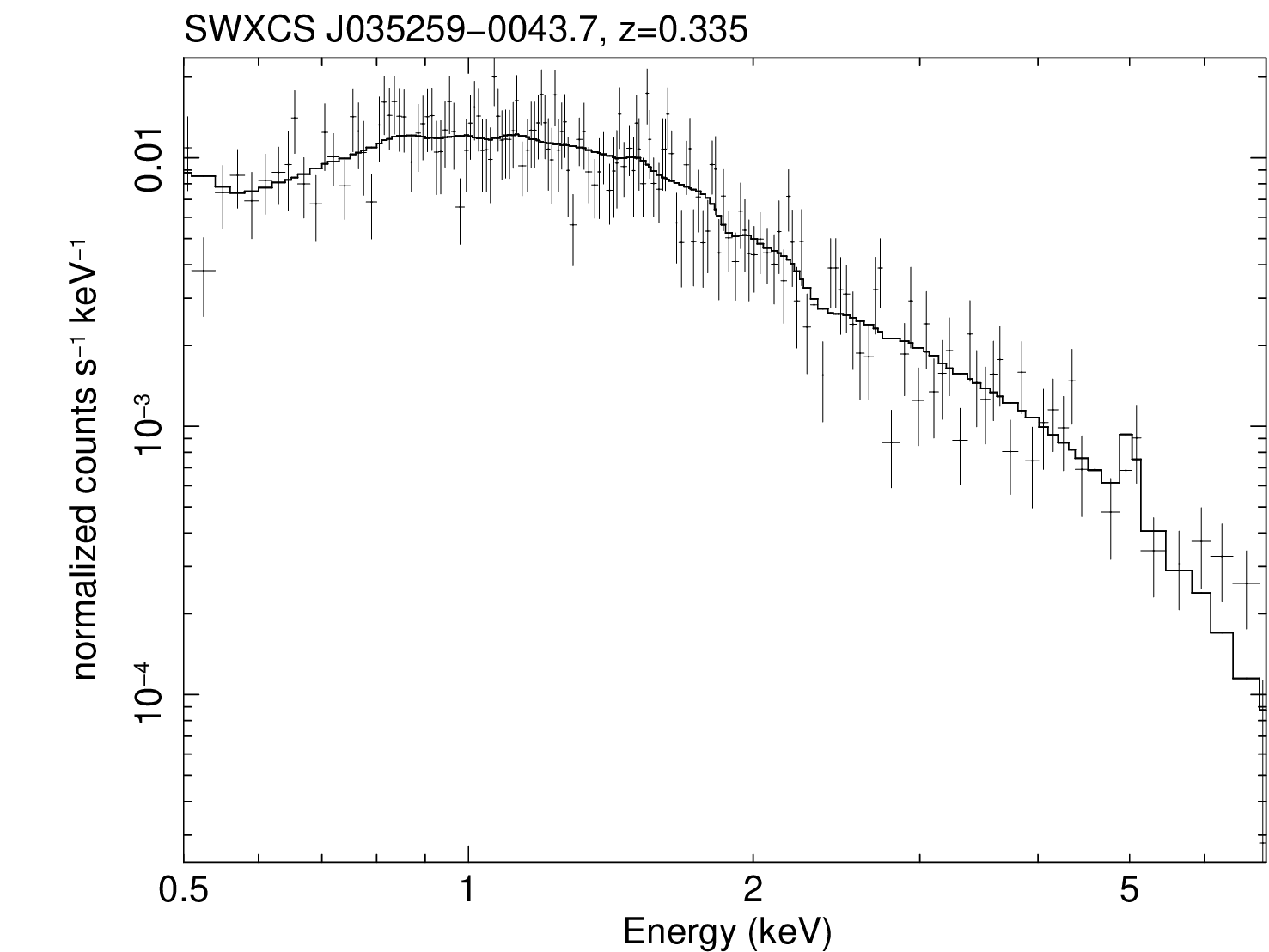}
\includegraphics[width=7cm,angle=-90]{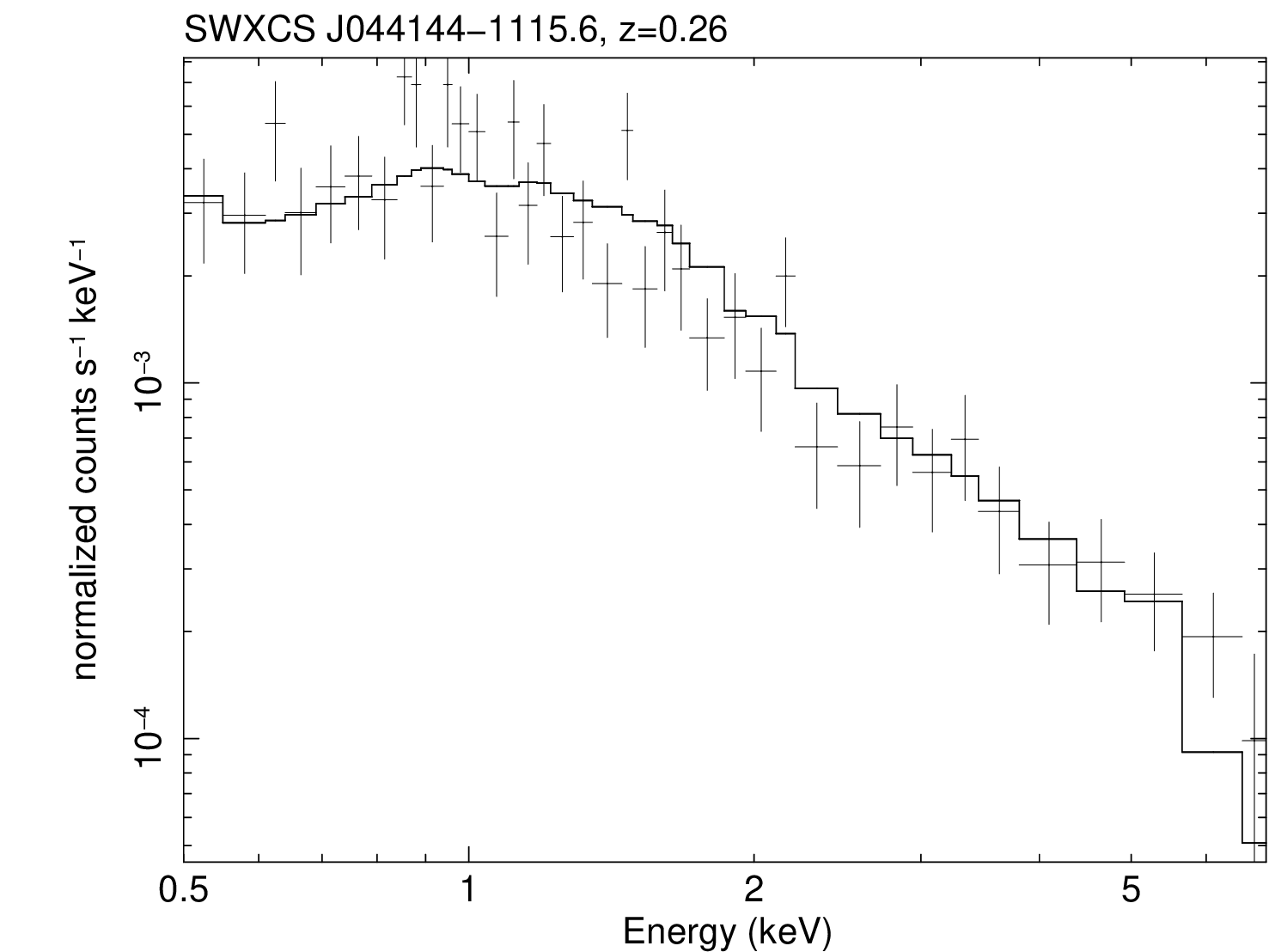}
\includegraphics[width=7cm,angle=-90]{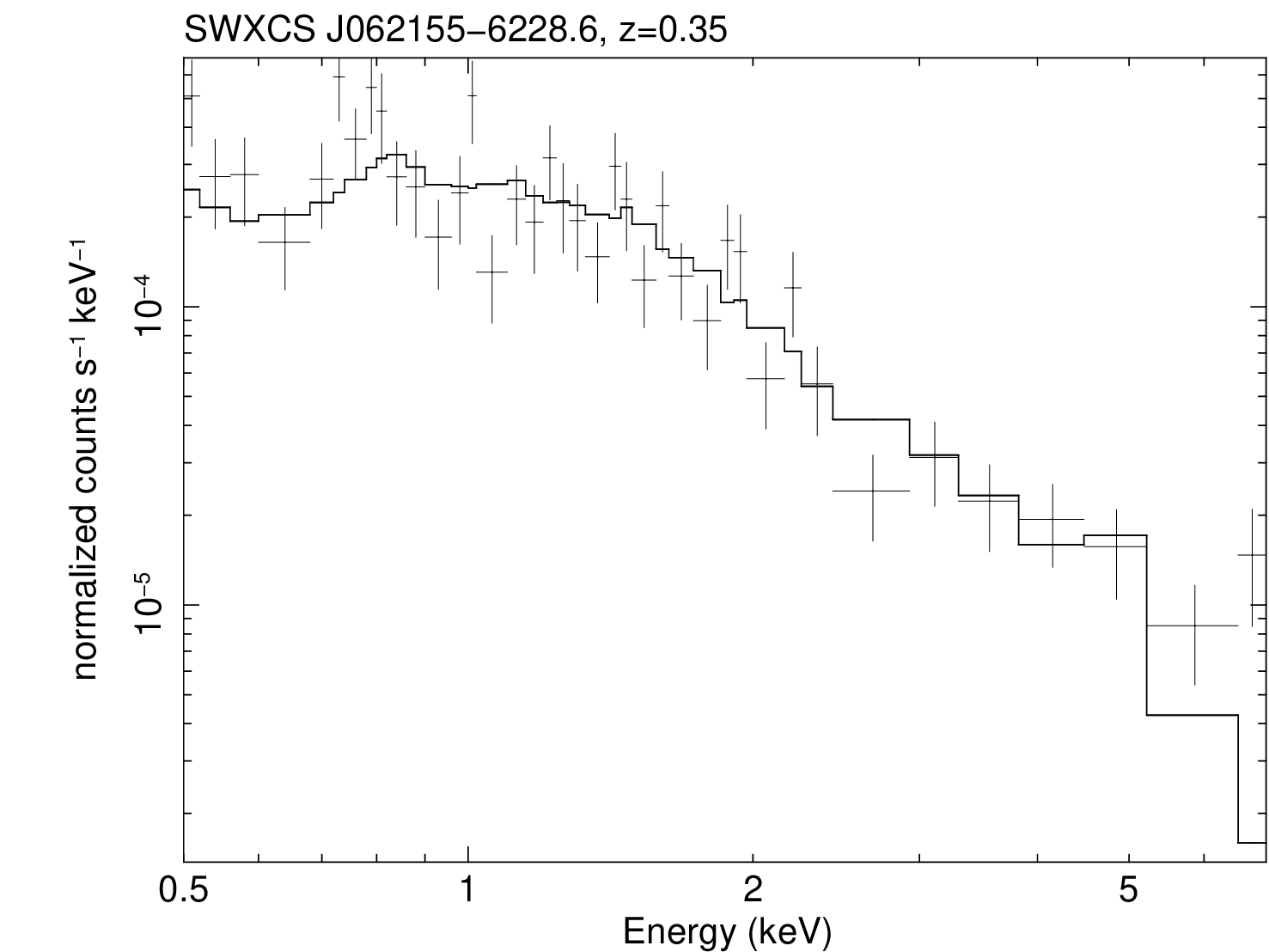}
\includegraphics[width=7cm,angle=-90]{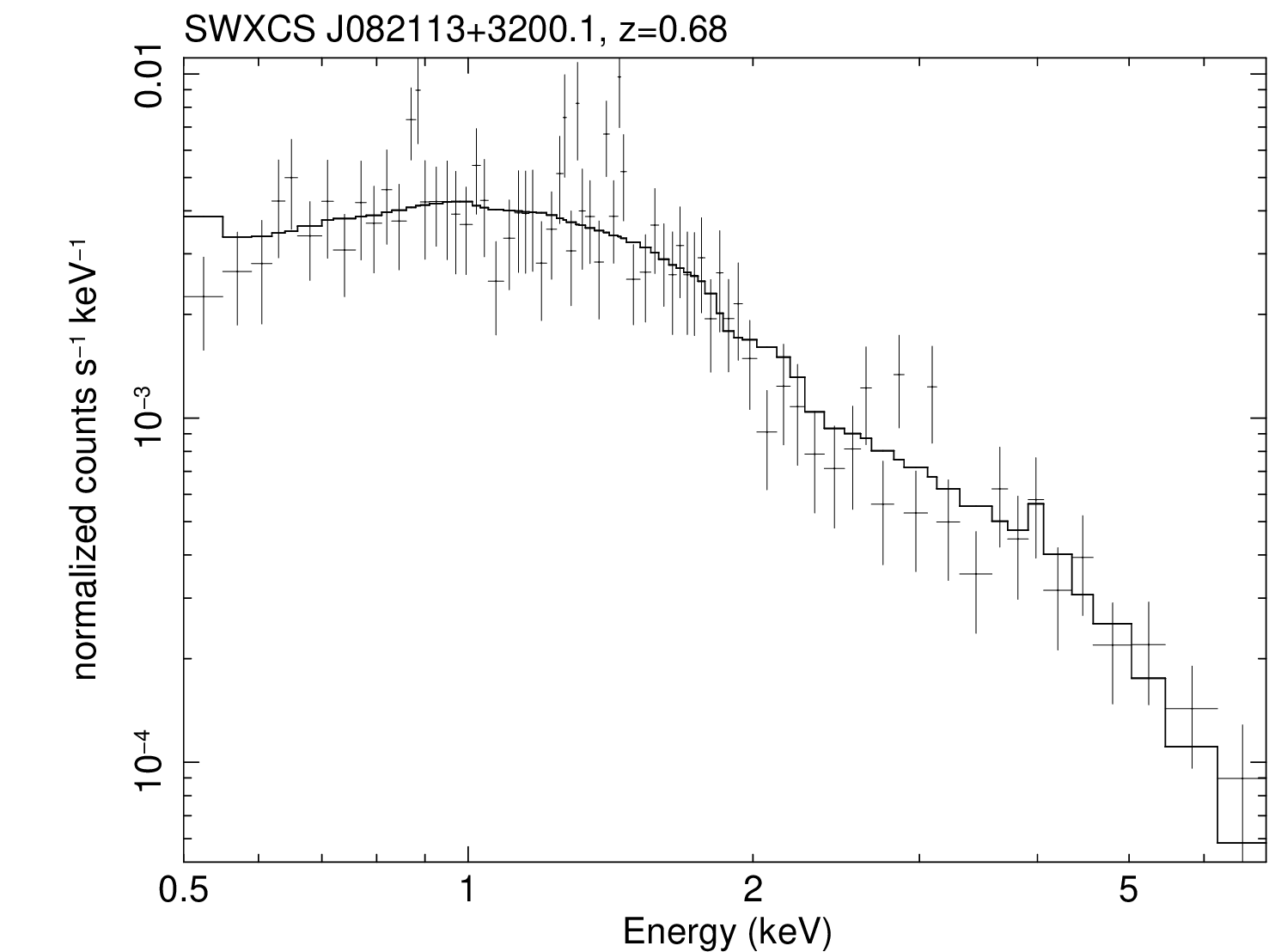}
\caption{continued.}
\end{figure*}

\renewcommand{\thefigure}{\arabic{figure}}

\renewcommand{\thefigure}{A.\arabic{figure}}
\addtocounter{figure}{-1}

\begin{figure*}
\includegraphics[width=7cm,angle=-90]{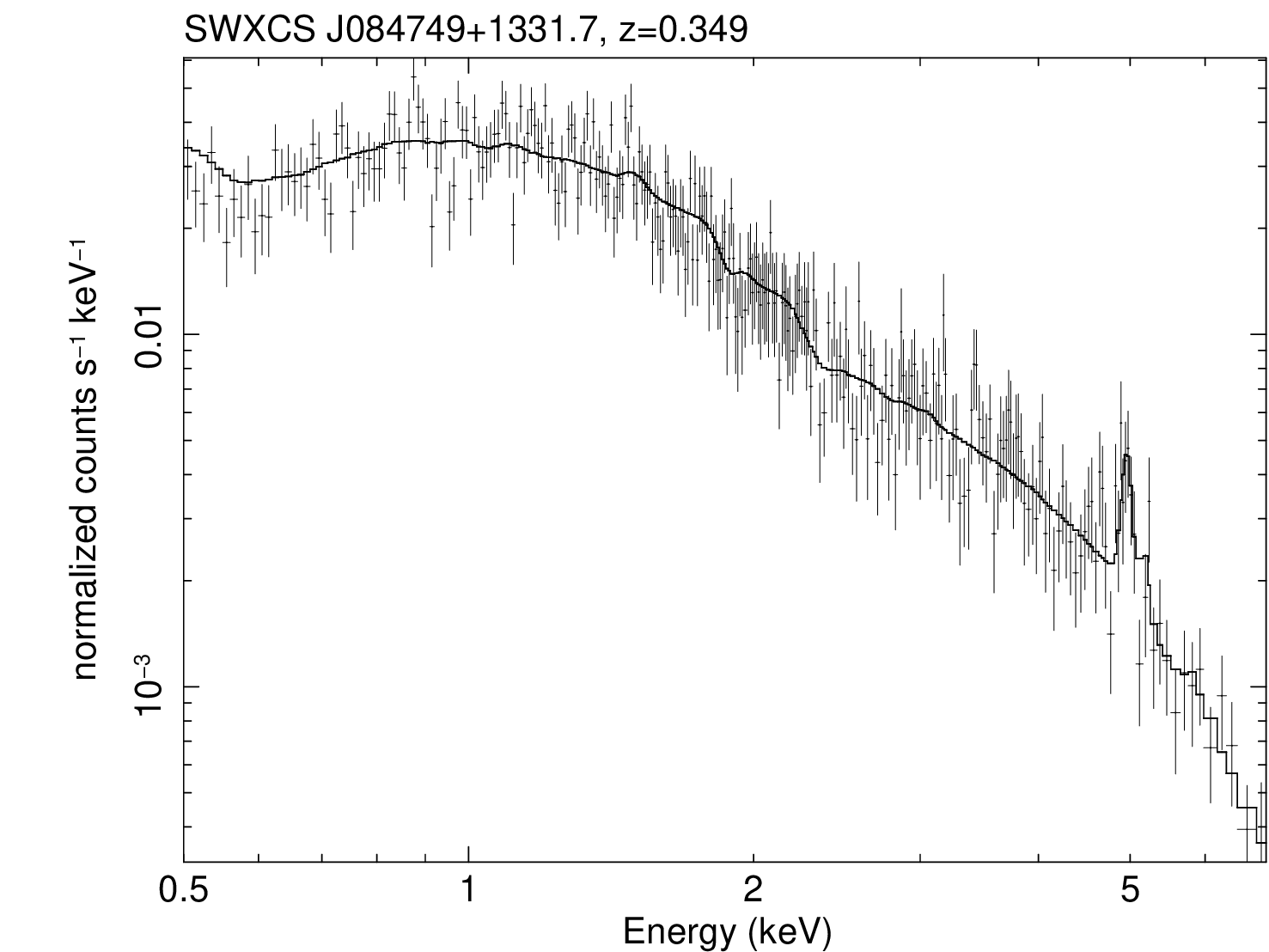}
\includegraphics[width=7cm,angle=-90]{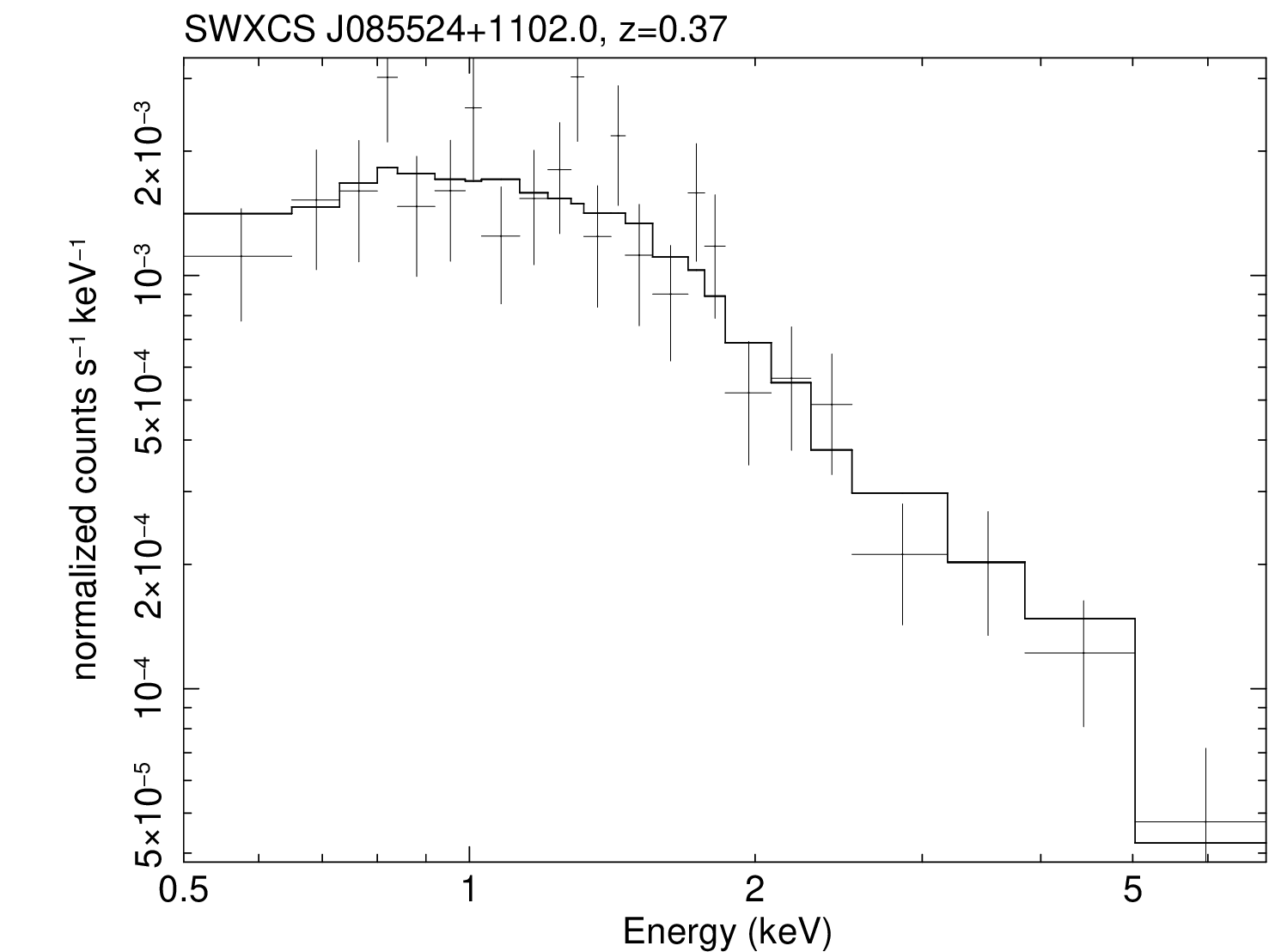}
\includegraphics[width=7cm,angle=-90]{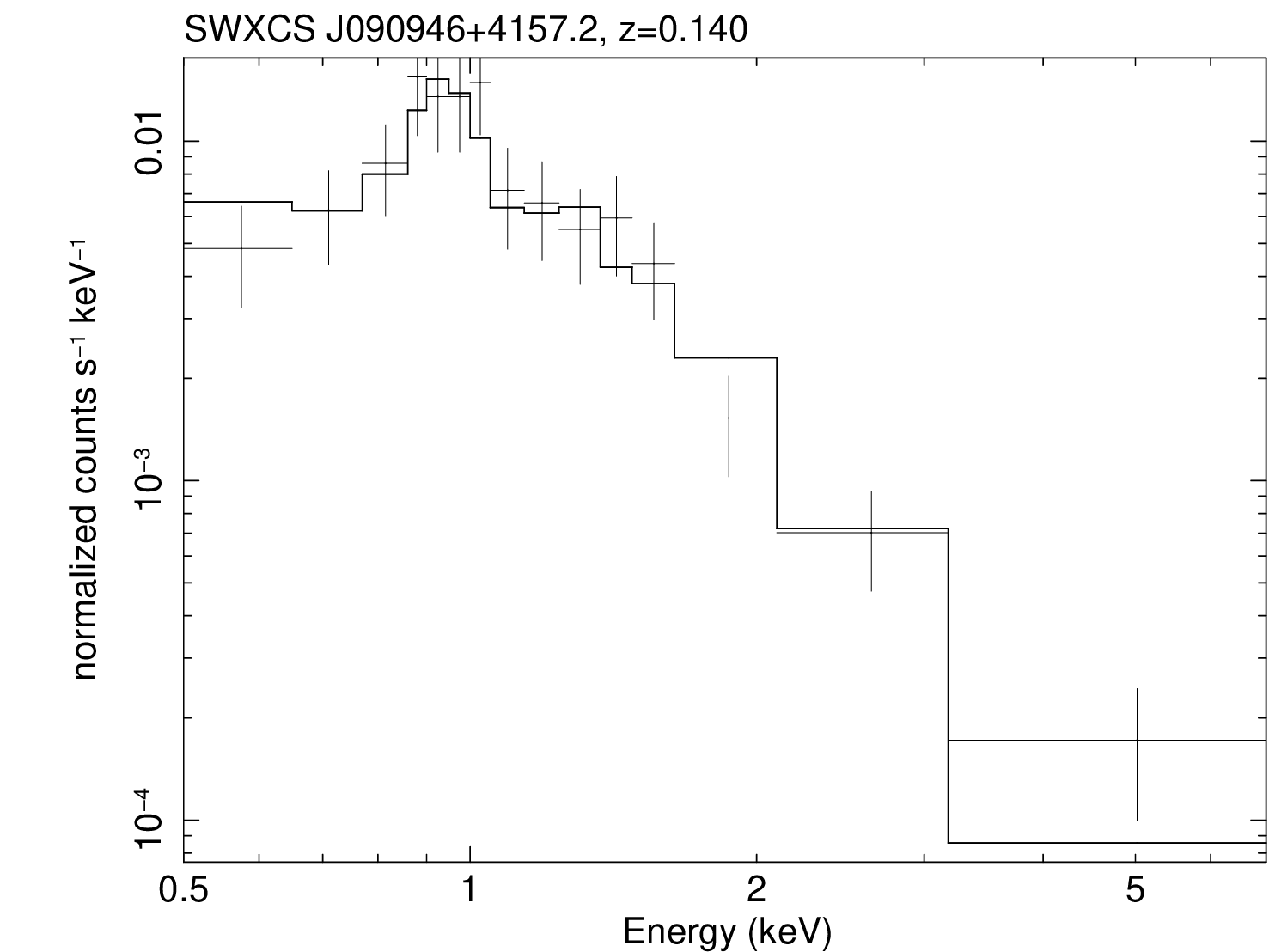}
\includegraphics[width=7cm,angle=-90]{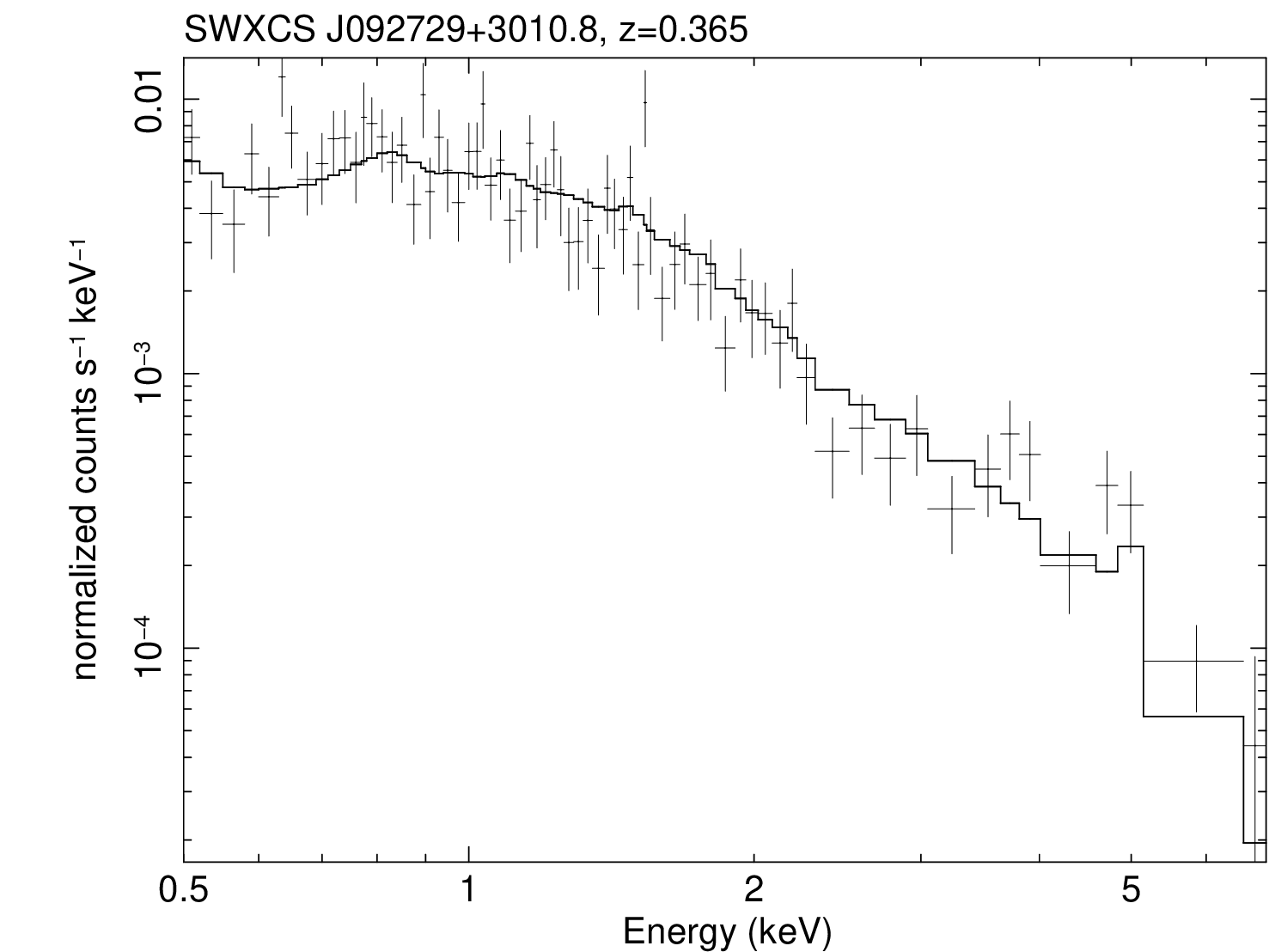}
\includegraphics[width=7cm,angle=-90]{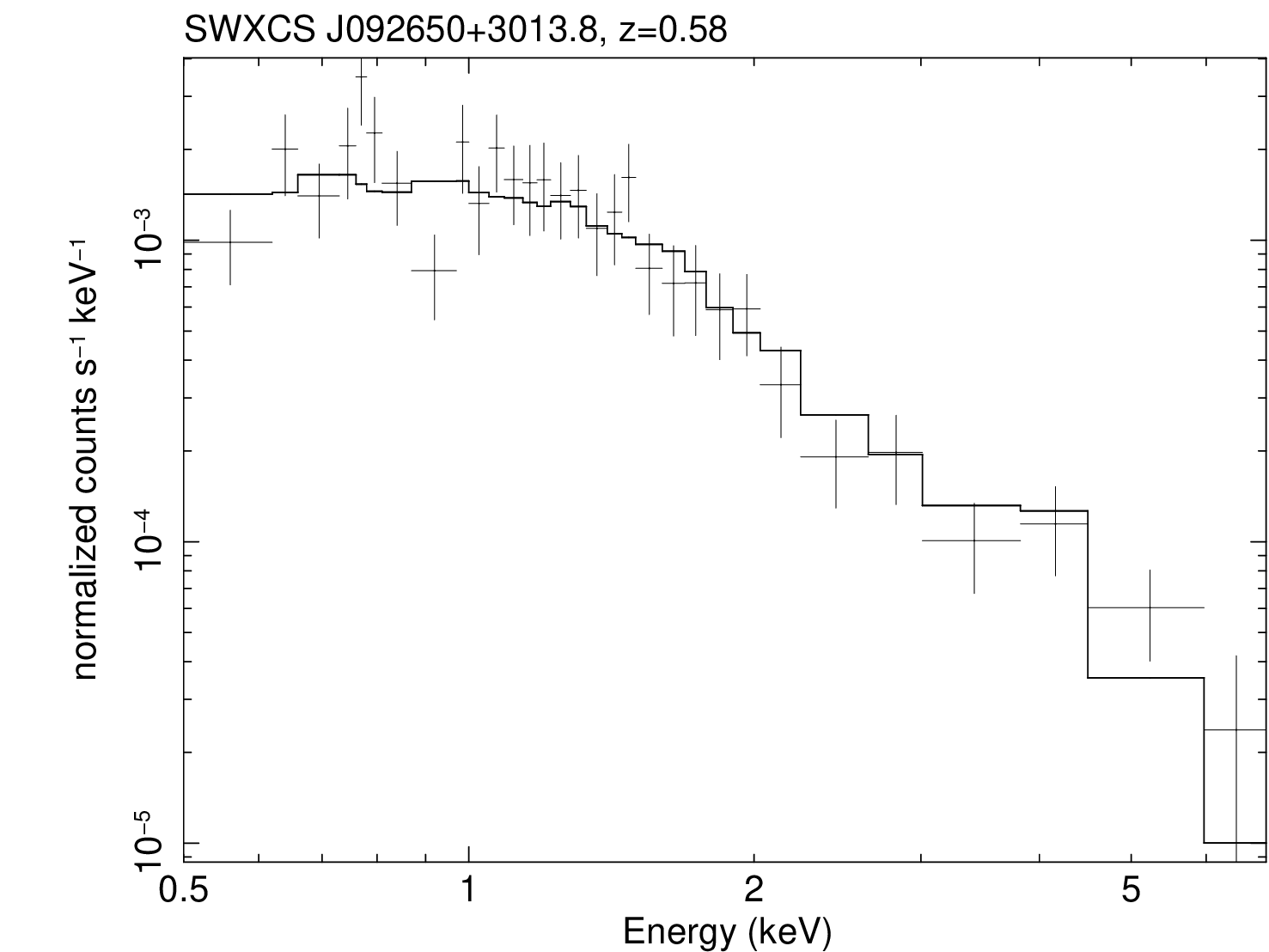}
\includegraphics[width=7cm,angle=-90]{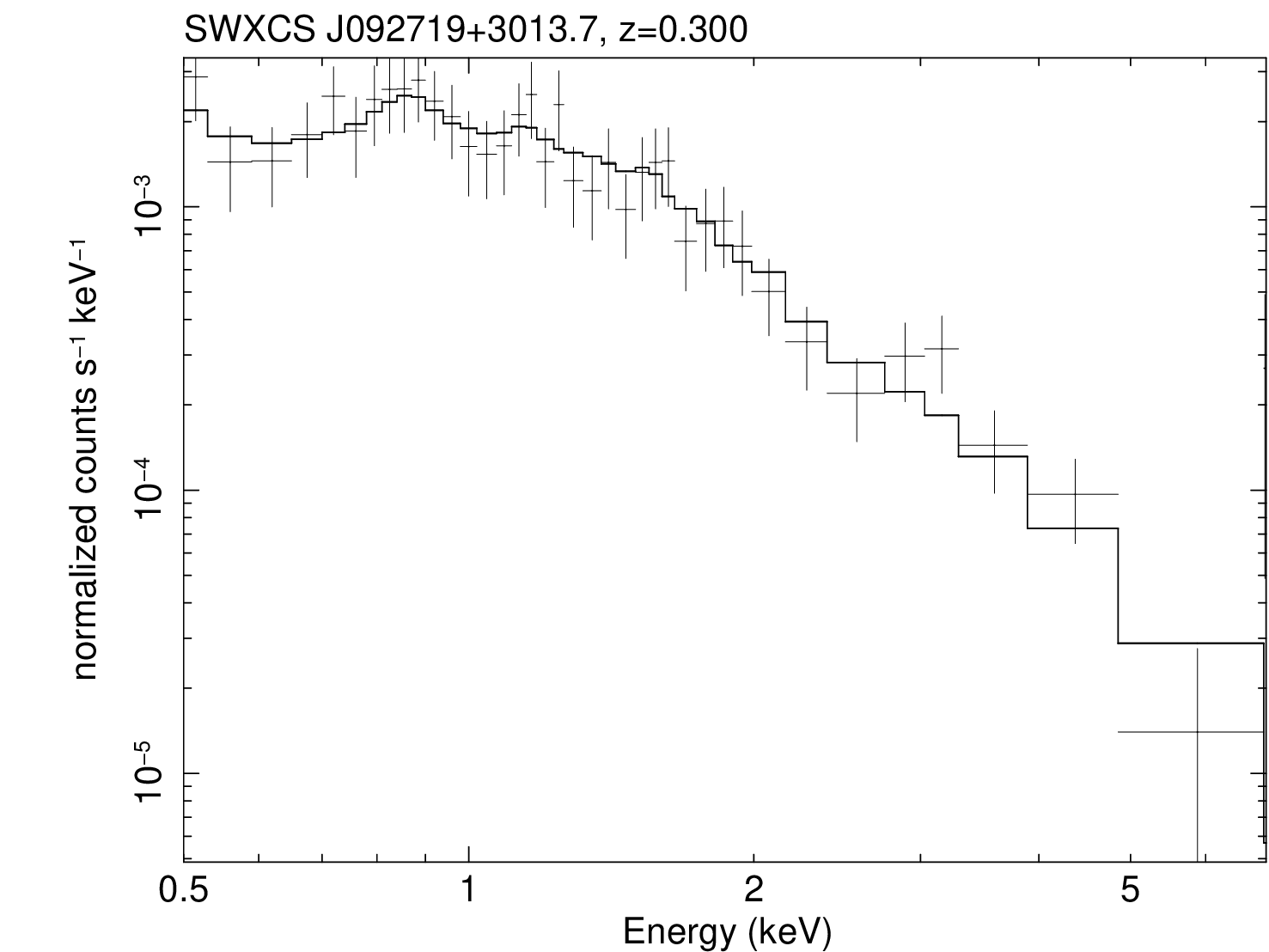}
\caption{continued.}
\end{figure*}

\renewcommand{\thefigure}{\arabic{figure}}

\renewcommand{\thefigure}{A.\arabic{figure}}
\addtocounter{figure}{-1}

\begin{figure*}
\includegraphics[width=7cm,angle=-90]{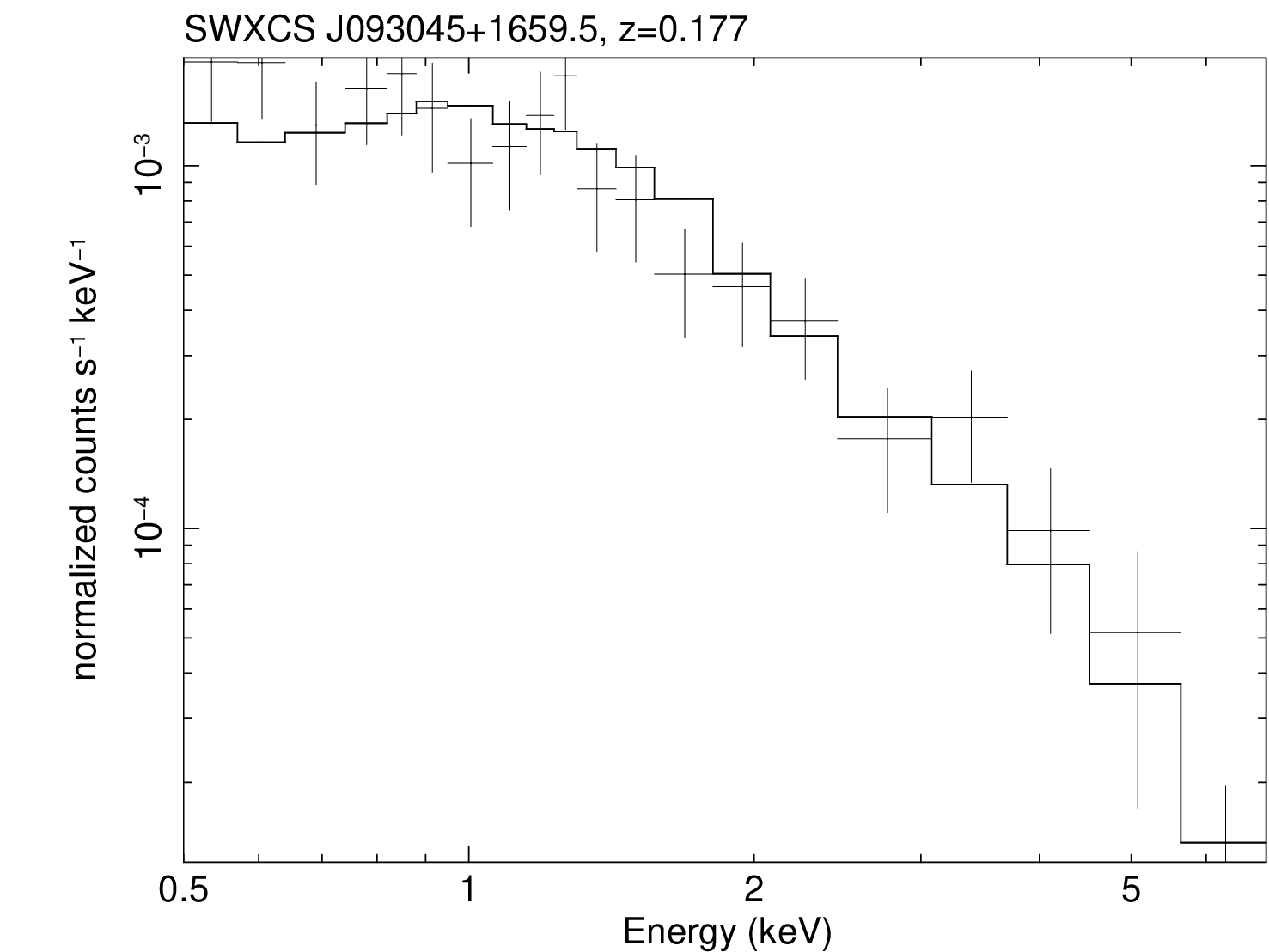}
\includegraphics[width=7cm,angle=-90]{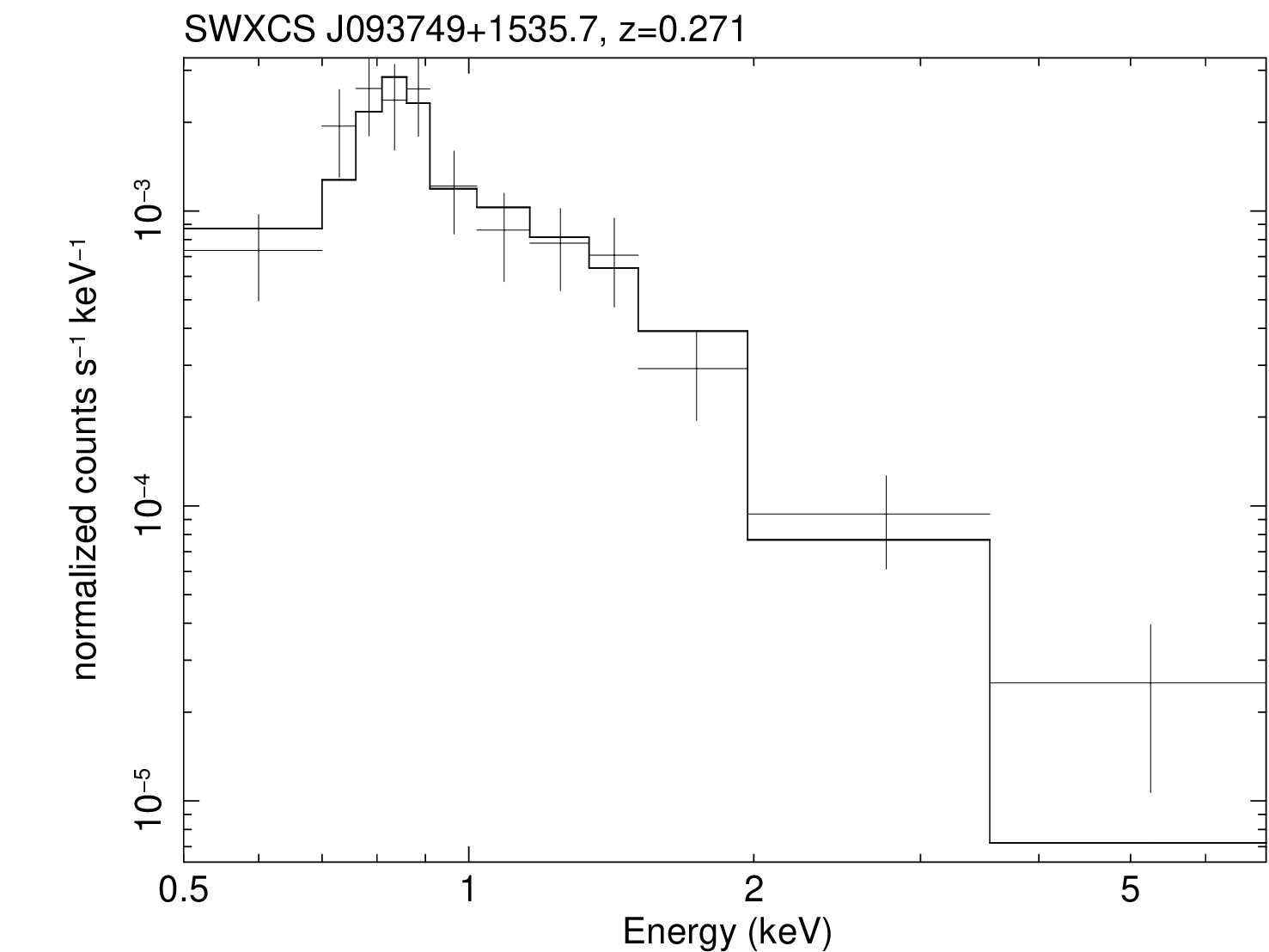}
\includegraphics[width=7cm,angle=-90]{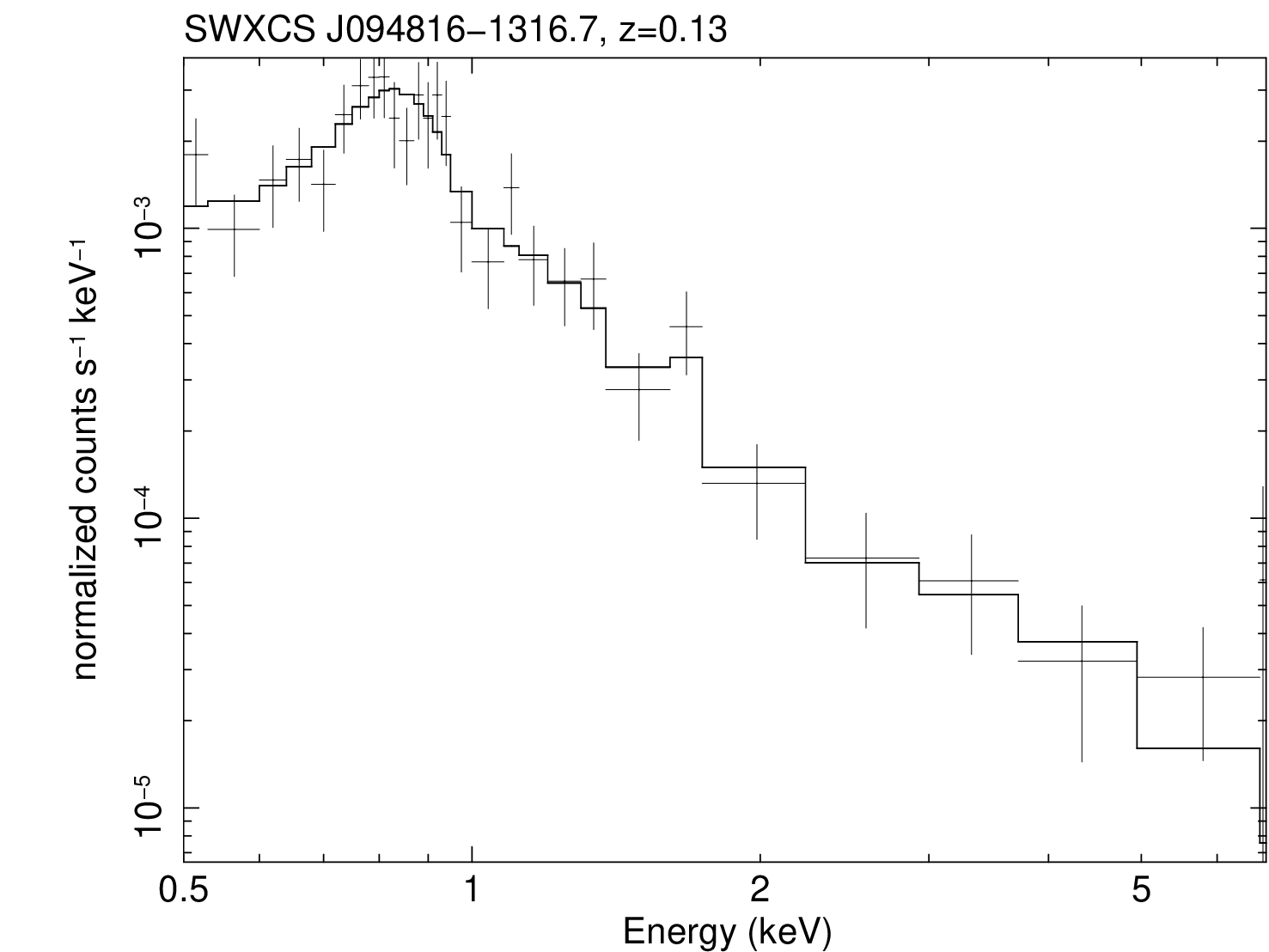}
\includegraphics[width=7cm,angle=-90]{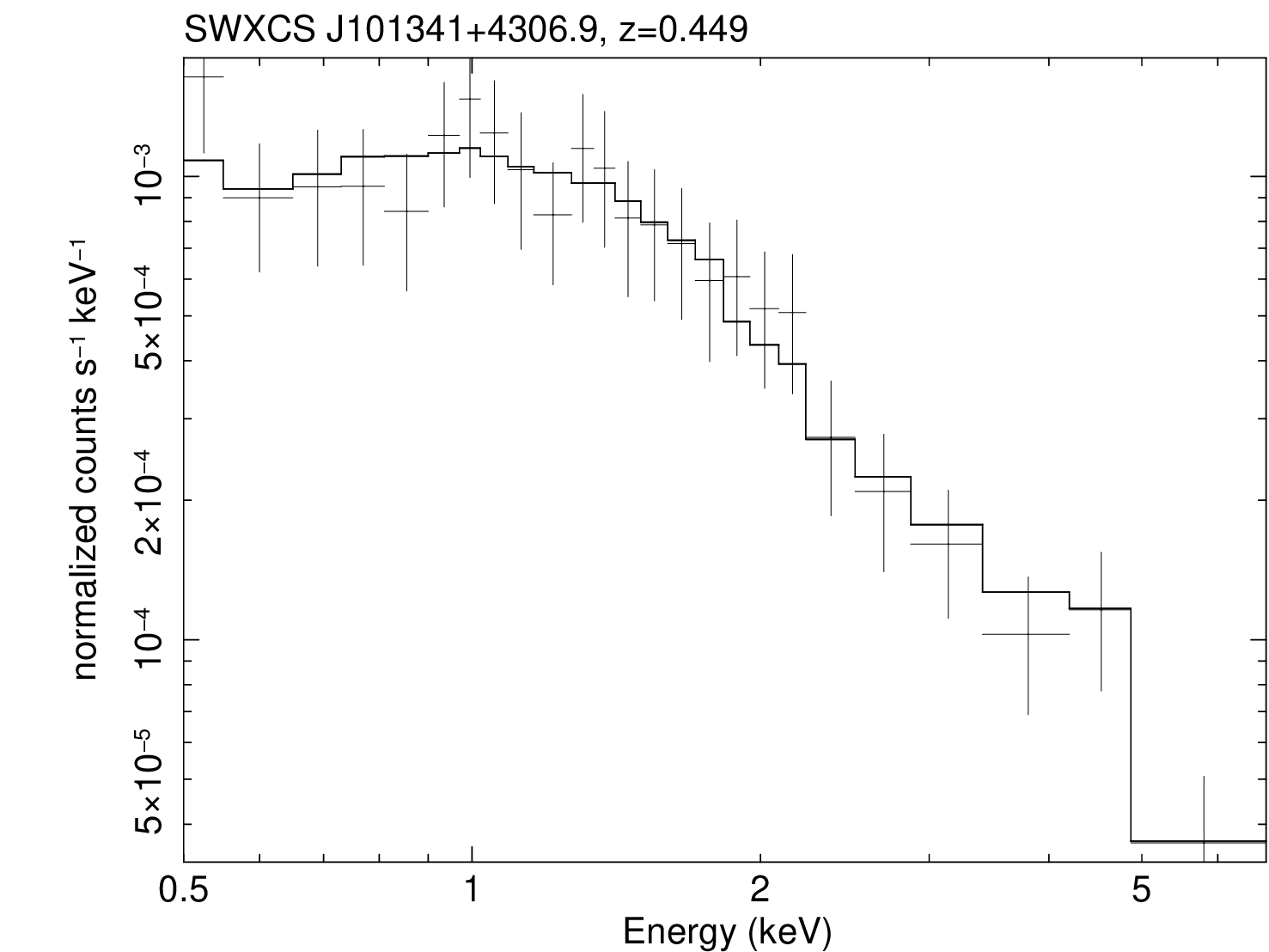}
\includegraphics[width=7cm,angle=-90]{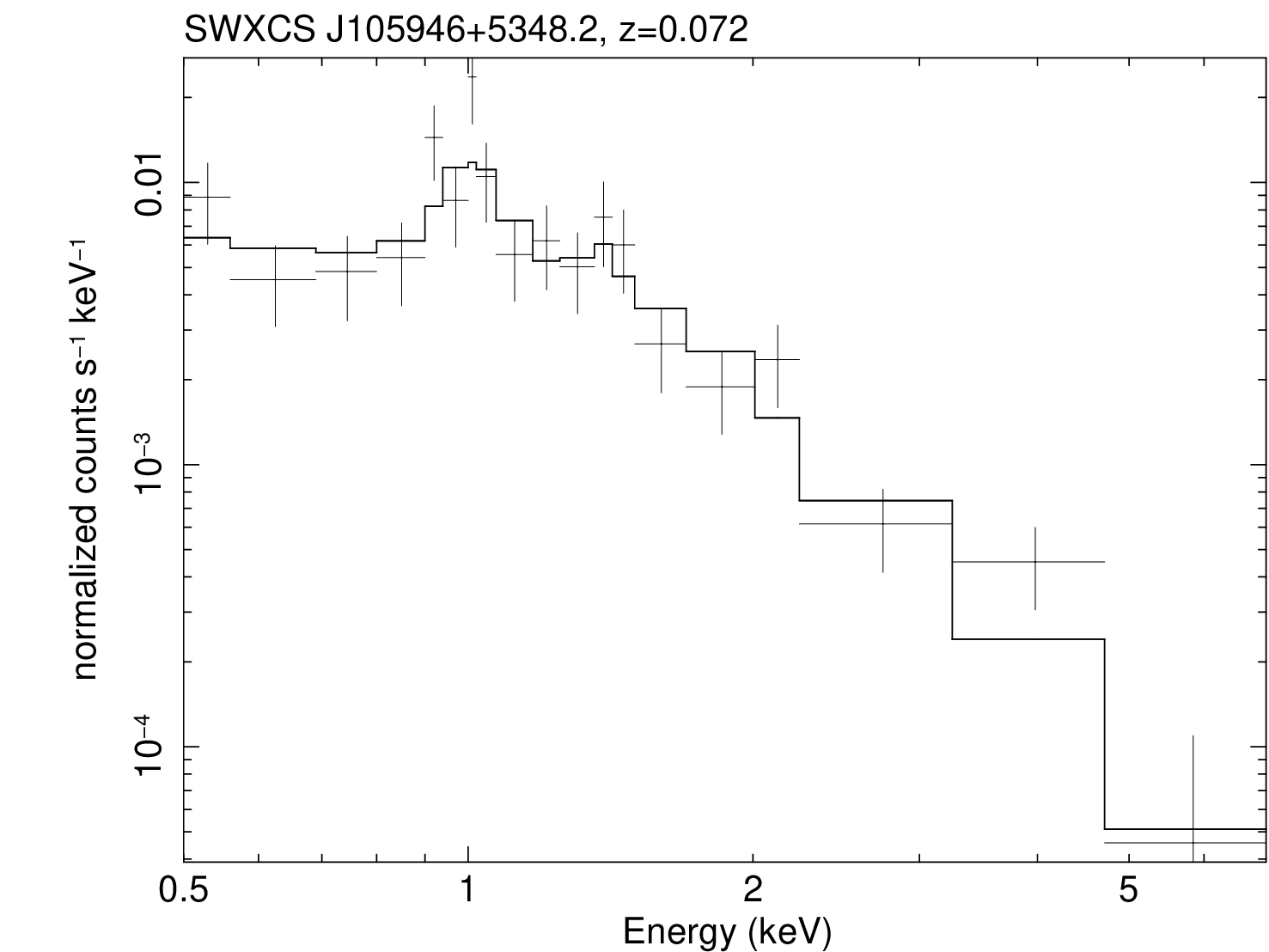}
\includegraphics[width=7cm,angle=-90]{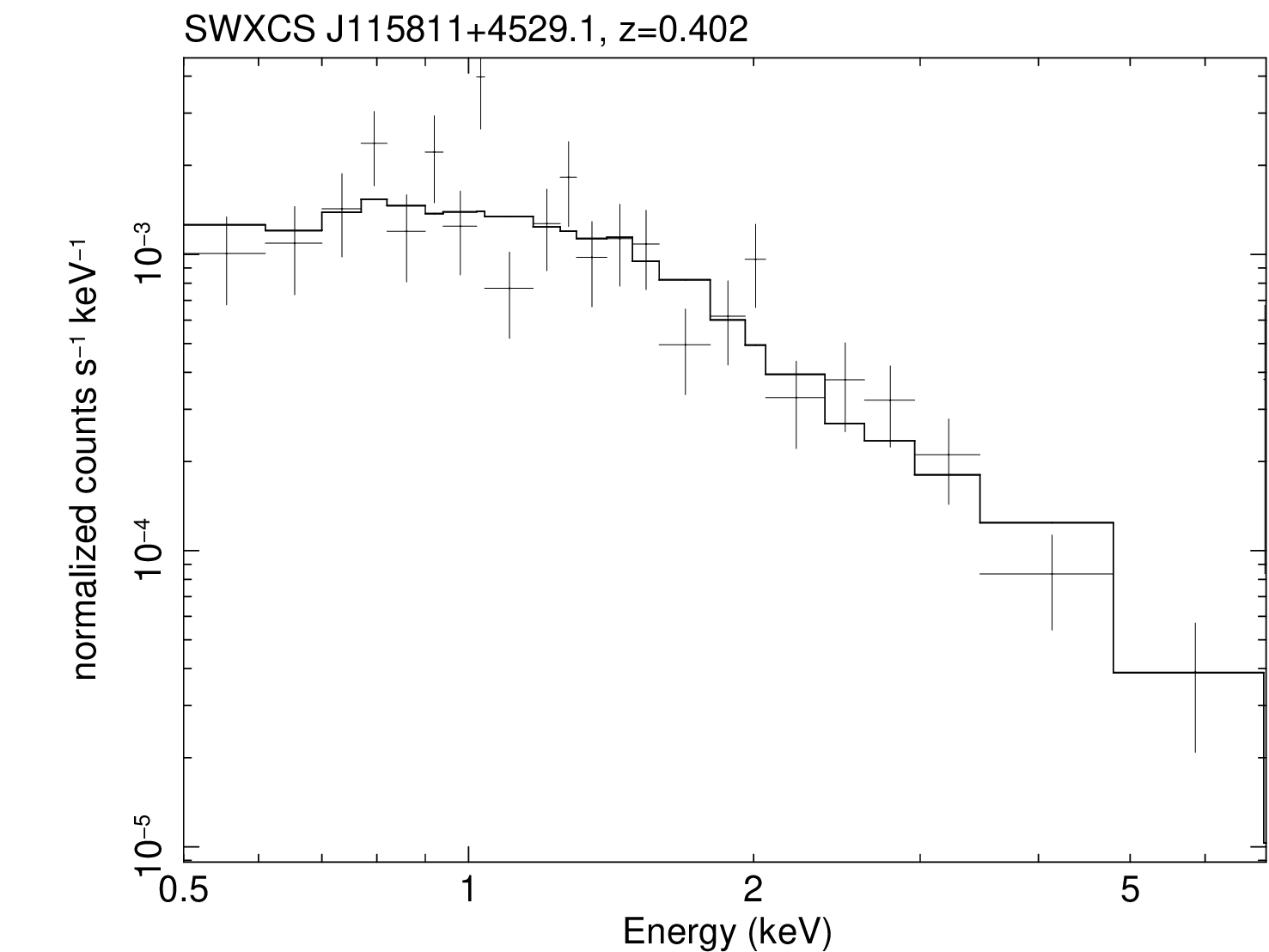}
\caption{continued.}
\end{figure*}

\renewcommand{\thefigure}{\arabic{figure}}

\renewcommand{\thefigure}{A.\arabic{figure}}
\addtocounter{figure}{-1}

\begin{figure*}
\includegraphics[width=7cm,angle=-90]{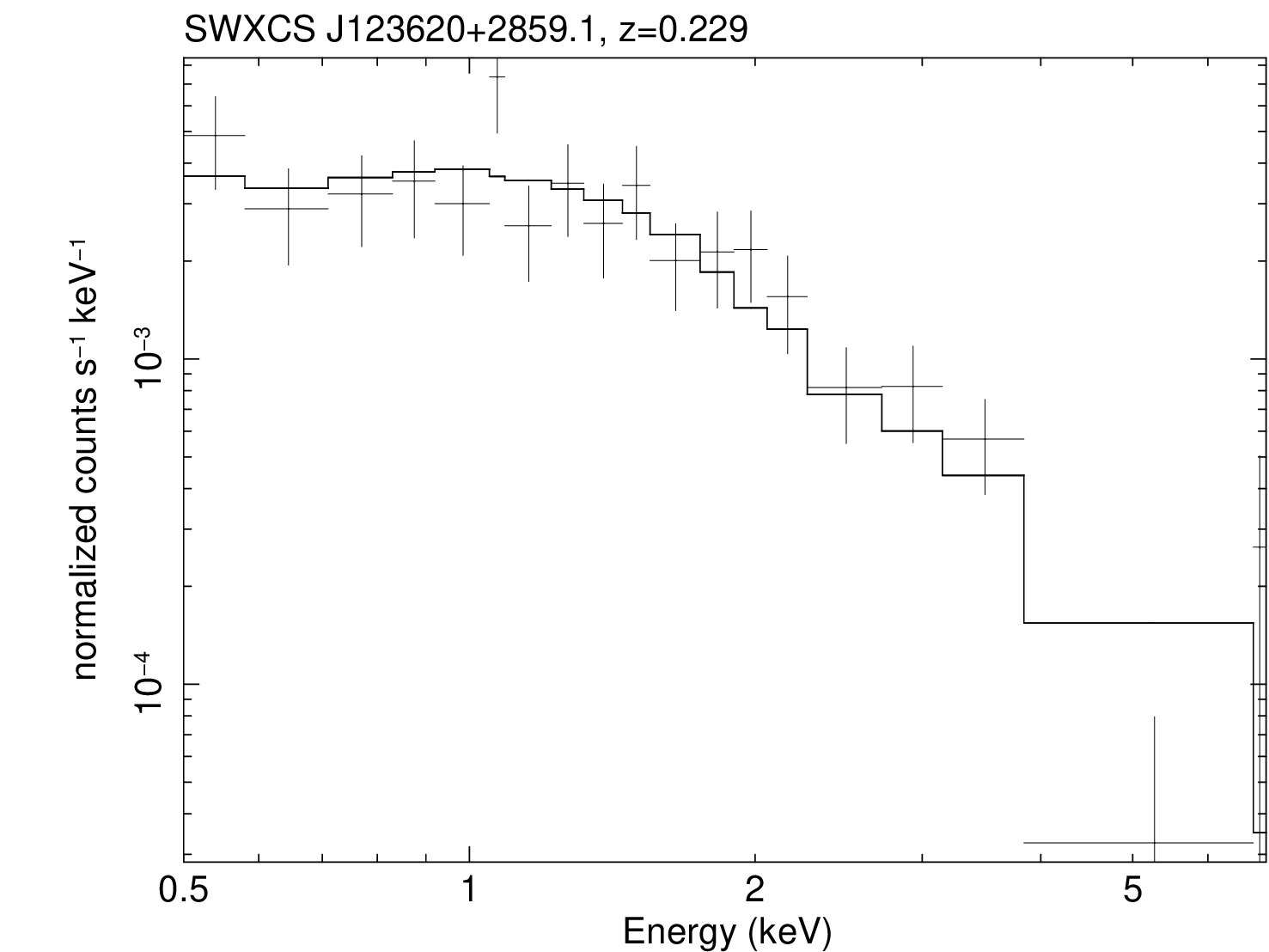}
\includegraphics[width=7cm,angle=-90]{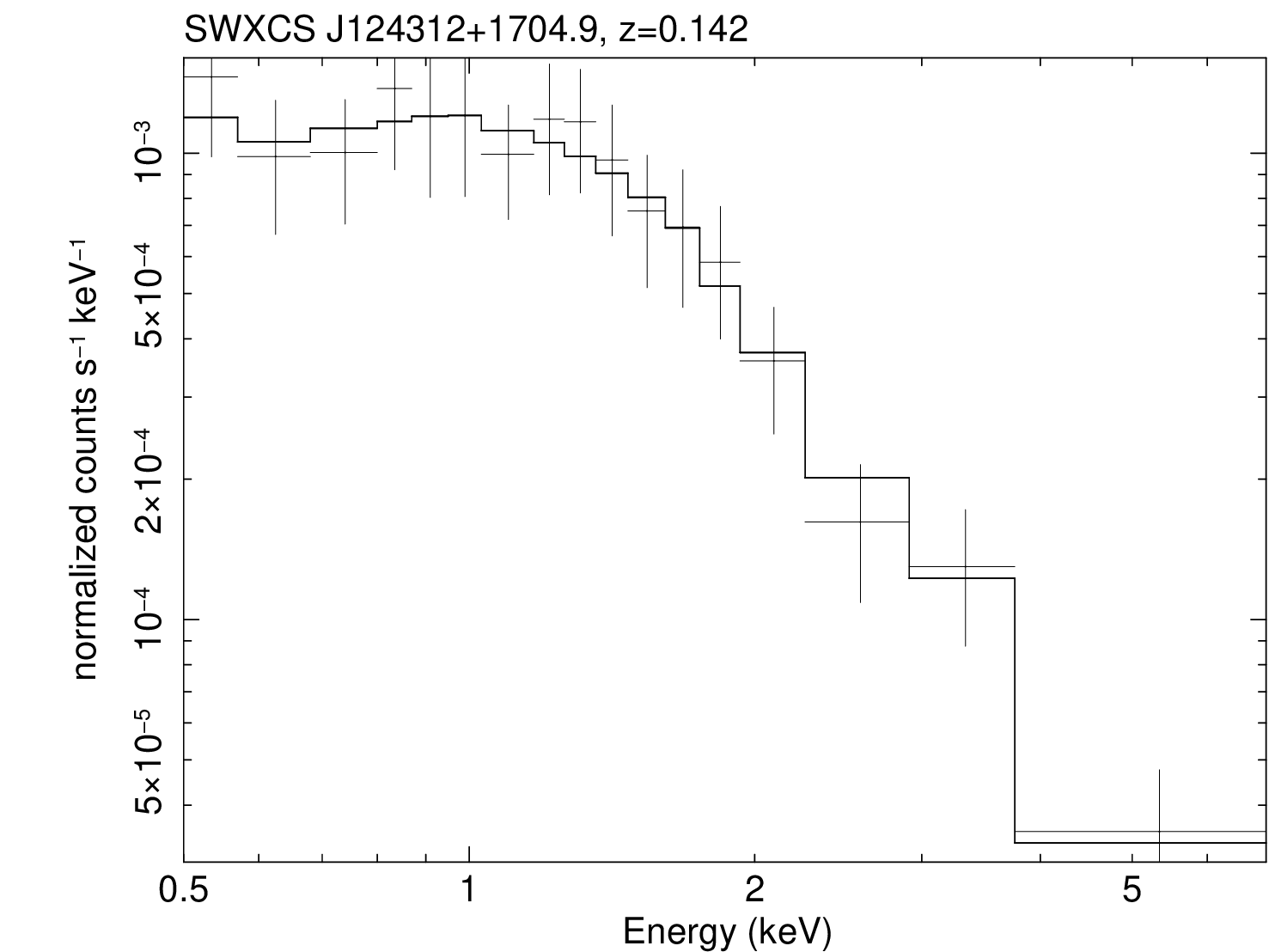}
\includegraphics[width=7cm,angle=-90]{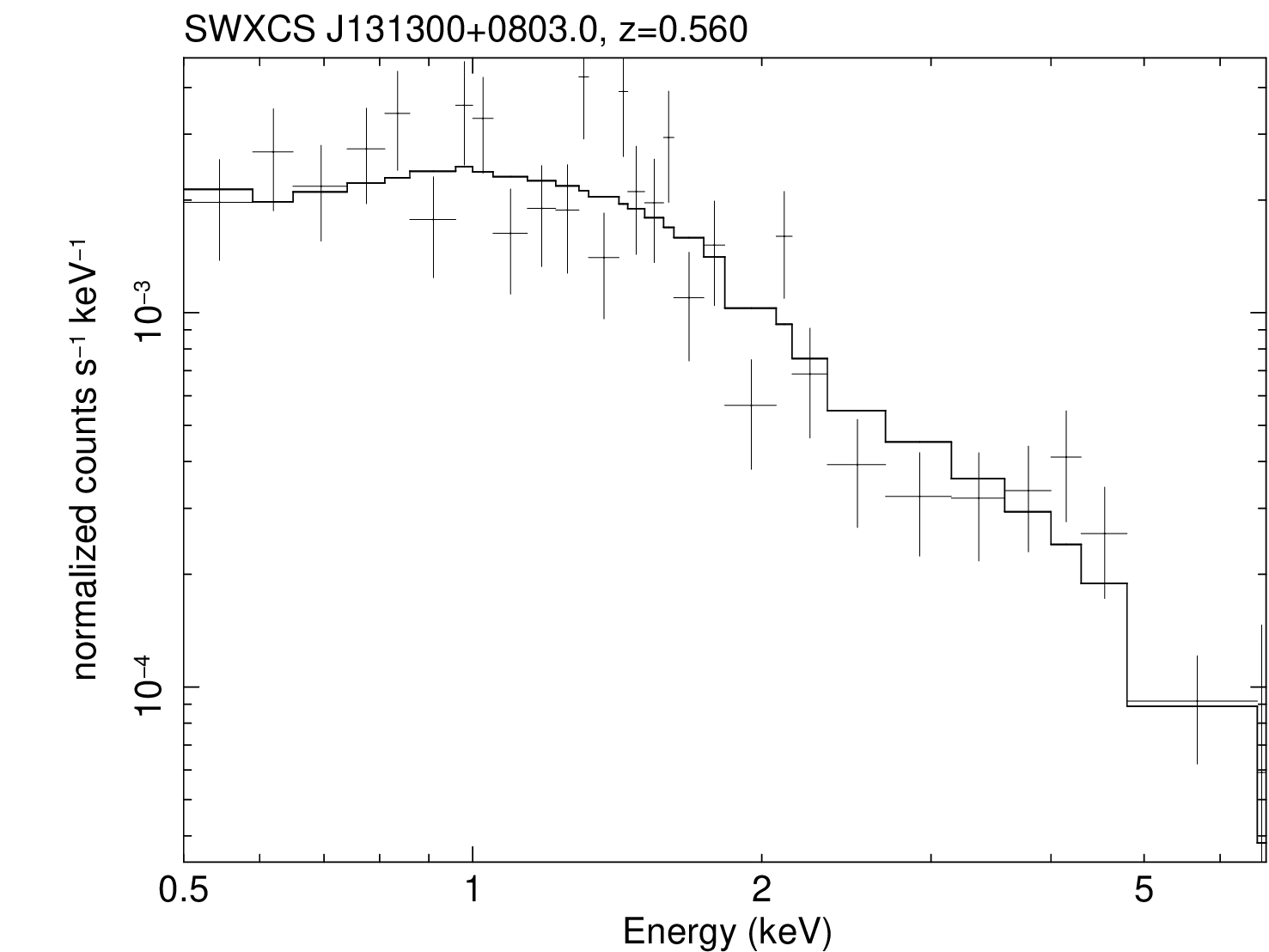}
\includegraphics[width=7cm,angle=-90]{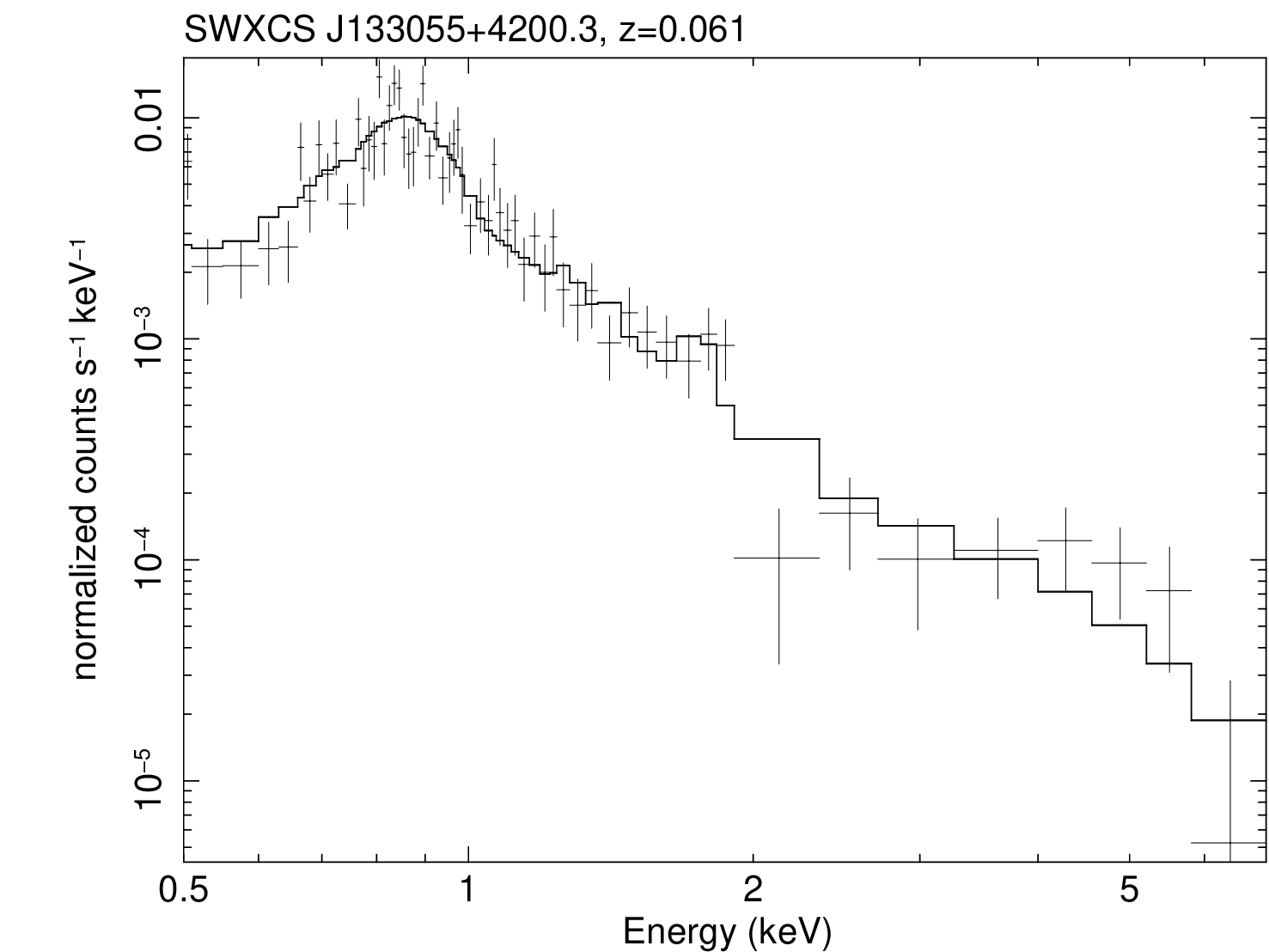}
\includegraphics[width=7cm,angle=-90]{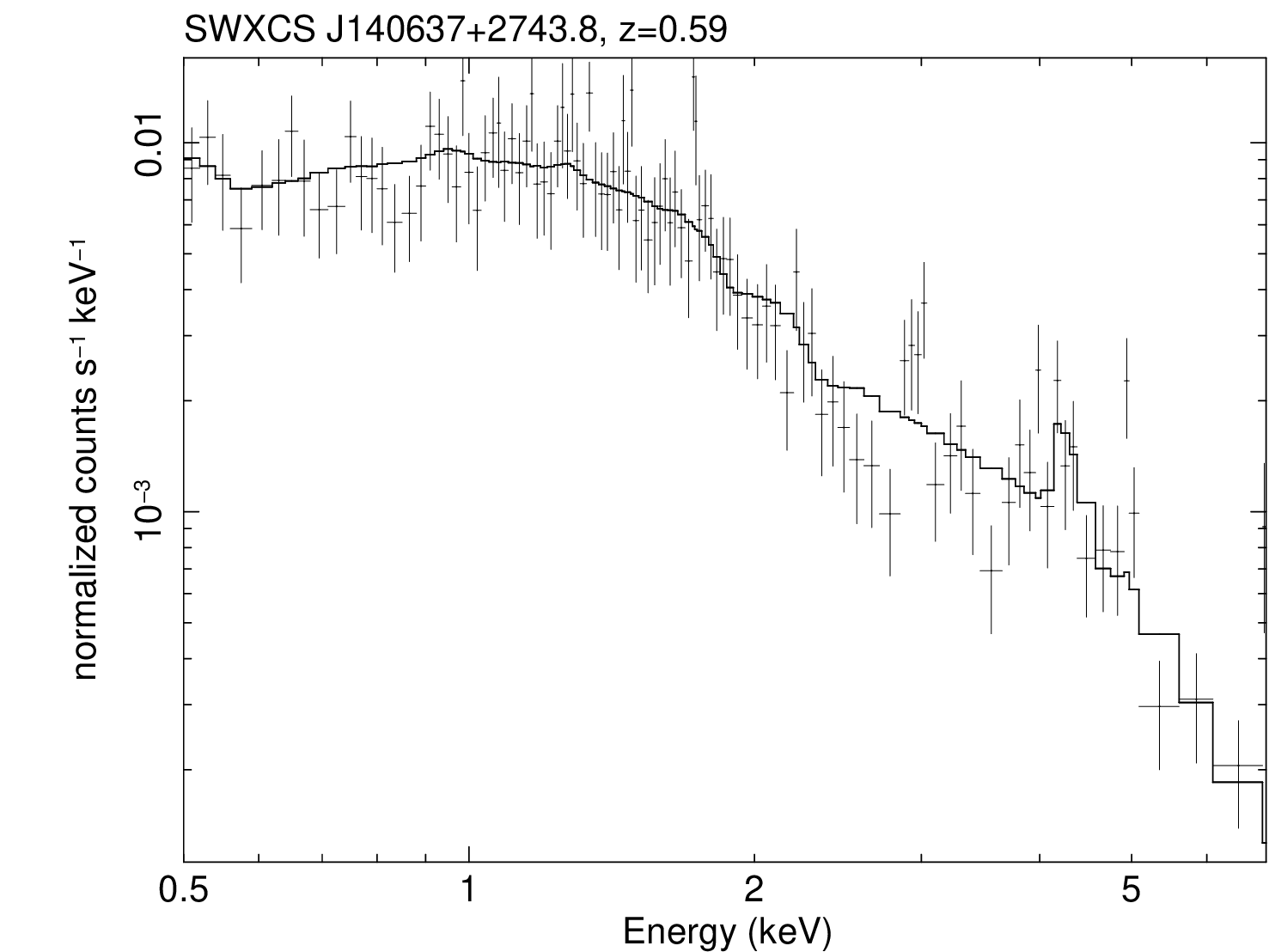}
\includegraphics[width=7cm,angle=-90]{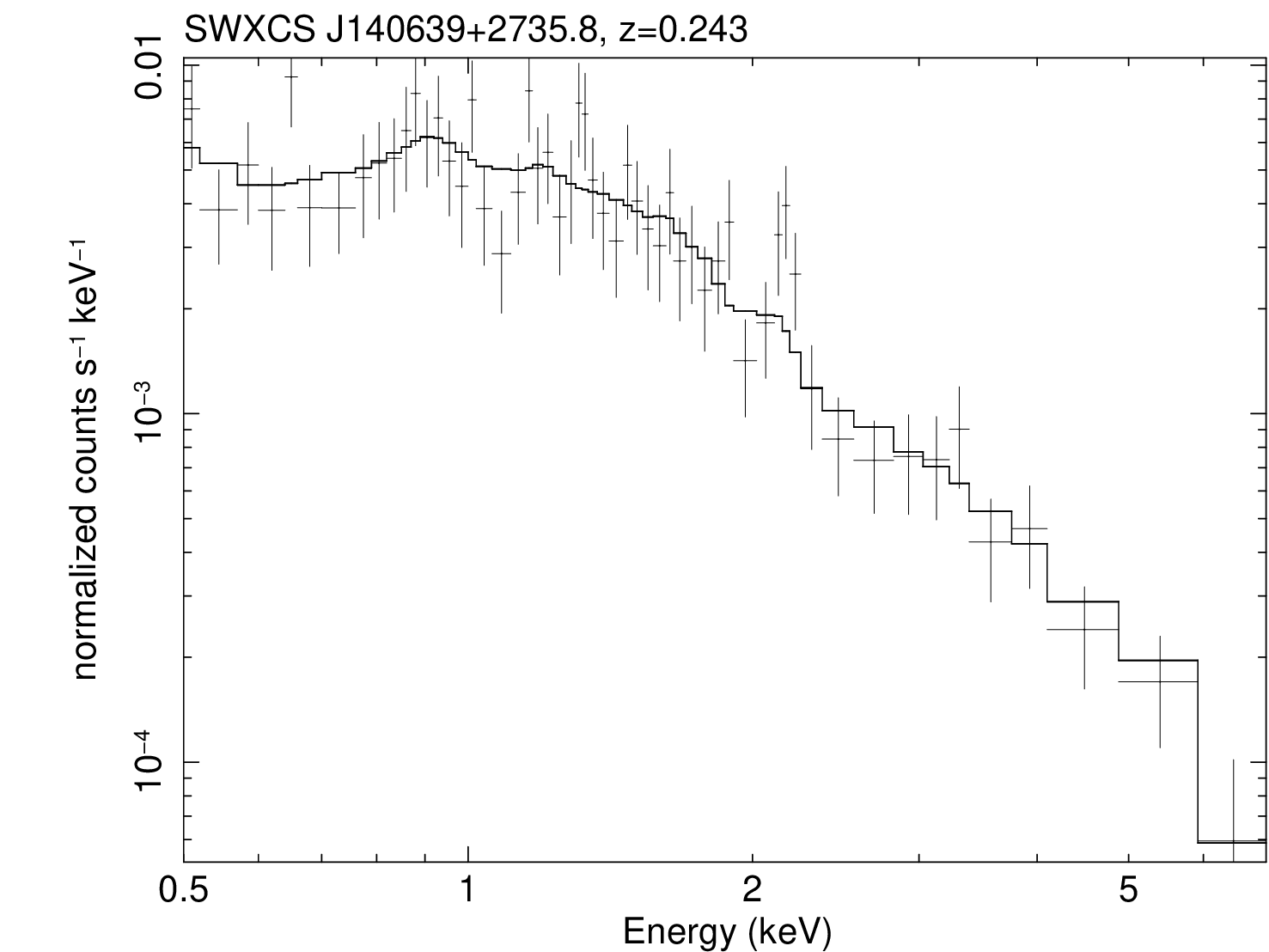}
\caption{continued.}
\end{figure*}

\renewcommand{\thefigure}{\arabic{figure}}

\renewcommand{\thefigure}{A.\arabic{figure}}
\addtocounter{figure}{-1}

\begin{figure*}
\includegraphics[width=7cm,angle=-90]{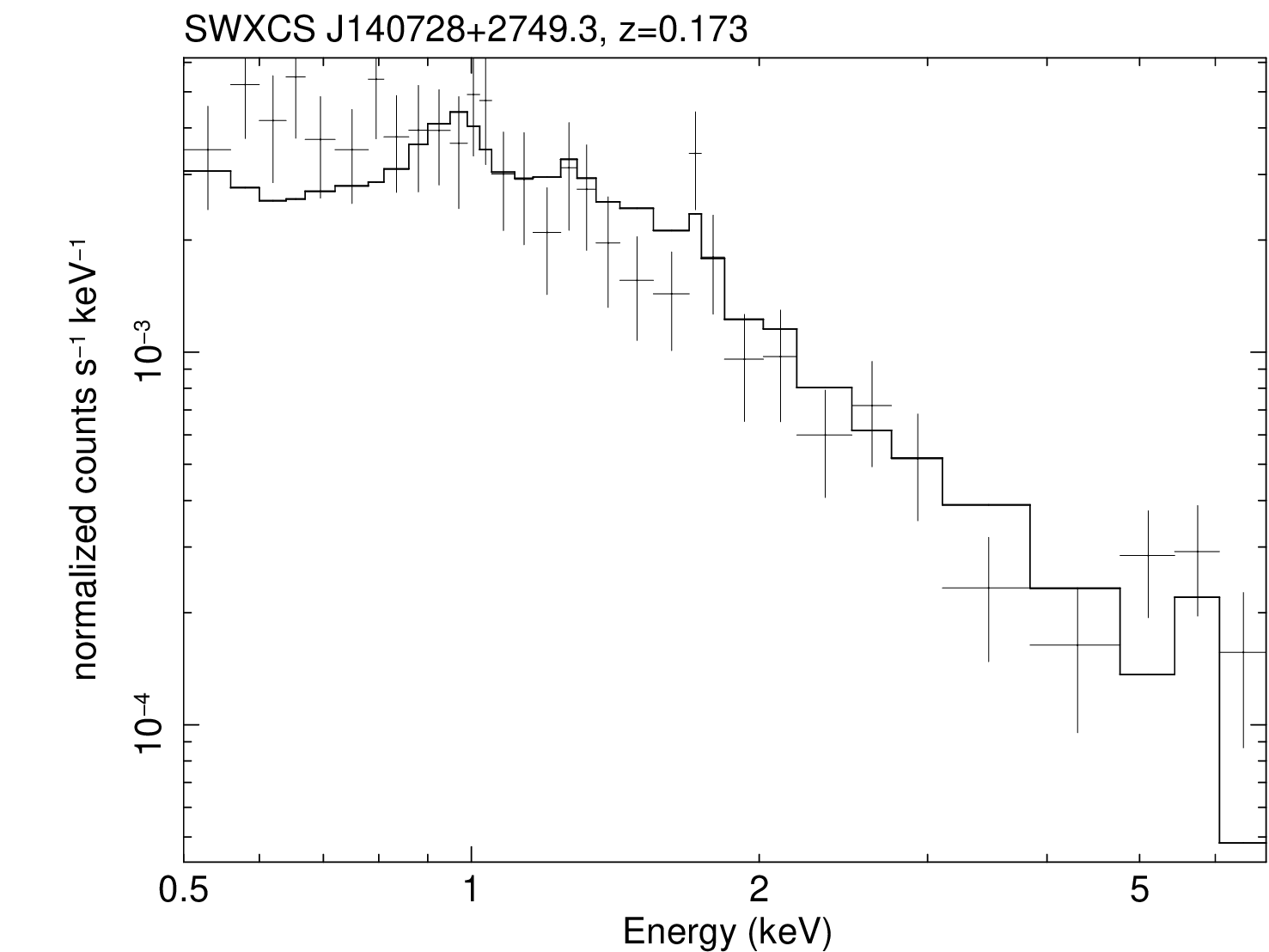}
\includegraphics[width=7cm,angle=-90]{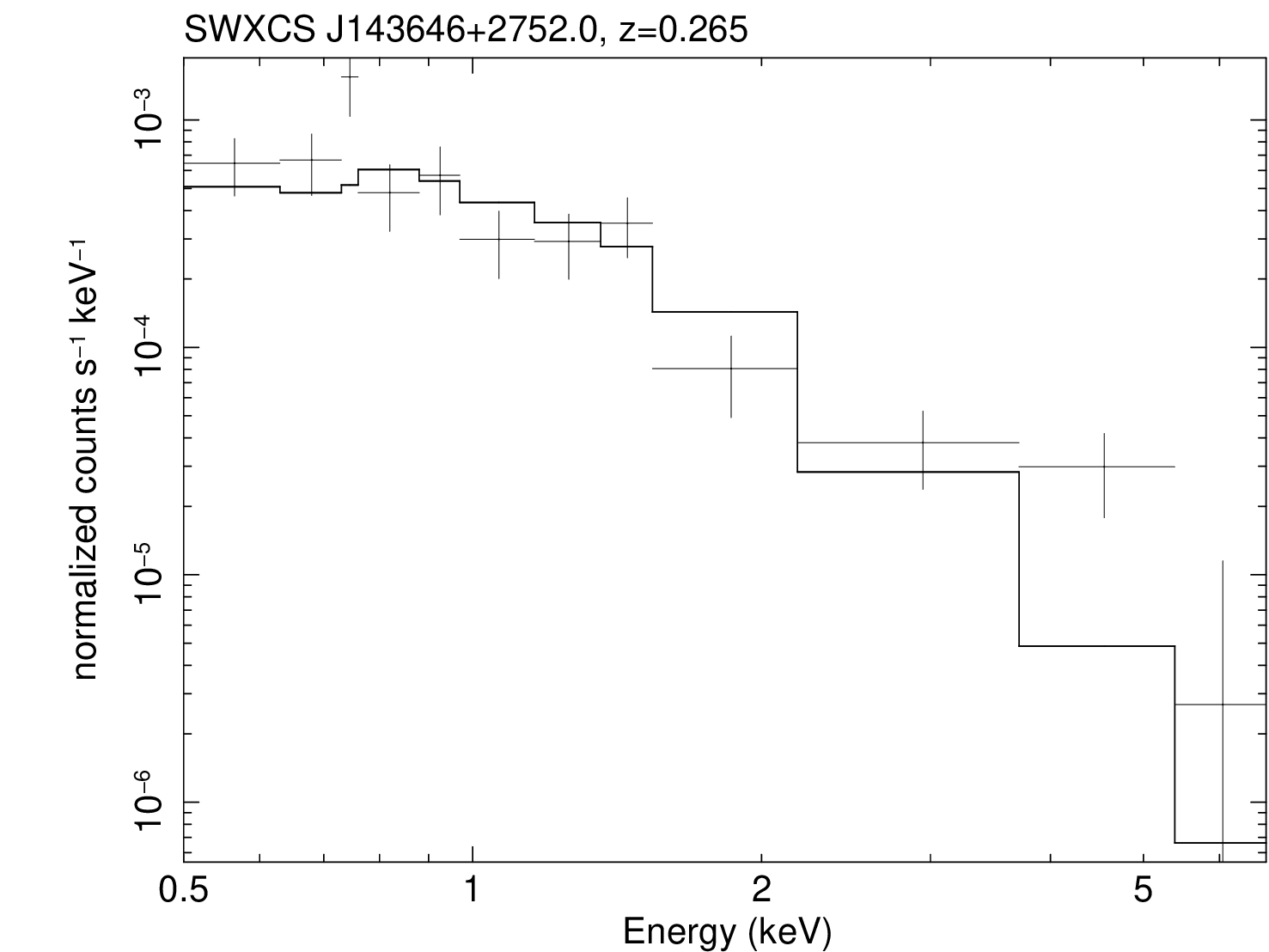}
\includegraphics[width=7cm,angle=-90]{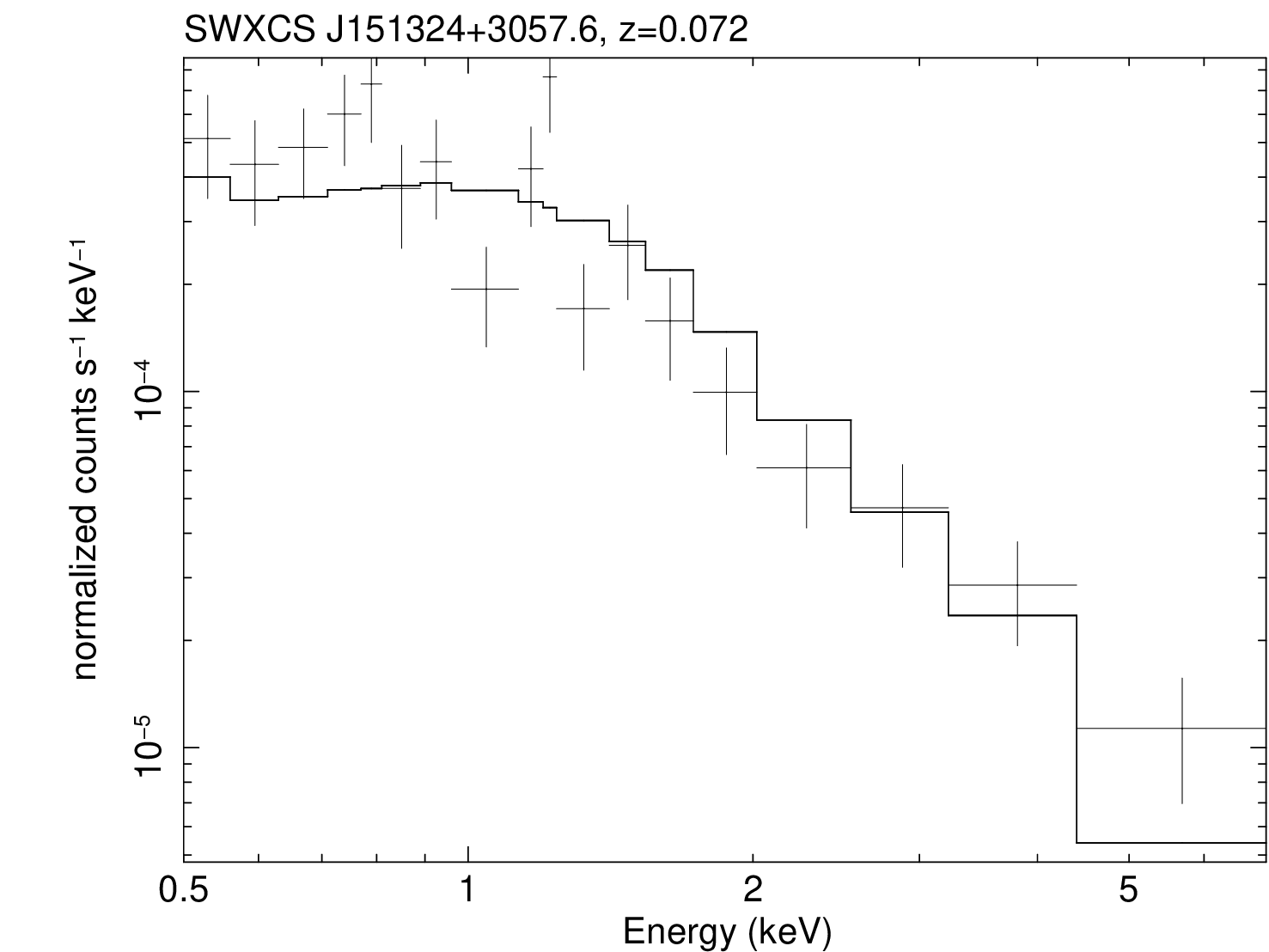}
\includegraphics[width=7cm,angle=-90]{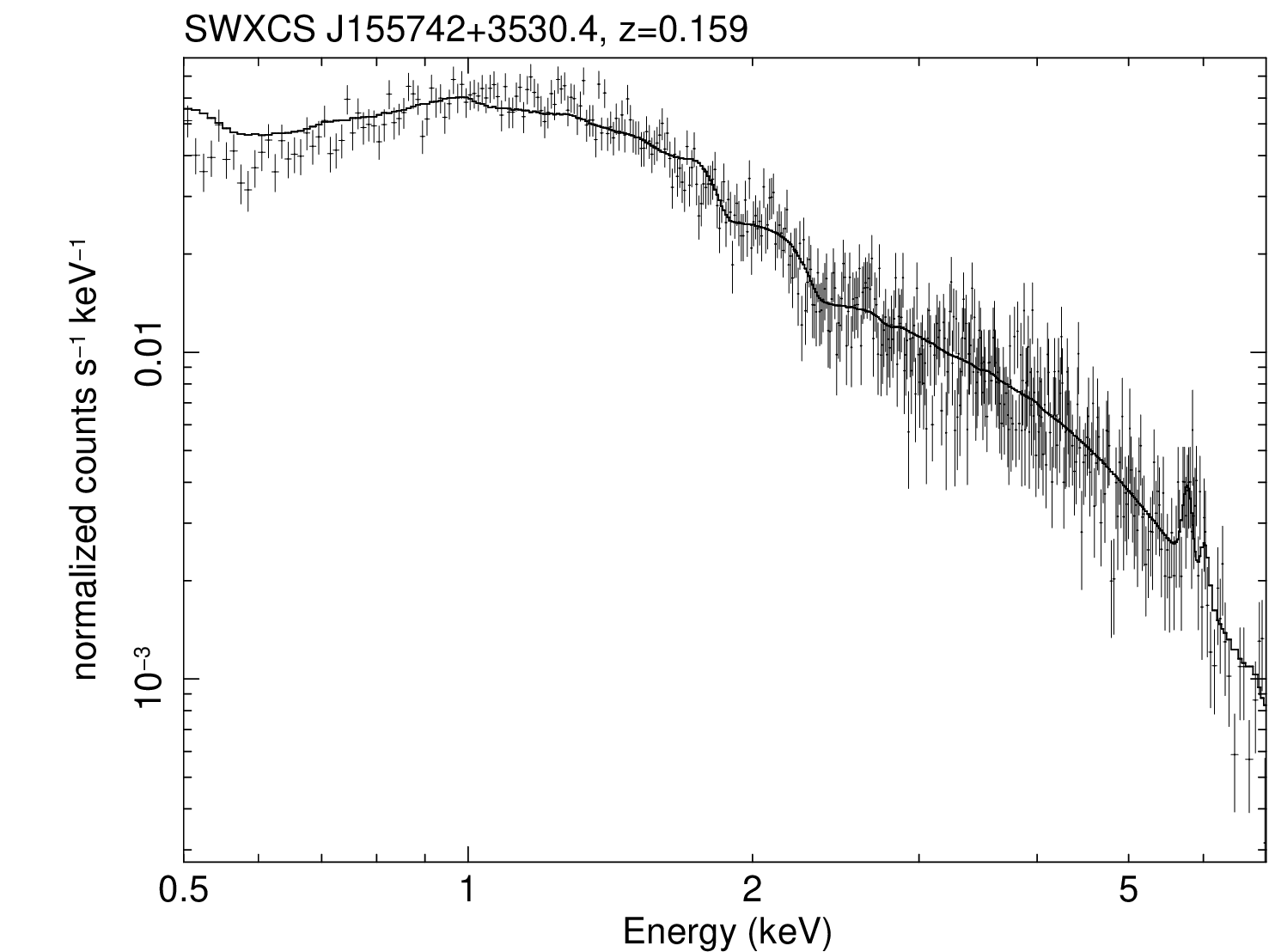}
\includegraphics[width=7cm,angle=-90]{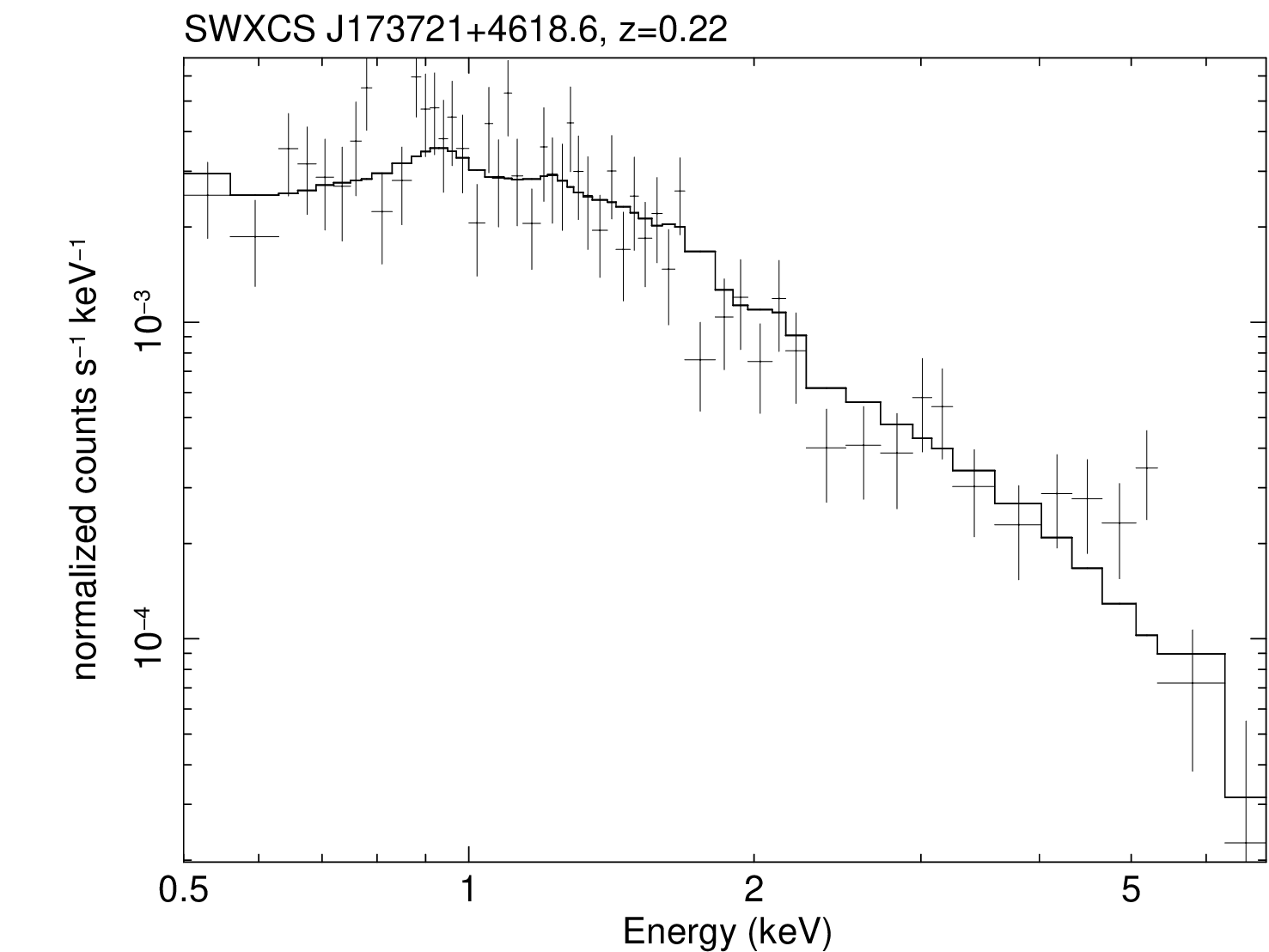}
\includegraphics[width=7cm,angle=-90]{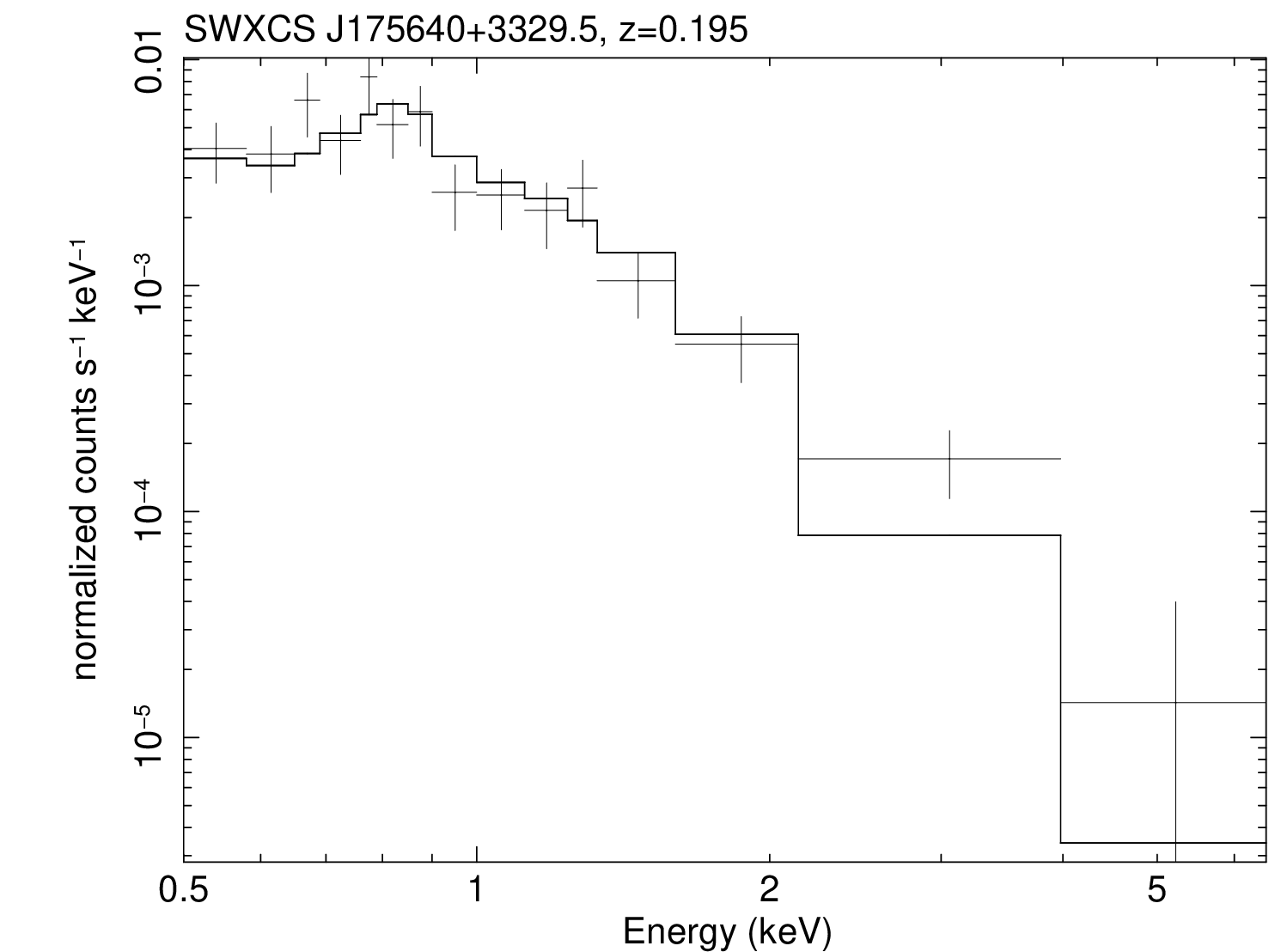}
\caption{continued.}
\end{figure*}

\renewcommand{\thefigure}{\arabic{figure}}

\renewcommand{\thefigure}{A.\arabic{figure}}
\addtocounter{figure}{-1}

\begin{figure*}
\includegraphics[width=7cm,angle=-90]{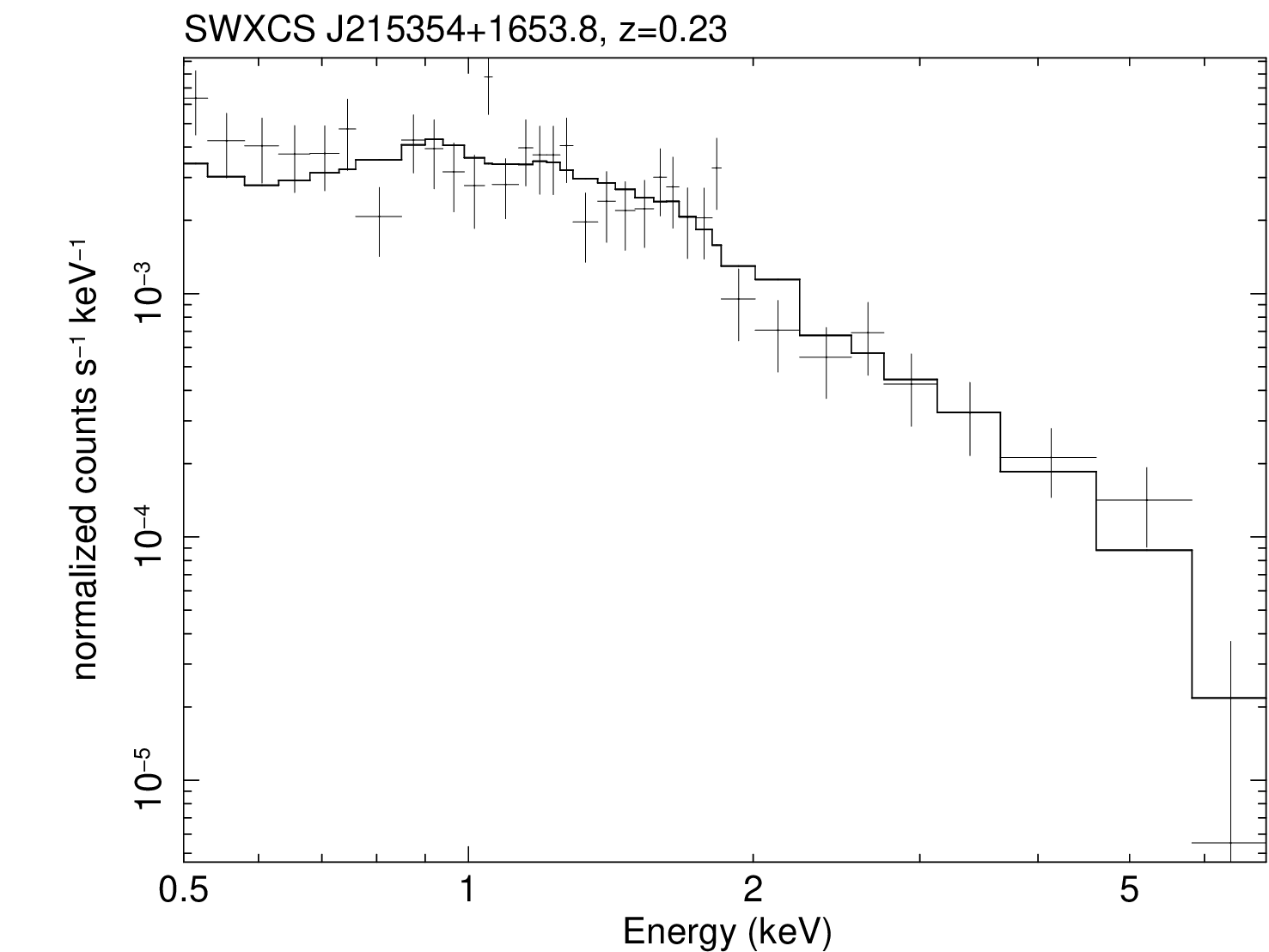}
\includegraphics[width=7cm,angle=-90]{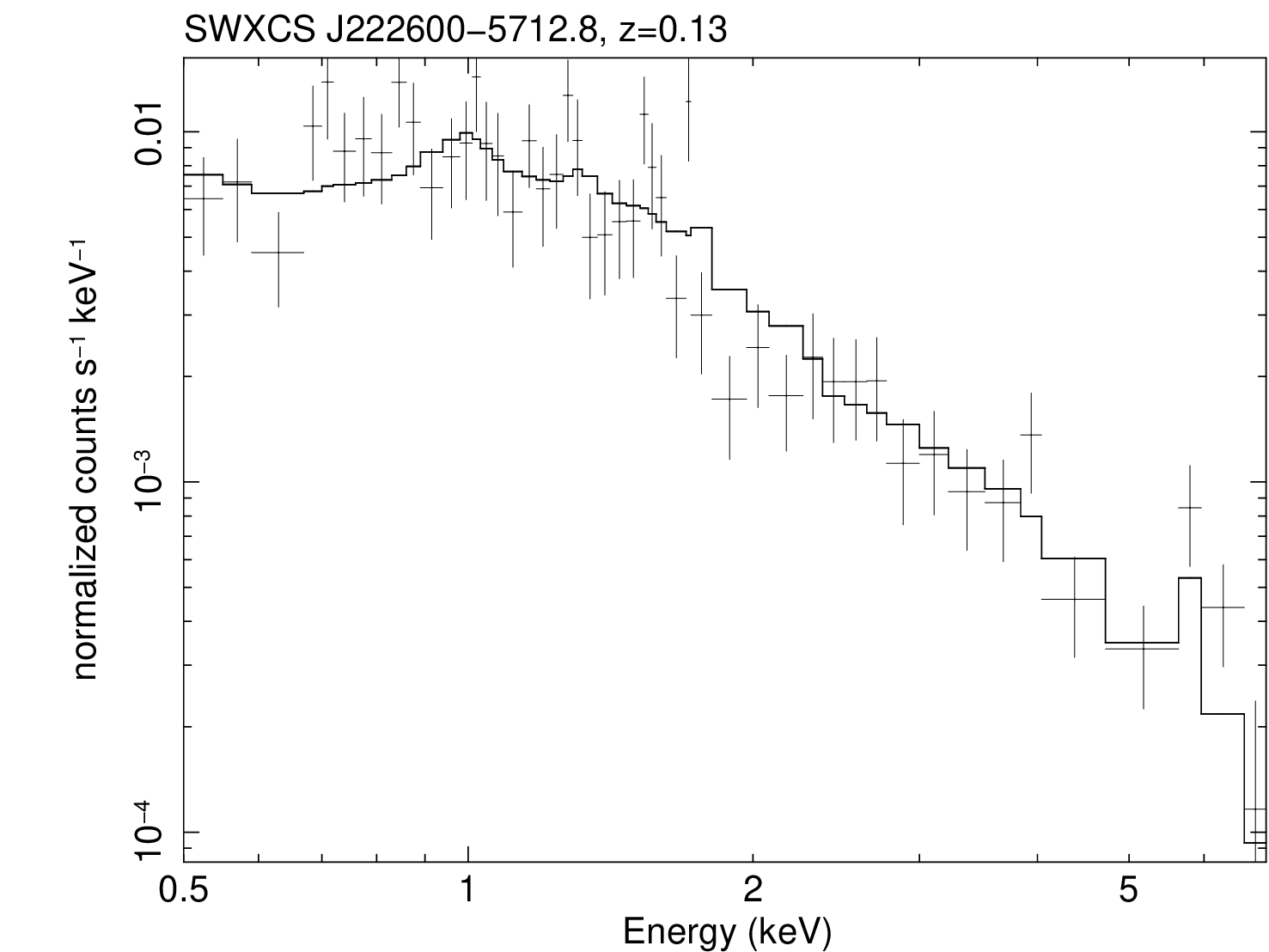}
\includegraphics[width=7cm,angle=-90]{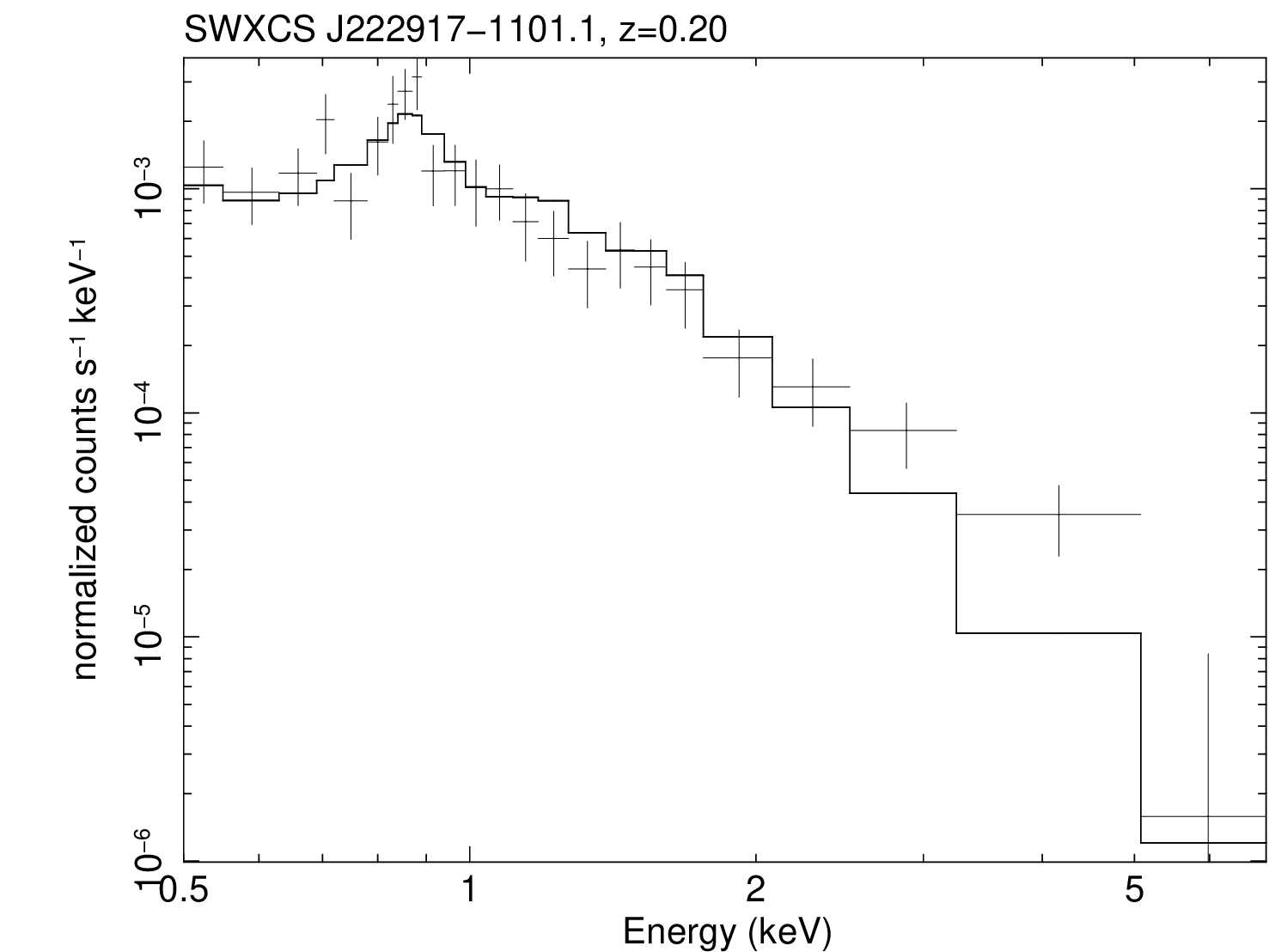}
\includegraphics[width=7cm,angle=-90]{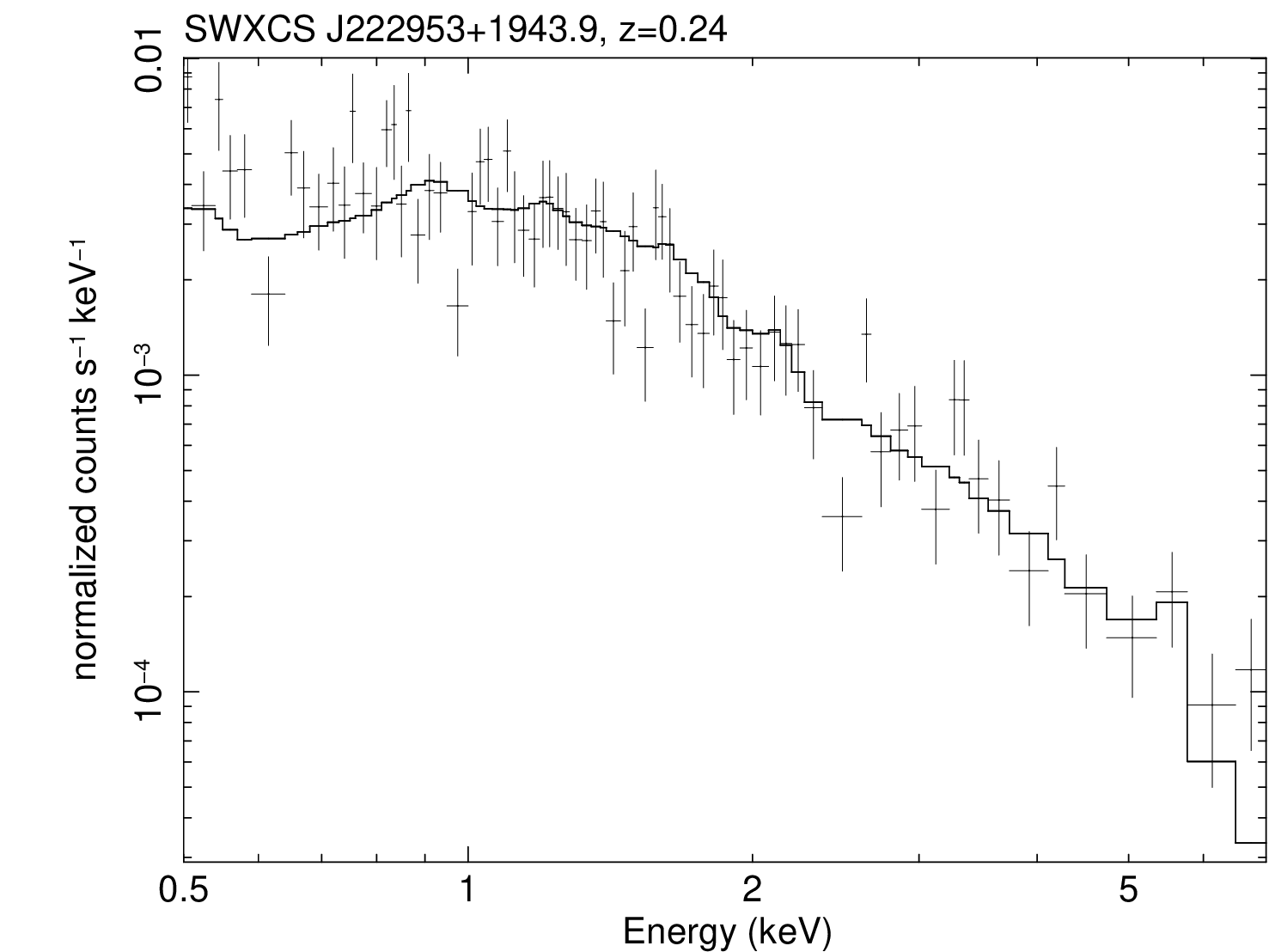}
\includegraphics[width=7cm,angle=-90]{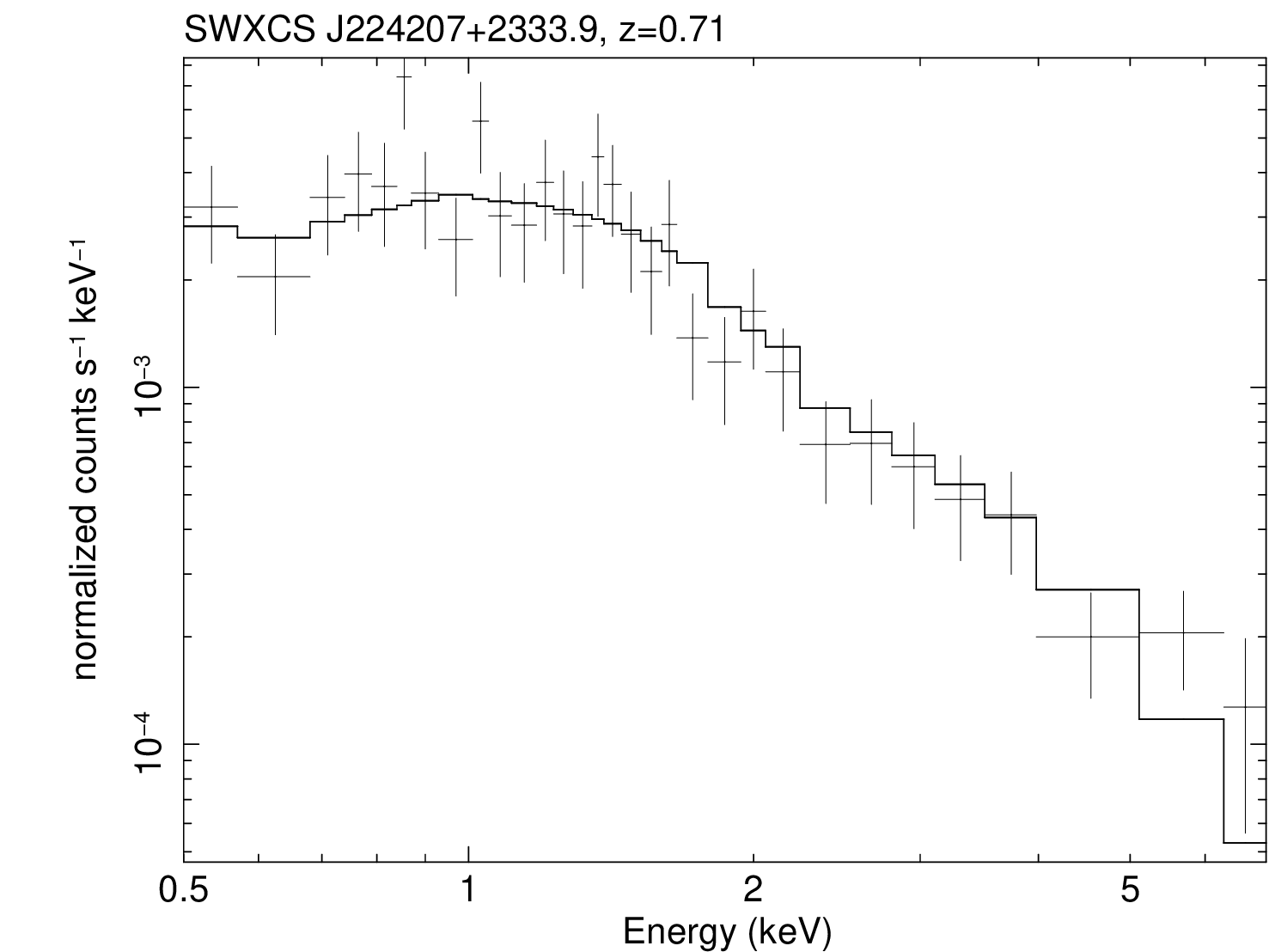}
\includegraphics[width=7cm,angle=-90]{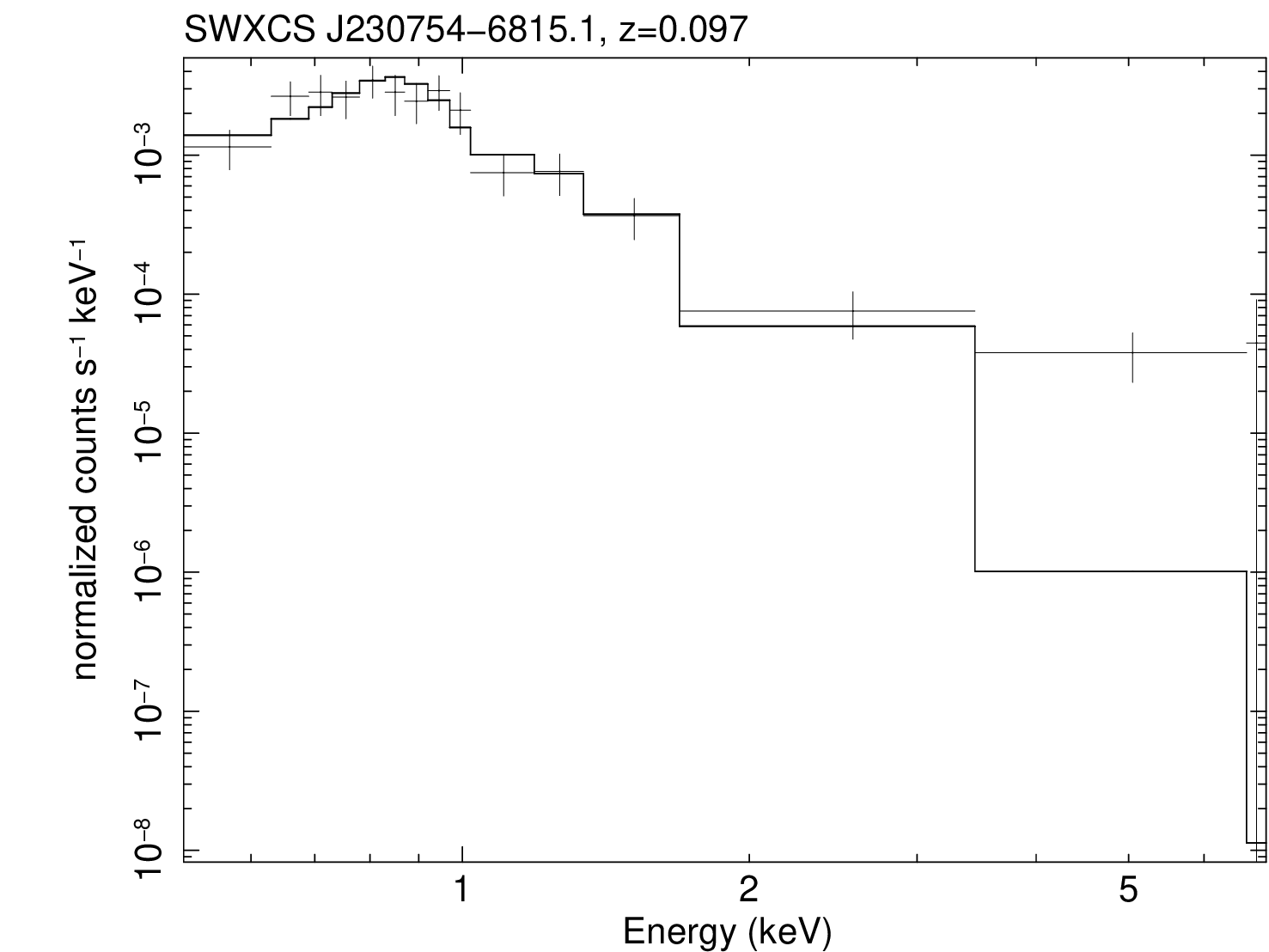}
\caption{continued.}
\end{figure*}

\renewcommand{\thefigure}{\arabic{figure}}

\renewcommand{\thefigure}{A.\arabic{figure}}
\addtocounter{figure}{-1}

\begin{figure*}
\includegraphics[width=7cm,angle=-90]{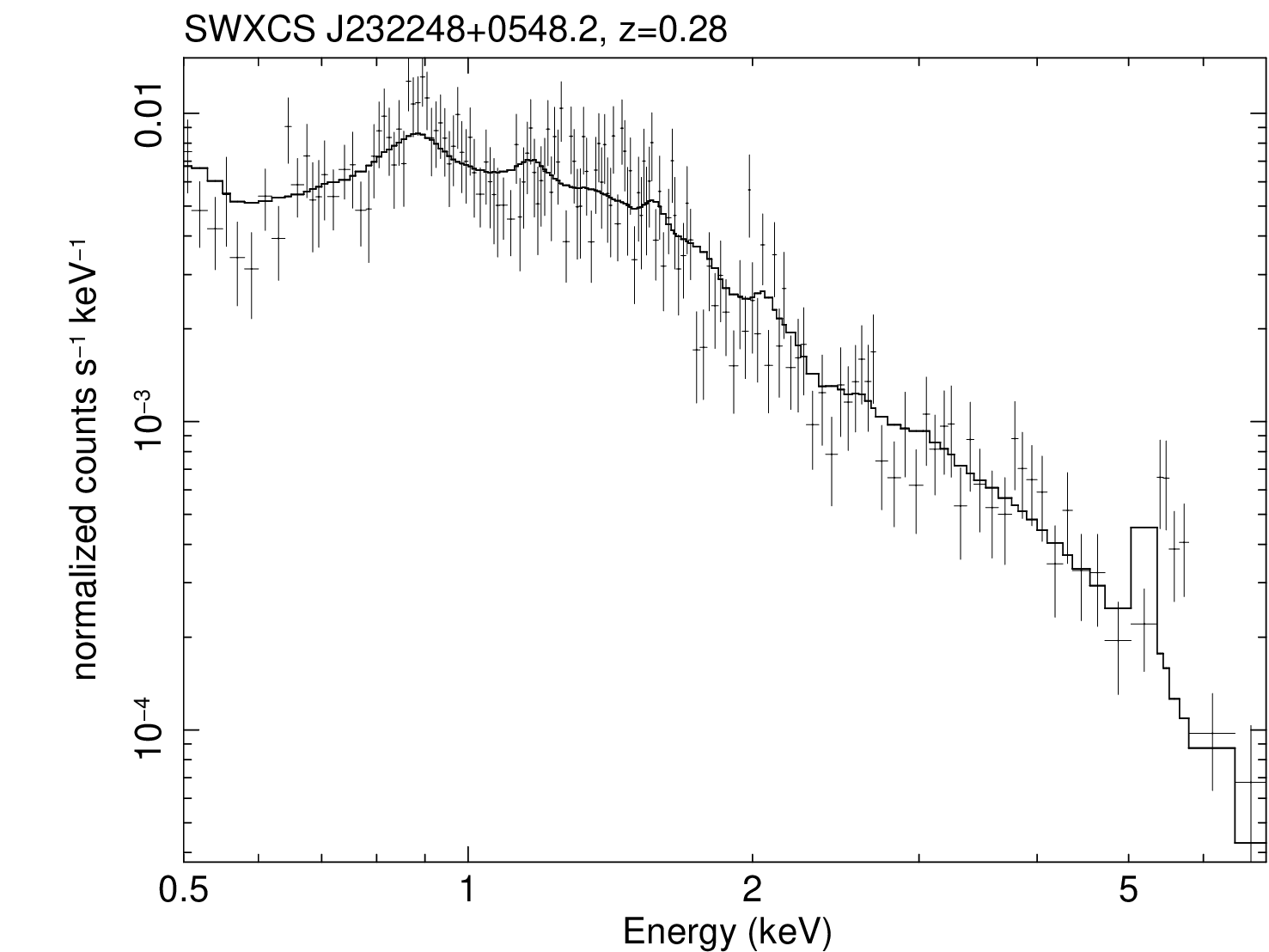}
\includegraphics[width=7cm,angle=-90]{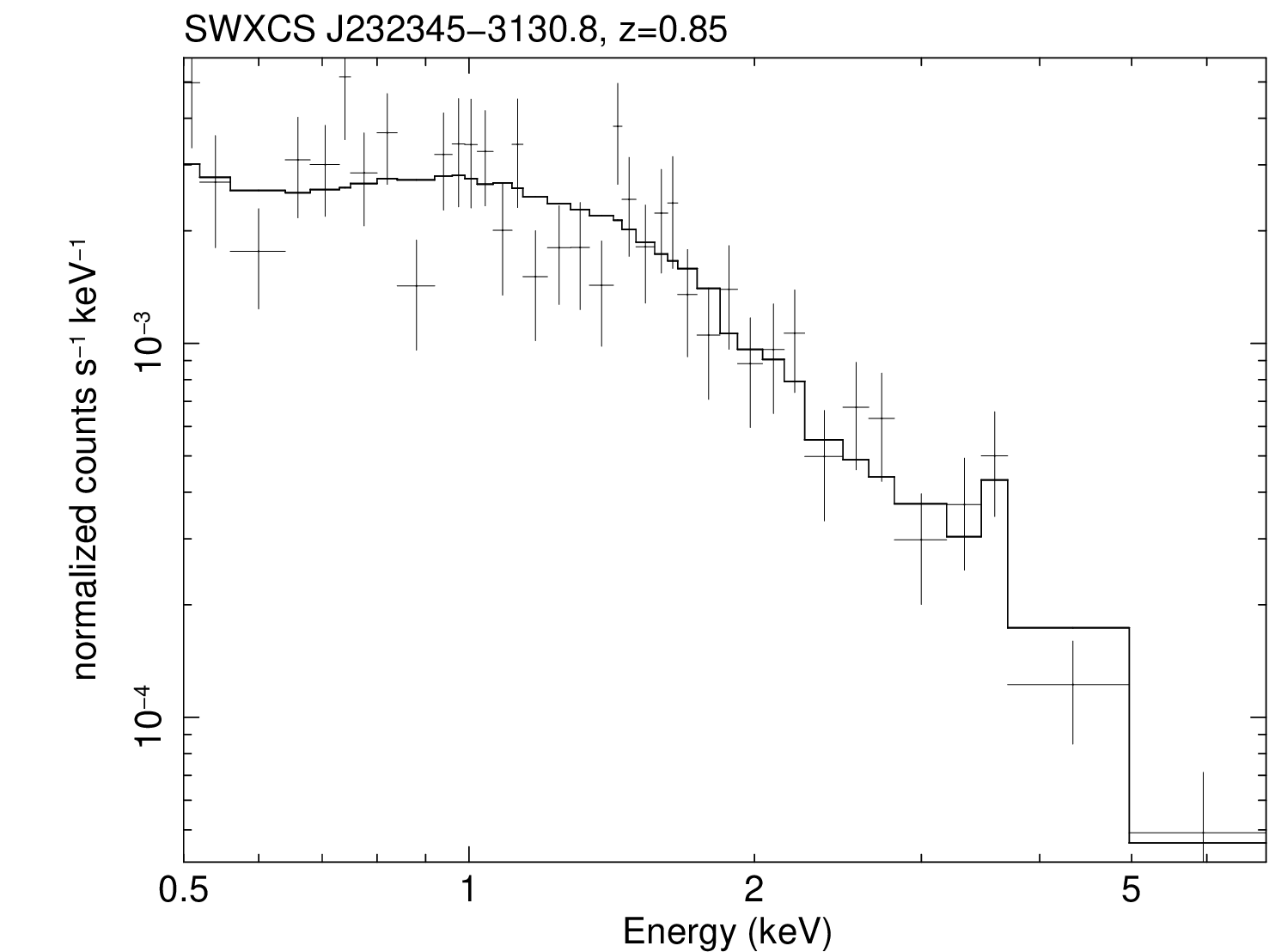}
\includegraphics[width=7cm,angle=-90]{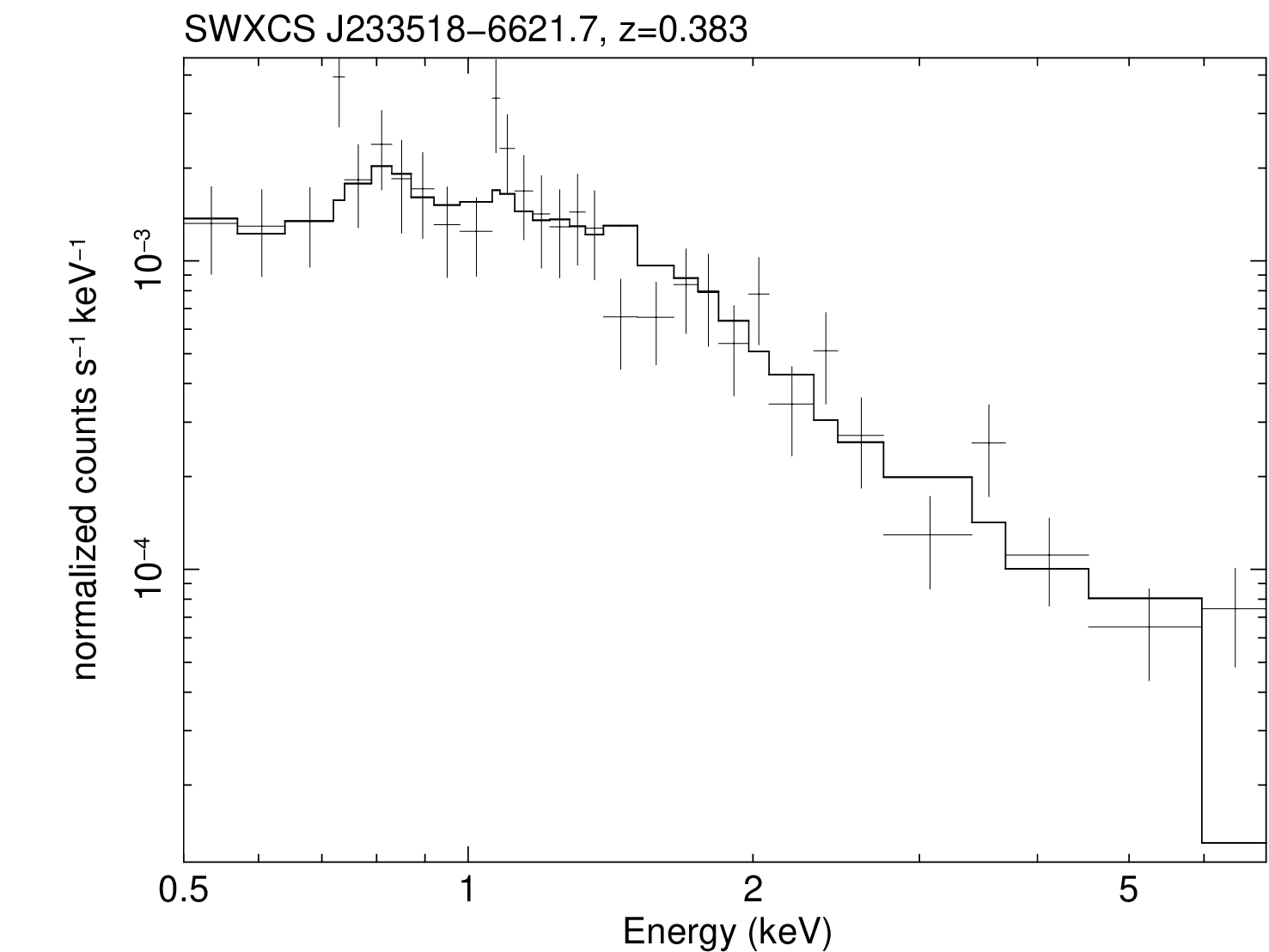}
\includegraphics[width=7cm,angle=-90]{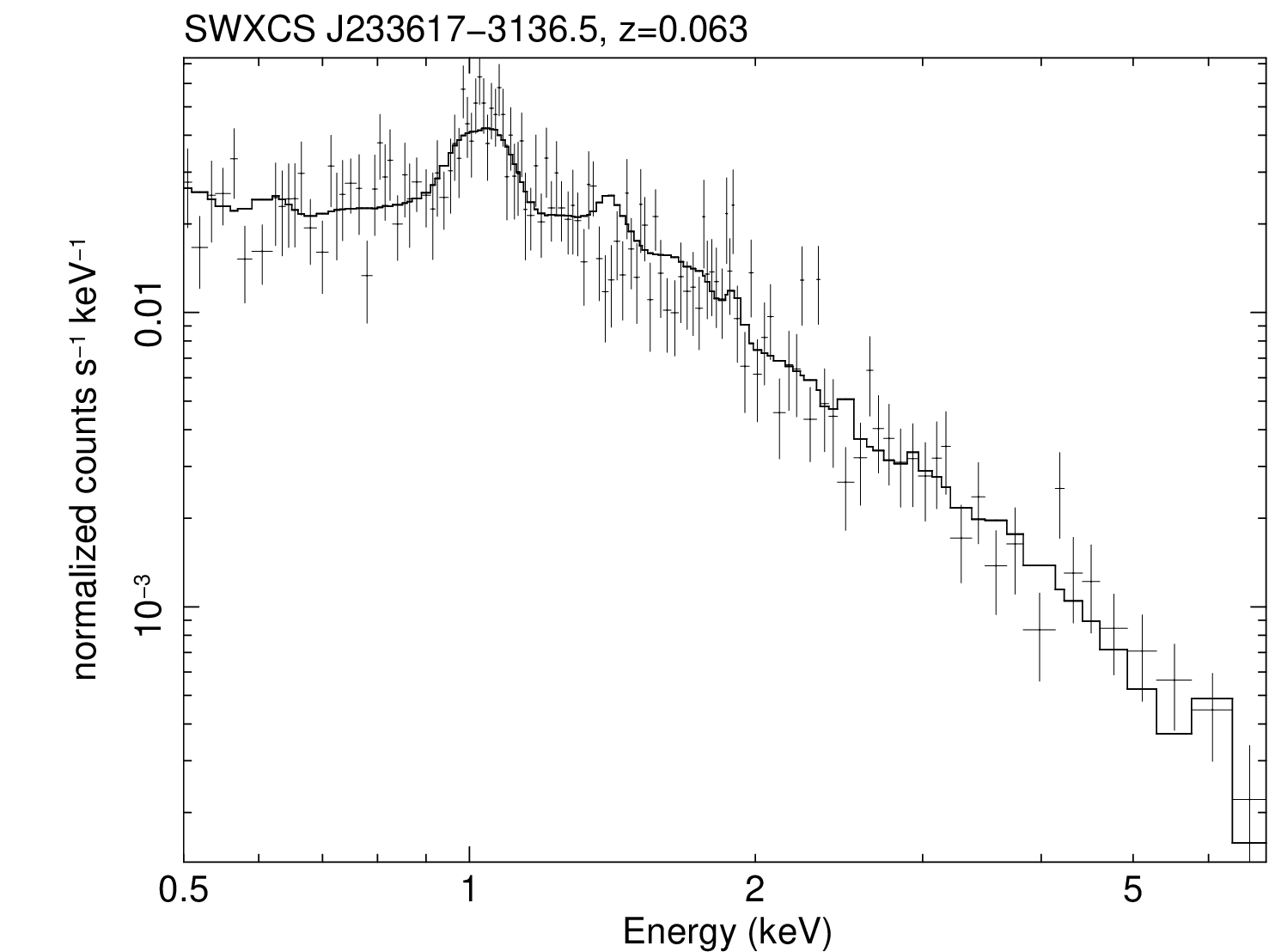}
\caption{continued.}
\end{figure*}

\end{document}